\documentclass[11pt,a4paper]{article}

\usepackage{amssymb,amsmath} 

\usepackage{graphicx}

\usepackage[sort&compress]{natbib}
\bibpunct{[}{]}{,}{n}{}{}

\newcommand{\fet}[1]{\mbox{\boldmath $#1$}}

\newcommand{\beq}{\begin{equation}}
\newcommand{\eeq}{\end{equation}}
\newcommand{\beqa}{\begin{eqnarray}}
\newcommand{\eeqa}{\end{eqnarray}}
\newcommand{\be}{\begin{equation}}
\newcommand{\ee}{\end{equation}}
\newcommand{\bea}{\begin{eqnarray}}
\newcommand{\eea}{\end{eqnarray}}
\newcommand{\nn}{\nonumber \\ }

\def\palka{\hspace{-7.0pt}/ \,}

\begin{document}

\hfill FZJ-IKP-TH-2010-2 

\vskip -0.2 true cm
\hfill HISKP-TH-10/4

\begin{center}
{\huge \bf Nuclear forces from chiral effective field theory\\[4pt]
-- a primer --}

\vskip 1 true cm 

{\large \it Evgeny Epelbaum \\[3pt] Forschungszentrum J\"ulich, Institut f\"ur Kernphysik  (Theorie) and
  J\"ulich Center for Hadron Physics, D-52425 J\"ulich, Germany; \\
Helmholtz-Institut f\"ur Strahlen- und Kernphysik (Theorie) and
Bethe
Center for Theoretical Physics, Universit\"at Bonn,  D-53115 Bonn, Germany}

\end{center}

\vskip 0.7 true cm 
{\bf ABSTRACT}\\[3pt]
This paper is a write-up of introductory lectures on the modern approach 
to the nuclear force problem based on chiral effective field theory 
given at the 2009 Joliot-Curie School, Lacanau, France, 27 September - 3 October 2009.

\vskip 0.8 true cm
\tableofcontents

\section{Introduction}
\def\theequation{\arabic{section}.\arabic{equation}}
\label{sec1}

One of the oldest but still actual problems in nuclear physics is related to 
the determination of the interaction between the nucleons. A quantitative understanding 
of the nuclear force is crucial in order to describe the properties of nuclei and 
nuclear matter in terms of hadronic degrees of freedom. The conventional way to 
parametrize the nuclear force utilizes the meson-exchange picture,
which goes back to the seminal work by Yukawa \cite{Yukawa:1935aa}. 
His idea, followed by the experimental discovery of $\pi$- and heavier mesons 
($\rho$, $\omega$, $\dots$), stimulated the development of boson-exchange models 
which still provide a basis for many modern, highly sophisticated phenomenological 
nucleon-nucleon (NN) potentials.

According to our present understanding, the nuclear force is due to residual 
strong interactions between the color-charge neutral hadrons. A direct derivation
of the nuclear force from QCD, the underlying theory of strong interactions, is not 
yet possible, see however Ref.~\cite{Ishii:2006ec} for a recent attempt using lattice QCD. 
In order to provide reliable input for few- and many-body calculations, 
a  (semi-)phenomenological approach has been followed over the past few decades aiming
to achieve the best possible description of the available low-energy NN data.
As will be discussed in section \ref{sec2}, the two-nucleon potential 
can be decomposed in only few different spin-space structures, so that  
the corresponding radial functions can be parameterized using an extensive set of data. 
Although the resulting models provide an excellent description of experimental 
data in many cases, there are certain major conceptual deficiencies that cannot 
be overcome. In particular, one important concern is related to the problem 
of the construction of \emph{consistent} many-body forces. These can 
only be meaningfully defined in a consistent scheme with a given two-nucleon  
interaction \cite{Gloeckle:1990aa}.
Notice that because of the large variety of different possible structures in the 
three-nucleon force, following the 
same phenomenological path as in the NN system and parametrizing its  most general structure 
seems not to be feasible without additional theoretical guidance.   
Clearly, the same problem of consistency arises in the context of reactions
with electroweak probes, whose description requires the knowledge of the corresponding 
consistent nuclear current operator. 
Further, one lacks within phenomenological treatments a method of systematically improving 
the theory of the nuclear force in terms of the dominant dynamical contributions. 
Finally, and most important, the phenomenological scheme 
provides only a loose connection to QCD.

Chiral perturbation theory (ChPT) is an effective field theory (EFT) of QCD which  
exploits its symmetries and symmetry-breaking pattern and allows to analyze the 
properties of hadronic systems at low energies in a systematic and model independent way. 
We will see in section \ref{sec3} that QCD with two flavors of the $u$- and $d$-quarks 
and, to a less extent, with three 
flavors of the $u$-, $d$- and $s$-quarks, exhibits an approximate chiral symmetry
which is explicitly broken due to non-vanishing (but small) quark masses. In
addition, the chiral symmetry is also spontaneously broken down to its vector
subgroup. These symmetry/symmetry-breaking pattern manifest themselves in the
hadron spectrum leading, in particular, to a natural explanation of the very
small masses  (compared to other hadrons) of pions which play the role of the 
corresponding Goldstone bosons. Most important, the Goldstone boson nature of the
pions implies that they interact weakly at low energy and allows to calculate
low-energy observables in the pion and single-nucleon sector in perturbation
theory. The situation in the few-nucleon sector is conceptually much  
more complicated due to the strong  nature of the nuclear force which
manifests itself in the appearance of self-bound atomic nuclei and invalidates
a naive application of perturbation theory. As pointed out by Weinberg, the
breakdown of perturbation theory in the few-nucleon sector can be traced back
to the infrared enhancement of reducible time-ordered diagrams which involve purely
nucleonic intermediate states and can be resummed by iterating the corresponding 
dynamical equation \cite{Weinberg:1990rz,Weinberg:1991um}.
These important observation made in Weinberg's seminal papers  
opened a new era in nuclear physics and
has triggered an intense research activity along these lines. In these lectures I will 
outline the basic concepts of chiral effective field theory and its application to 
nucleon-nucleon scattering and the derivation of the nuclear force.  

The manuscript is organized as follows. In section \ref{sec2} I discuss 
the general structure of the nuclear force and outline the main ingredients of the 
conventional NN potentials. Section \ref{sec3} provides an elementary introduction 
to chiral perturbation theory. Generalization of EFT to strongly interacting nuclear 
systems is discussed in section \ref{sec4}. Derivation of the nuclear forces in chiral 
EFT  is outlined in section \ref{sec5}. 
A brief summary is given in section \ref{sec6}.

\section{Nuclear potentials and nucleon-nucleon scattering}
\def\theequation{\arabic{section}.\arabic{equation}}
\label{sec2}

The most general structure of a non-relativistic two-nucleon potential is expressible 
in terms of just a few operators. The potential can be viewed as an operator 
acting in the position, spin and isospin spaces of the nucleons. It is instructive to 
discuss its isospin structure separately from the operators acting in the position-spin space.  

The isospin structure of the two-nucleon force falls into the four different classes 
according to the classification of Ref.~\cite{Henley:1979aa}:
\beq
\label{2NF_classes}
\begin{array}{lcl}
\mbox{Class I:} & \mbox{\hskip 1 true cm} & V_{\rm I} = \alpha_{\rm I}  + \beta_{\rm I} \,\fet \tau_1 \cdot \fet \tau_2 \,,\\ [0.6ex]
\mbox{Class II:} &                        & V_{\rm II} = \alpha_{\rm II}  \, \tau_1^3 \, \tau_2^3  \,,\\[0.6ex]
\mbox{Class III:} &                        & V_{\rm III} = \alpha_{\rm III}  \,( \tau_1^3 + \tau_2^3)  \,,\\[0.6ex]
\mbox{Class IV:} &                        & V_{\rm IV} = \alpha_{\rm IV}  \, (\tau_1^3 - \tau_2^3) + \beta_{\rm IV} \, [\fet \tau_1 \times \fet \tau_2]^3  \,. 
\end{array}
\eeq
Here, $\alpha_{\rm i}$, $\beta_{\rm i}$ are position-spin operators and $\fet \tau_i$ are 
Pauli isospin matrices of a nucleon $i$.  
The operator $\beta_{\rm IV}$ has to be odd under a time reversal transformation.
While class (I) forces are isospin-invariant, all other classes (II), (III) and (IV) are isospin-breaking. 
Class (II) forces, $V_{\rm II}$, maintain charge symmetry but break charge independence. They are usually referred 
to as charge independence breaking (CIB) forces. Charge symmetry represents invariance under reflection about the 
1-2 plane in charge space.
The charge symmetry operator $P_{cs}$ transforms proton and neutron states into each other and 
is given by $P_{cs} = e^{i \pi T_2}$ with $\fet T \equiv \sum_i \fet \tau_i/2$ being the total isospin operator. 
Class (III) forces break charge symmetry but do  not lead to isospin mixing in the NN system, i.e.~they do not 
give rise to transitions between isospin-singlet and isospin-triplet two-nucleon states. 
Finally, class (IV) forces break charge symmetry and cause isospin mixing in the NN system.

\begin{minipage}{\textwidth}
\vskip 0 true cm
\rule{\textwidth}{.2pt}
{\it
Exercise:  show that class-III two-nucleon forces do not lead to isospin mixing in the 
two-nucleon system,  i.e.~they commute with the operator 
$T^2$. 
Does this still hold true for systems with three and more nucleons?
} \\
\vskip -0.8 true cm
\rule{\textwidth}{.2pt}
\end{minipage}

\medskip
Let us now discuss the position-spin structure of the potential. 
For the sake of simplicity, I restrict myself to the isospin-invariant case. 
The available vectors are given by the position, 
momentum and spin operators for individual nucleons: 
$\vec r_1, \, \vec r_2, \, \vec p_1, \, \vec p_2, \, \vec \sigma_1, \, \vec \sigma_2$. 
The translational and Galilean invariance of the potential 
implies that it may only depend on the relative distance between the nucleons,
$\vec r \equiv \vec r_1 - \vec r_2$, and the relative momentum, 
$\vec p \equiv (\vec p_1 - \vec p_2)/2$. 
Further constraints due to (i) rotational invariance, (ii) invariance 
under a parity operation, (iii) time reversal invariance,  (iv) hermiticity 
as well as (v) invariance with respect to interchanging the nucleon labels, 
$1 \leftrightarrow 2$, lead to the following operator form of the 
potential \cite{Okubo:1958aa}:
\beq
\label{pot_operat}
\left\{ \fet 1_{\rm spin}, \; \vec \sigma_1 \cdot \vec \sigma_2, \; 
S_{12} (\vec r \, ), \; S_{12} (\vec p \, ), \; \vec L \cdot \vec S, \; 
(\vec L \cdot \vec S\, )^2 \right\} \times
\left\{ \fet 1_{\rm isospin}, \; \fet \tau_1 \cdot \fet \tau_2  \right\}\,,
\eeq
where $\vec L \equiv \vec r \times \vec p$, $\vec S \equiv 
(\vec \sigma_1 + \vec \sigma_2 )/2$ and $S_{12} ( \vec x \,) \equiv
3 \vec \sigma_1 \cdot \hat x \, \vec \sigma_2 \cdot \hat x - \vec \sigma_1 
\cdot \vec \sigma_2$ with $\hat x \equiv \vec x/| \vec x \, |$. 
The operators entering the above equation are multiplied by scalar operator-like 
functions that depend on $r^2$, $p^2$ and $L^2$. 

Throughout this work, two-nucleon observables will be computed  by 
solving the Lippmann-Schwinger equation in momentum space. It is, therefore, 
instructive to look at the momentum-space representation of the potential, 
$V (\vec p \, ', \, \vec p \, ) \equiv \langle \vec p\, ' | V |   \vec p \, \rangle$, 
with $\vec p$ and $\vec p \, '$ denoting the two-nucleon center of mass momenta before and 
after the interaction takes place. Following the same logic as above, the 
most general form of the potential potential in momentum space can be shown to be: 
\beq
\label{pot_mom}
\left\{ \fet 1_{\rm spin}, \; \vec \sigma_1 \cdot \vec \sigma_2, \; 
S_{12} (\vec q \, ), \; S_{12} (\vec k \, ), \; i \vec S \cdot \vec q \times \vec k, \; 
\vec \sigma_1 \cdot \vec q \times \vec k \, \vec \sigma_2 \cdot \vec q \times \vec k \right\} \times
\left\{ \fet 1_{\rm isospin}, \; \fet \tau_1 \cdot \fet \tau_2  \right\}\,,
\eeq
where $\vec q \equiv \vec p \, ' - \vec p$ and $\vec k \equiv (\vec p \, ' + \vec p \, )/2$. 
The operators are multiplied with the scalar functions that depend on $p^2$, ${p '}^2$ and 
$\vec p \cdot \vec p \, '$. 
Notice that contrary to  Eq.~(\ref{pot_operat}) which involves the \emph{operator} $\vec p$, 
$\vec p$ and $\vec p \, '$ that enter Eq.~(\ref{pot_mom}) 
denote the corresponding eigenvalues. It should also be emphasized that further   
spin-momentum operators contribute in the case of class-IV isospin-breaking interactions. 

For low-energy processes I will be focused in here, it is convenient to switch 
to the partial wave basis $| \vec p \, \rangle \to | p l m_l \rangle$. A two-nucleon 
state $| p (l s) j m_j \rangle$ in the partial-wave basis depends on the orbital 
angular momentum $l$, spin $s$, the total angular momentum $j$ and the corresponding 
magnetic quantum number $m_j$.  The partial wave decomposition of the potential in 
Eq.~(\ref{pot_mom}) is given by:
\beq
\label{partw}
\langle p ' (l' s') j' m_j' | V |   p  (l s) j m_j \rangle
\equiv \delta_{j' j} \, \delta_{m_j' m_j} \, \delta_{s' s}\, 
V^{sj}_{l' l} (p ', \, p) \,,
\eeq
with 
\beqa
\label{partw1}
V^{sj}_{l' l} (p ', \, p) 
&=& \sum_{m_l' , \, m_l} \int d \hat p ' \, d \hat p \,  
c(l', s, j; m_l', m_j-m_l ', m_j) \, c(l, s, j; m_l, m_j-m_l , m_j)\nn
&& {} \times 
Y^\star_{l' m_l'} (\hat p ') \, Y_{l m_l} (\hat p ) \,  
\langle s \, m_j-m_l ' | V (\vec p \, ', \, \vec p \, ) | s \, m_j-m_l \rangle  \,,
\eeqa
where $c(l, s, j; m_l, m_j - m_l, m_j)$ are Clebsch-Gordan coefficients and 
$Y_{l m_l} (\hat p) $ denote the spherical harmonics. The first two Kronecker 
$\delta$'s on the right-hand side of the first line in Eq.~(\ref{partw}) reflect 
the conservation of the total angular momentum. Rotational invariance of the 
potential prevents the dependence of the matrix elements on the magnetic quantum 
number $m_j$. The conservation of the total spin of the nucleons can be easily 
verified explicitly for all operators entering Eq.~(\ref{pot_mom}). I stress, however,
that transitions between the spin-singlet and spin-triplet channels are possible 
in a more general case of the broken isospin symmetry.  For each individual operator 
entering Eq.~(\ref{pot_mom}), the expression 
(\ref{partw1}) can be simplified and finally expressed as an integral 
over $\hat p \cdot \hat p \, '$ with the integrand being written
in terms of the corresponding scalar function and Legendre polynomials. 
Explicit formulae can be found e.g.~in \cite{Epelbaum:2004fk}, see also 
Ref.~\cite{Golak:2009ri} for a 
recent work on this topic. 

The Lippmann-Schwinger (LS) equation for the half-shell $T$-matrix in the partial wave 
basis has the form 
\beq\label{LSeq}
T^{sj}_{l'l} (p',p) = V^{sj}_{l'l} (p',p) +  \sum_{l''} \,
\int_0^\infty \frac{dp'' \, {p''}^2}{(2 \pi )^3} \,  V^{sj}_{l'l''} (p',p'')
\frac{m}{p^2-{p''}^2 +i\eta} T^{sj}_{l''l} (p'',p)~,
\eeq
with $m$ denoting the nucleon mass and $\eta \to 0^+$.
In the uncoupled case, $l$ is conserved. 
The relation between the on-shell $S$- and $T$-matrices is given by 
\beq
S_{l' l}^{s j} (p) = \delta_{l' l} - \frac{i}{8 \pi^2} 
\, p \, m \,  T_{l' l}^{s j} (p)~.
\eeq
The phase shifts in the uncoupled cases can be obtained from the
$S$-matrix via
\beq
S_{jj}^{0j} = \exp{ \left( 2 i \delta_{j}^{0j} \right)} \; , \quad 
S_{jj}^{1j} = \exp{ \left( 2 i \delta_{j}^{1j} \right)} \;,
\eeq
where I  use the notation $\delta^{sj}_l$.
The so-called Stapp parametrization
of the $S$-matrix in the coupled channels ($j>0$) is defined as:
\beq
S = \left( \begin{array}{cc} S_{j-1 \, j-1}^{1j} &  S_{j-1 \, j+1}^{1j} \\[3pt]
S_{j+1 \, j-1}^{1j} &  S_{j+1 \, j+1}^{1j} \end{array} \right) 
= 
\left( \begin{array}{cc} \cos{(2 \epsilon)} \exp{(2 i \delta^{1j}_{j-1})} &
i \sin{(2 \epsilon)} \exp{(i \delta^{1j}_{j-1} +i \delta^{1j}_{j+1})} \\[3pt]
i \sin{(2 \epsilon)} \exp{(i \delta^{1j}_{j-1} +i \delta^{1j}_{j+1})} &
\cos{(2 \epsilon)} \exp{(2 i \delta^{1j}_{j+1})} \end{array} \right)~\nonumber ,
\eeq
and is related to another frequently used parametrization due to Blatt and Biedenharn  
in terms of $\tilde{\delta}$ and $\tilde{\epsilon}$ via 
the following equations:
\beq
\label{blattb}
\delta_{j-1} + \delta_{j+1} =  \tilde{\delta}_{j-1} + \tilde{\delta}_{j+1}\,, \quad\quad
\sin ( \delta_{j-1} - \delta_{j+1} ) = \frac{\tan ( 2 \epsilon)}{\tan (2 \tilde{\epsilon})}\,,
\quad\quad
\sin (\tilde{\delta}_{j-1} - \tilde{\delta}_{j+1}) = 
\frac{\sin ( 2 \epsilon)}{\sin (2 \tilde{\epsilon})}\,.
\eeq

The appearance of the electromagnetic interaction requires special care when calculating 
scattering observables due to its long-range nature. In particular, the S-matrix has to be formulated in terms of 
asymptotic Coulomb states.  The electromagnetic interaction between the nucleons is driven 
by the Coulomb force and, to a lesser extent, magnetic moment interactions and vacuum polarization. 
It should also be emphasized that the 
expansion of the scattering amplitude in partial waves converges very slowly in the presence 
of the magnetic moment interactions.
For explicit expressions and a detailed discussion on their implementation when 
calculating nucleon-nucleon observables the reader is referred to \cite{Stoks:1993tb}.

The deuteron wave function and binding energy $E_d$ are 
obtained from the homogeneous part of Eq.~(\ref{LSeq}):
\beq\label{LSb}
\phi_l (p) = \frac{1}{E_d -p^2/m} \, \sum_{l'} \, \int_0^\infty \frac{dp'
\, {p'}^2}{(2 \pi )^3} \,  V^{sj}_{ll'} (p,p') \, \phi_{l'} (p')~,
\eeq
where $s=j=1$, $l=l'=0,2$. Once phase shifts are calculated, nucleon-nucleon 
scattering observables can be computed in a standard way, see 
\cite{Bystricky:1976jr,Gloeckle:1983aa}. 

\begin{figure*}
\vspace{0.3cm}
\centerline{
\includegraphics[width=0.32\textwidth]{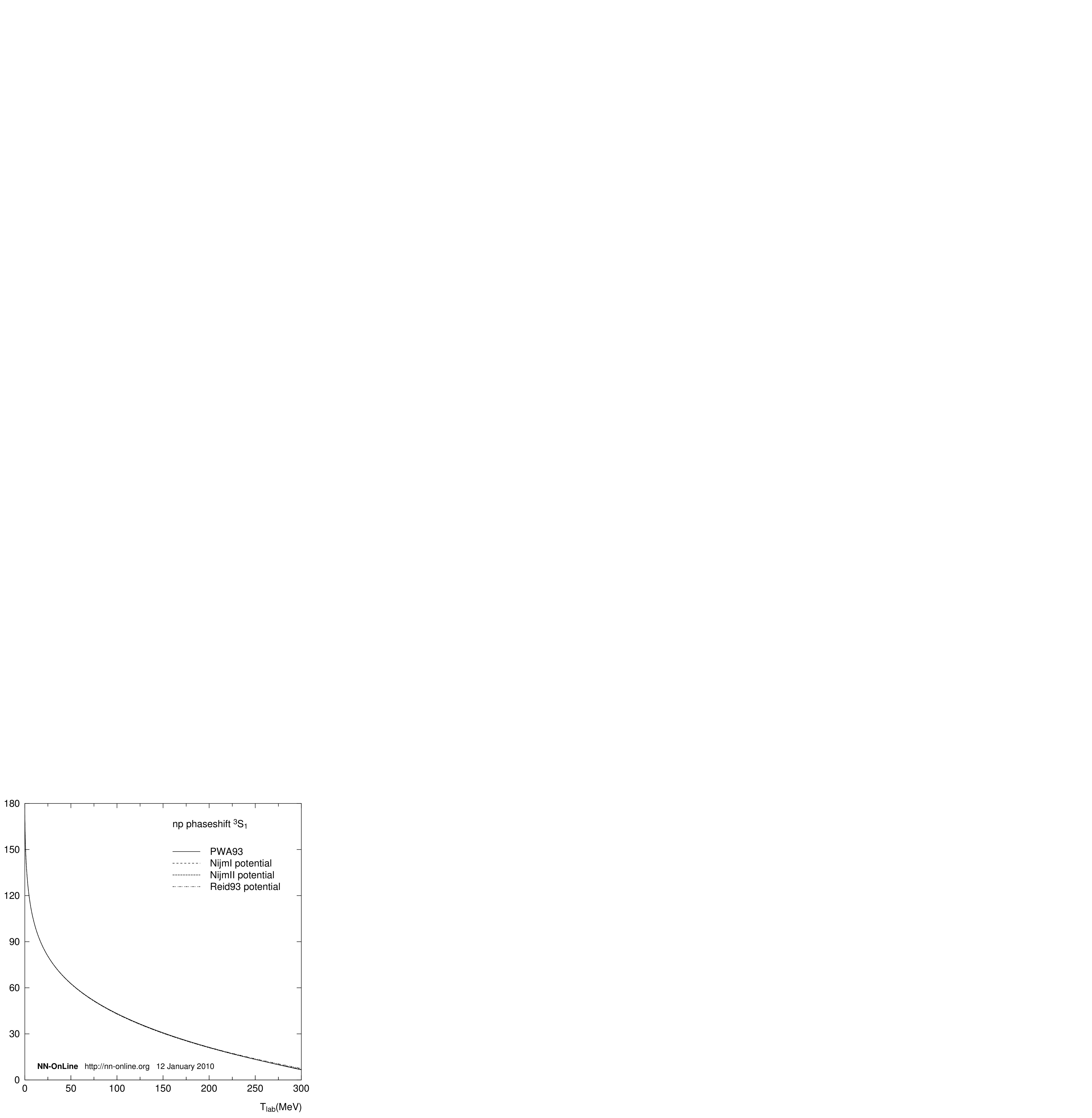} \hfill
\includegraphics[width=0.315\textwidth]{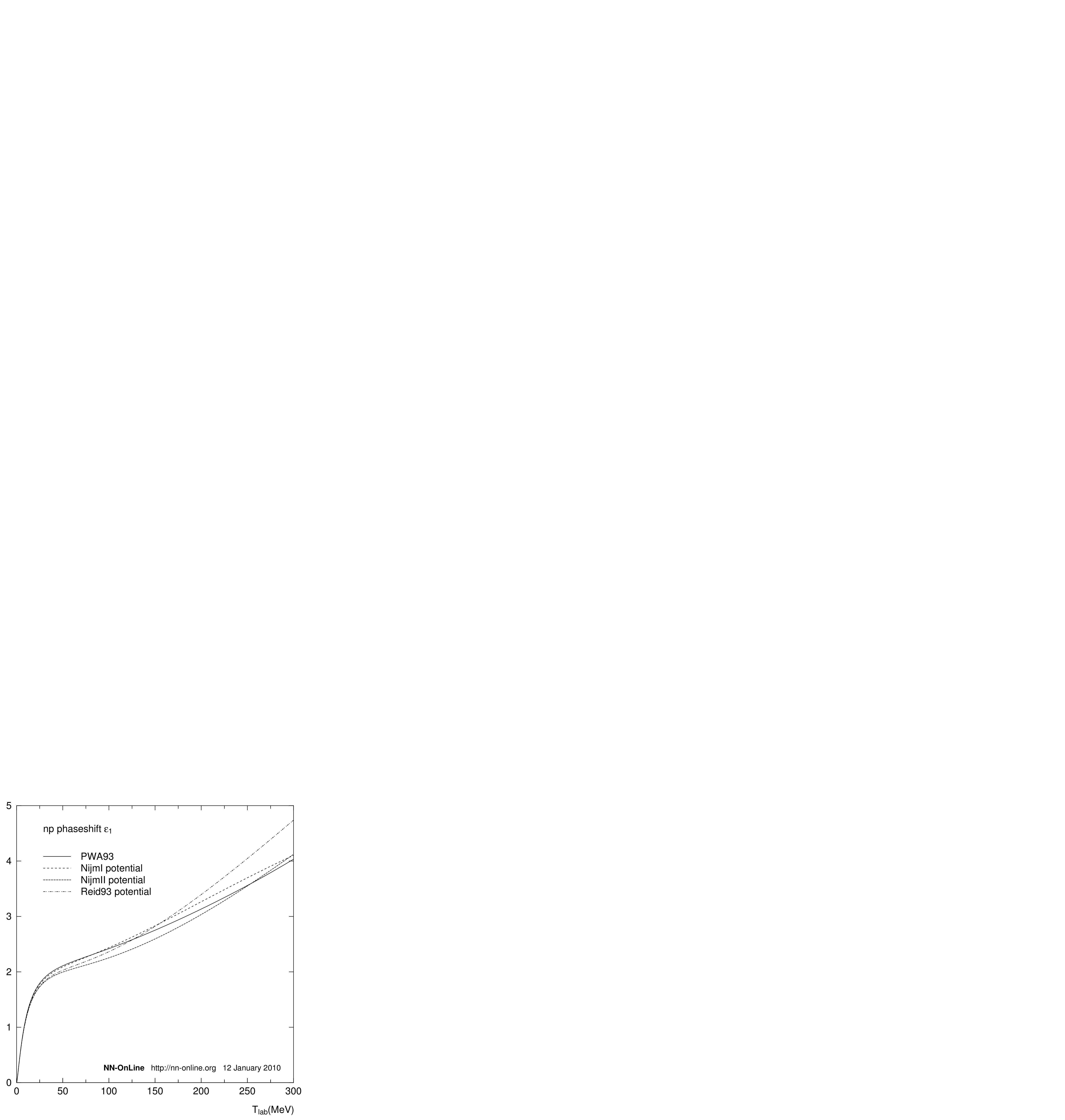} \hfill
\includegraphics[width=0.32\textwidth]{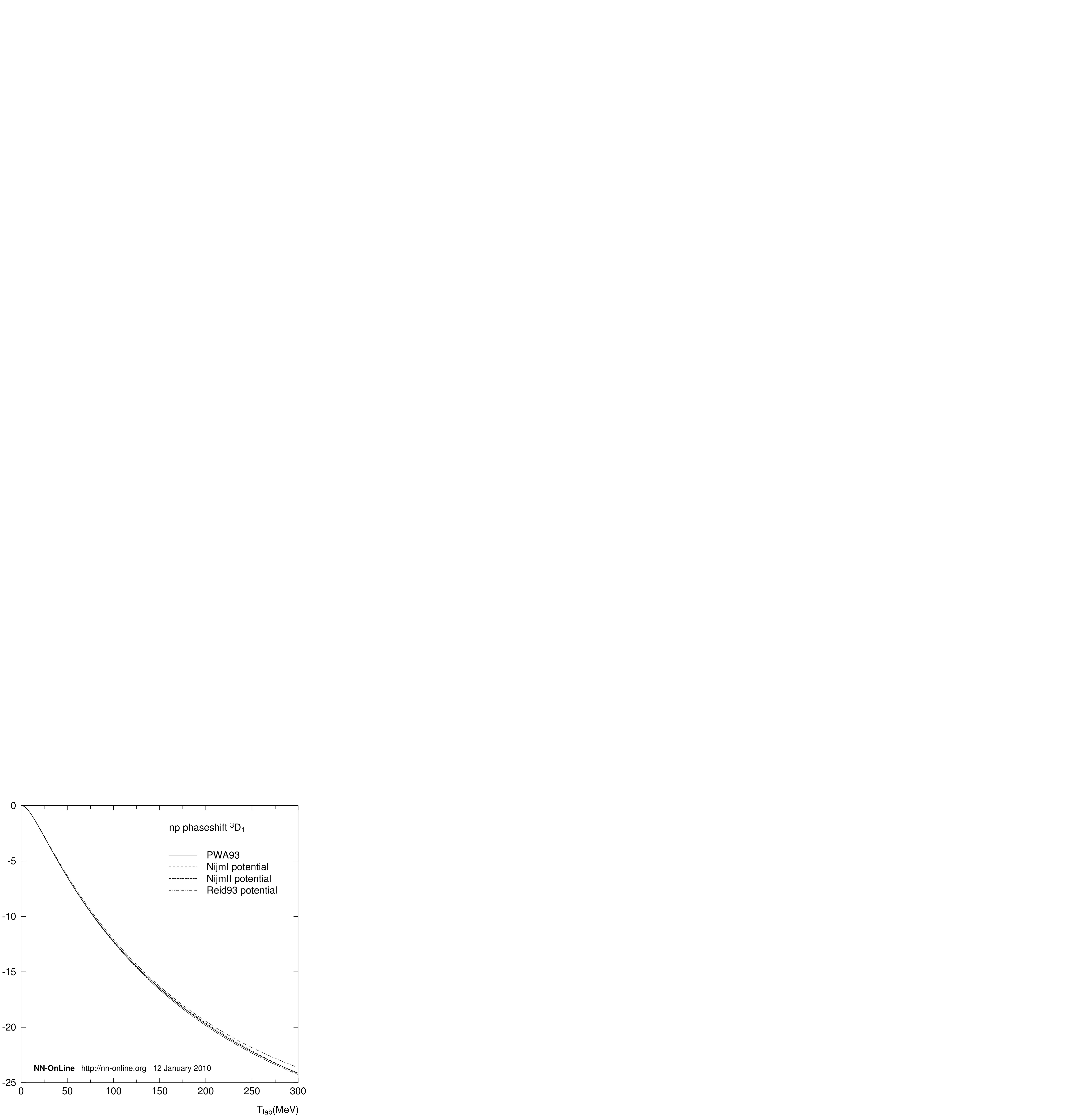} 
}
\vspace{-0.2cm}
\caption[fig4aa]{\label{phases_nijm} $^3S_1$ (left panel) and $^3D_1$ (right panel) phase shifts 
and the mixing angle $\epsilon_1$ (middle panel) calculated from several modern high-precision 
potentials in comparison with the results of the Nijmegen PWA. The phase shifts and the mixing 
angle are shown in degrees.  Plots are generated through the NN-Online web site {\sl http://nn-online.org}.   
}
\vspace{0.2cm}
\end{figure*}

The appearance of only a few structures in the most general expression for the two-nucleon force,  
see Eq.~(\ref{pot_mom}), and the large amount of available 
low-energy nucleon-nucleon scattering data motivated and enabled the development of 
modern high-precision phenomenological potential models such as e.g.~the 
CD-Bonn 2000 \cite{Machleidt:2000ge}, Argonne $V_{18}$
(AV18) \cite{Wiringa:1994wb} and 
Nijmegen I, II potentials \cite{Stoks:1994wp}.  The general strategy involves incorporating 
the proper long-range behavior due to the electromagnetic interaction and the 
one-pion exchange potential which is important to correctly describe the low-energy behavior  
of the amplitude, cf.~section \ref{sec:analyt}, 
and parametrizing the medium- and short-range 
contributions in a general way. AV18, a local $r$-space potential, can be viewed 
as a representative example. It 
includes (i) electromagnetic interactions multiplied by short-range functions to account for the 
finite size of the nucleon, (ii) regularized one-pion exchange potential including 
isospin-breaking corrections due to different masses of the charged and neutral pions, (iii) 
some additional phenomenological isospin-breaking terms of a shorter range, (iv)   
medium-range (short-range) contributions of Yukawa-type (Woods-Saxon type) multiplying 
the operators in Eq.~(\ref{pot_operat}). With about 40 
adjustable parameters, it describes the proton-proton and neutron-proton scattering data 
with $\chi_{\rm datum}^2 = 1.09$. Other high-precision potentials are constructed in 
a similar way and allow to reproduce the data or phase shifts from e.g.~the Nijmegen 
partial wave analysis (PWA) with a comparable accuracy. This is visualized in 
Fig.\ref{phases_nijm}. I  refer the reader to Ref.~\cite{Machleidt:2001rw}
for a recent review article on the modern high-precision potentials.
 
While various phenomenological potentials
provide an accurate representation of the nucleon-nucleon phase shifts and
most of the deuteron properties, the situation is much less satisfactory
when it comes to the much weaker but necessary three-nucleon forces.
Such three-body forces are needed to describe the nuclear binding
energies and levels, as most systematically shown by the
Urbana-Argonne group~\cite{Pieper:2001mp}. Systematic studies of 
the dynamics and reactions of systems with three or four-nucleons
further sharpen the case for the necessity of including three-nucleon
forces,  see e.g.~\cite{Gloeckle:1995jg}. A phenomenological path to modeling 
the three-nucleon force following the same strategy as in the two-nucleon case
seems to be not feasible (at least, at present).  Indeed, in the case of two nucleons,
the potential can be decomposed in only a few different spin-space structures, and 
the corresponding radial functions can be adjusted to the extensive set of data. 
Such an approach would, however, fail for the three-nucleon force due to the large variety 
of different possible structures, a scarcer data base and considerably more time consuming
calculations required. 

While the conventional approach based on the high-precision two-nucleon potentials 
accompanied with the existing three-nucleon force models enjoyed
many successes and is frequently used in e.g.~nuclear structure and reaction
calculations, it remains incomplete as there are certain deficiencies
that can only be overcome based on EFT approaches. These are: (i) it is
very difficult - if not impossible - to assign a trustworthy theoretical
error, (ii) gauge and chiral symmetries are difficult to implement, (iii)
none of the three-nucleon forces is consistent with the underlying 
nucleon-nucleon interaction models/approaches and (iv) the connection
to QCD is not at all obvious.

\section{Chiral perturbation theory: An elementary introduction}
\def\theequation{\arabic{section}.\arabic{equation}}
\label{sec3}

Effective field theories have proved to be an important and very useful tool in nuclear and particle physics. 
One understands under an effective (field) theory an approximate theory whose scope is to describe 
phenomena which occur at a chosen length (or energy) range. 
The main idea of this method can be illustrated with the following example from classical electrodynamics.   
\begin{figure*}
\vspace{0.3cm}
\centerline{
\includegraphics[width=0.6\textwidth]{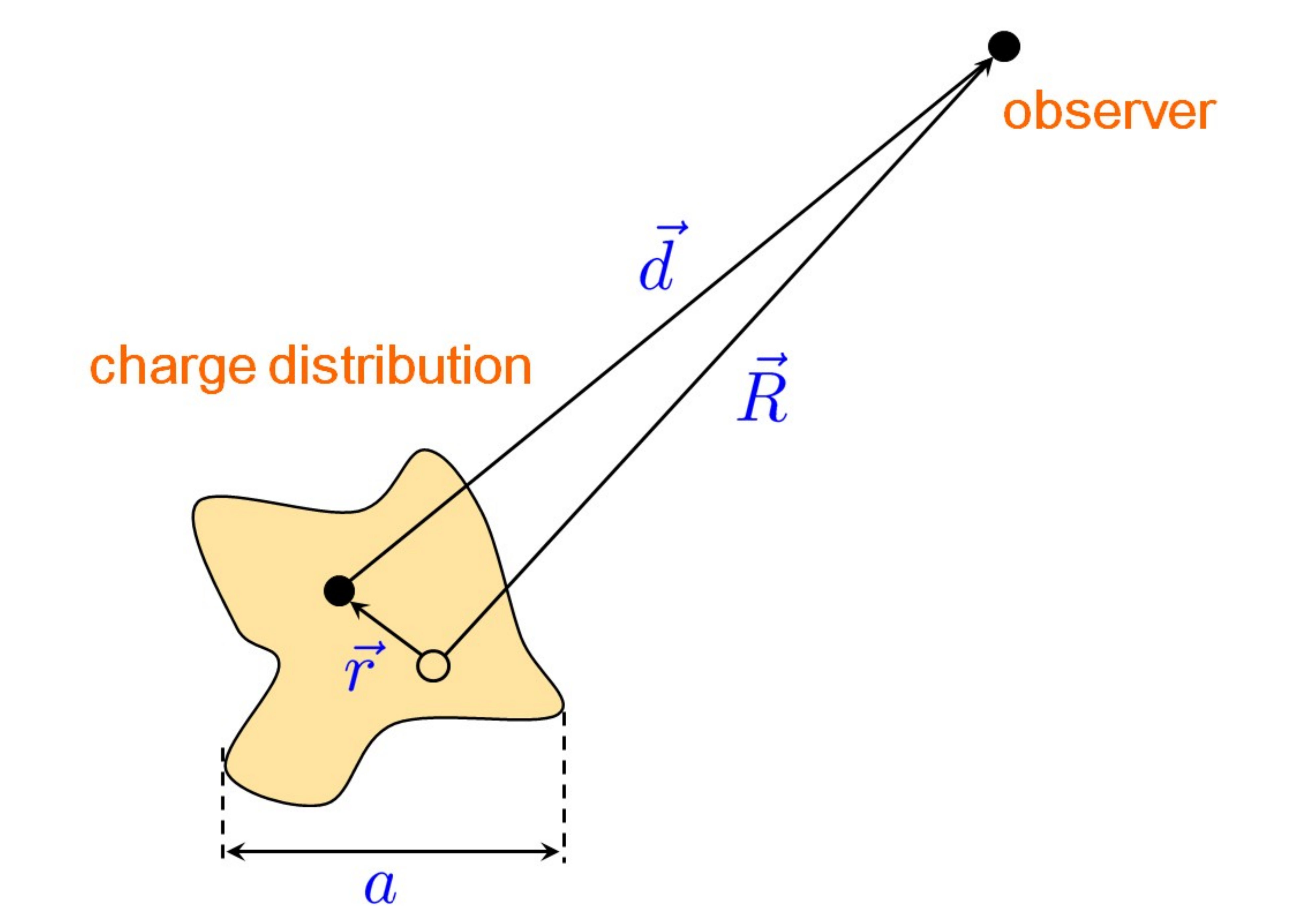}
}
\vspace{-0.2cm}
\caption[fig4aa]{\label{fig0} A localized charge distribution generates an electrostatic potential which can be 
described in terms of the multipole expansion.  
}
\vspace{0.2cm}
\end{figure*}
Consider a localized charge distribution in space of a size $a$. The resulting electrostatic potential 
at any given position $\vec R$ can be calculated by integrating over the elementary charges and using the 
familiar expression for the Coulomb potential generated by a point charge:
\beq
\label{multipole0}
V (\vec R ) \propto \int d^3 r \, \frac{\rho (\vec r \, ) }{| \vec R - \vec r \,| } 
\eeq
Expanding $1/| \vec R - \vec r \, |$ for $r \ll R$, 
\beq
\frac{1}{| \vec R - \vec r \,| } = \frac{1}{R} + \sum_i r_i\frac{R_i}{R^3}
+ \frac{1}{2} \sum_{ij} r_i r_j \frac{3 R_i R_j - \delta_{ij} R^2}{R^5} + \ldots\,,
\eeq
with $i, \, j$ denoting the Cartesian components 
allows to rewrite the integral as 
\beq
\label{multipole}
\int d^3 r \, \frac{\rho (\vec r \, ) }{ | \vec R - \vec r \,| } =
\frac{q}{R} +   \frac{1}{R^3} \sum_i R_i P_i + 
\frac{1}{6R^5} \sum_{ij} (3 R_i R_j - \delta_{ij} R^2 ) Q_{ij} + \ldots
\eeq
where the total charge $q$, dipole moment $P_i$ and quadrupole moment $Q_{ij}$ are 
defined via
\beq
q =  \int d^3 r \, \rho (\vec r \, ), \quad \quad
P_i =  \int d^3 r \, \rho (\vec r \, ) \, r_i, \quad \quad
Q_{ij} =  \int d^3 r \, \rho (\vec r \, ) (3 r_i r_j - \delta_{ij} r^2 ) \,.
\eeq
The expression in Eq.~(\ref{multipole})
represents the well-known multipole expansion for the electrostatic potential. When truncated, 
it provides an approximation to the ``underlying theory'' given by the exact expression (\ref{multipole0}).  
The multipoles entering every term in this expansion contain certain amount of information 
about the charge distribution and can, of course, be calculated provided $\rho( \vec r \, )$ is known.  
The multipole expansion is, however, particularly useful if $\rho( \vec r \, )$ is unknown 
(except for the fact that it is localized). It then allows to describe the electrostatic potential 
at every point in space far from the charge distribution with, in principle, an arbitrarily high 
accuracy provided one has enough data (e.g.~experimentally measured values of the electrostatic 
potential at some points) to determine the desired number of the multipoles.  

Chiral Perturbation Theory (CHPT) is the effective theory of QCD (more generally,
of the Standard Model) which was formulated by Weinberg \cite{Weinberg:1978kz} 
and developed in to a systematic tool for analyzing low-energy hadronic observables 
by Gasser and Leutwyler \cite{Gasser:1983yg,Gasser:1984gg}. In this section,
I give a brief overview of the foundations of this approach. My main purpose here is to  
outline the logical 
steps which are needed in order to set up this theoretical framework. 
I will also give references to the existing extensive literature on this subject
which is suitable for further reading. 

\subsection{Chiral symmetry of QCD}
Symmetries provide powerful constraints on effective interactions and thus 
play the crucial role for effective field theories. In the following, I will discuss the 
symmetries of QCD which are relevant in the context of ChPT.   
Consider the QCD Lagrangian in the two-flavor case of the light up and down quarks 
\beq
\label{LQCD}
\mathcal{L}_{\rm QCD} = \bar q \, (  i \gamma_\mu D^\mu - \mathcal{M} ) q  -  \frac{1}{4}  
G^a_{\mu \nu} G^{a\, \mu \nu}\,,
\eeq
where $D_\mu = \partial_\mu - i g_s G_\mu^a T^a$ with $T^a$, (with $a = 1 \ldots 8$)
are the  SU(3)$_{\rm color}$ Gell-Mann matrices and $q$
the quark fields. Further, $G_{\mu \nu}^a$ are the 
gluon field strength tensors, and the quark mass matrix is given by $\mathcal{M} = \mbox{diag} (m_u, \, m_d)$.  
I do not show in Eq.~(\ref{LQCD}) the 
$\theta$- and gauge fixing terms 
which are not relevant for our consideration. It is instructive to write the QCD Lagrangian 
in terms of the left- and right-handed quark field components defined by 
$q_{R} = (1/2) (1 + \gamma_5 ) q$, $q_{L} = (1/2) (1 - \gamma_5 ) q$:
 \beq
\label{qcdumgeschr}
\mathcal{L}_{\rm QCD} = \bar{q}_L i D \palka q_L + \bar{q}_R i D \palka q_R - \bar{q}_L  \mathcal{M} q_R 
- \bar{q}_R  \mathcal{M} q_L 
- \frac{1}{4} G^\alpha_{\mu \nu} G^{\alpha, \mu \nu}\,.
\eeq
We see that the left- and right-handed quark fields are only connected through the mass term. 
Given the smallness of the light quark masses \cite{Amsler:2008zzb}
\footnote{The following values correspond to the $\overline{\rm MS}$ scheme at scale $\mu = 2$ GeV.}, 
\beq
m_u \simeq 1.5 \ldots 3.3 \mbox{ MeV,} \quad  \quad
m_d \simeq 3.5 \ldots 6.0 \mbox{ MeV,}
\eeq
as compared to the typical hadron masses of the order of $1$ GeV, the quark mass term can, to 
a good approximation, be neglected. The Lagrangian in Eq.~(\ref{qcdumgeschr}) is, therefore,
approximately invariant under independent global flavor rotations of the left- and right-handed 
quark fields:
\beq
\label{trafo_quarks}
q_L \longrightarrow q_L'=  L q_L =   \exp{\left(- i \fet{\theta}_L \cdot  \fet{\tau} /2 \right)} q_L , \quad \quad  
q_R \longrightarrow q_R'= R q_R = \exp{\left(- i \fet{\theta}_R \cdot  \fet{\tau} /2 \right)} q_R \,,
\eeq
where $\fet \tau$ denote the Pauli matrices in the flavor space and $\fet \theta_{L,R}$ are the corresponding 
rotation angles. The corresponding symmetry group $SU(2)_L \times SU(2)_R$ is referred to as 
the $SU(2)$ chiral group. According to Noether's theorem, there are six conserved currents 
\beq
L_\mu^i = \bar q_L \gamma_\mu  \frac{\tau^i}{2} q_L \,, \quad \quad
R_\mu^i = \bar q_R \gamma_\mu  \frac{\tau^i}{2} q_R \,, 
\eeq
which can equally well be expressed in terms of the vector and axial-vector currents 
$V_\mu^i = L_\mu^i + R_\mu^i$ and $A_\mu^i = R_\mu^i - L_\mu^i$.  The corresponding conserved  
charges generate the algebra of the chiral group
\beq
\label{lee1}
\left[ Q_{I}^i, \; Q_{I}^j \right] = i \epsilon^{ijk} Q_I^k \quad \mbox{with $I=L,R$,}
\quad \quad 
\left[ Q_{L}^i, \; Q_{R}^j \right] = 0\,,
\eeq
or, equivalently, 
\beq
\label{lee2}
\left[ Q_{V}^i, \; Q_{V}^j \right] = i \epsilon^{ijk} Q_V^k\,,  \quad \quad
\left[ Q_{A}^i, \; Q_{A}^j \right] = i \epsilon^{ijk} Q_V^k\,, \quad \quad
\left[ Q_{V}^i, \; Q_{A}^j \right] = i \epsilon^{ijk} Q_A^k\,.
\eeq
Application of the above commutation relations to hadronic reactions was at the heart of the 
current algebra calculations in  the early seventies of the last century. 

The Lagrangian for massless u- and d-quarks is, in fact, invariant under even a larger 
group of transformations in the flavor space, namely  
$SU(2)_L \times SU(2)_R \times U(1)_V \times U(1)_A$. While the vector $U(1)$ corresponds to 
quark number conservation, the axial $U(1)_A$ current is known to be broken by quantum 
effects (the so-called $U(1)_A$ anomaly) and thus does not represent a symmetry of the quantum 
theory.  

In spite of the fact that QCD for two light flavors is approximately chiral invariant, 
its ground state is not symmetric with respect to $SU(2)_L \times SU(2)_R$ but only with 
respect to its vector subgroup $SU(2)_V \subset SU(2)_L \times SU(2)_R$ generated by 
the charges $\{Q_V^i \}$. This means that the axial charges do not annihilate the vacuum, 
that is $Q_V^i |0 \rangle = 0$ while $Q_A^i |0 \rangle \neq 0$. 
Evidence of the spontaneous breakdown of the chiral symmetry comes from various sources. 
For example, hadrons occur in nearly degenerate isospin multiplets corresponding to $SU(2)_V$ which 
implies that this group is realized in the usual Wigner-Weyl mode. If this were the case 
for the chiral group, one would observe larger chiral multiplets containing  
particles of opposite parity since the charges $Q_{V}^i$ 
and $Q_{A}^i$ have opposite parity. Generally, no such parity doubling is observed in the 
hadron spectrum. Another strong argument in favor of the spontaneous breakdown of 
the chiral symmetry comes from the existence of unnaturally light (in comparison with 
other hadrons) pseudoscalar mesons (pions) being  natural candidates 
for the corresponding Nambu-Goldstone bosons. Pions are not exactly massless but 
acquire a small mass due to the explicit chiral symmetry breaking by the nonvanishing quark
masses. These and further arguments coming from both the theory and experiment indicate 
undoubtly that the chiral $SU(2)_L \times SU(2)_R$ group is spontaneously broken
down to $SU(2)_V$.  

I now pause to summarize the content of this section. QCD Lagrangian in the two-flavor 
case of the up- and down-quarks is approximately invariant under global chiral $SU(2)_L \times SU(2)_R$
transformations. The chiral symmetry is broken explicitly due to the nonvanishing quark masses,
$m_u \neq 0$, $m_d \neq 0$. In addition to this explicit symmetry breaking, $SU(2)_L \times SU(2)_R$
is also broken spontaneously down to the isospin group $SU(2)_V$. The three corresponding 
pseudoscalar Goldstone bosons are identified with pions whose small masses emerge due to nonvanishing 
quark masses. There exist a large mass gap in the hadron spectrum: $M_\rho \simeq 770\mbox{ MeV} \; \gg
M_\pi \simeq 140\mbox{ MeV}$. 

\subsection{Effective Lagrangian for Goldstone bosons}
We now turn to the \emph{effective} description of low-energy QCD dynamics. The simplest possible case 
emerges when the energy is chosen so small that only pions need to be treated  
as explicit degrees of freedom. All other hadrons are much heavier and can be  
integrated out from the theory. The main ingredient of any effective field theory 
is the most general effective Lagrangian that involves \emph{all} possible terms which are consistent 
with the symmetries of the underlying theory. Let us, for the time being, consider the so-called chiral 
limit of QCD, i.e.~the idealized world in which quarks are massless and the chiral symmetry of 
$\mathcal{L}_{QCD}$ is exact. The task is then to construct the most general chiral invariant 
Lagrangian for pion fields. In order to do that we first need to figure out how pions transform 
with respect to chiral rotations. Our knowledge of the pion transformation properties with respect 
to  $SU(2)_L \times SU(2)_R$ can be summarized by the following two observations:
\begin{itemize}
\item
Pions build an isospin triplet and thus transform linearly 
under  $SU(2)_V \subset SU(2)_L \times SU(2)_R$ 
according to the corresponding irreducible representation;
\item
The chiral group must be realized \emph{nonlinearly}. This follows immediately 
from the geometrical argument based on the 
fact that the Lie algebra of $SU(2)_L \times SU(2)_R$ in Eq.~(\ref{lee1}) is isomorphic to that of $SO(4)$. 
We know that one needs three coordinates in order to construct  the smallest non-trivial 
representation, the so-called fundamental representation, of the three-dimensional rotation group. 
Similarly, the smallest nontrivial representation of the four-dimensional rotation group $SO(4)$
is four-dimensional. We have, however, only three ``coordinates'' at our disposal 
(the triplet of the pion fields)!
\end{itemize}
To construct a non-linear realization of $SO(4)$ we begin with the usual  
representation describing four-dimensional rotations of a vector 
$( \fet \pi , \, \sigma ) \equiv ( \pi_1 ,  \,  \pi_2 ,  \,  \pi_3 ,  \,  \sigma )$. 
For an infinitesimal rotation parametrized by six angles $\{ \theta_i^{V,A} \}$, with $i =1, 2, 3$, we have:
\beq
\label{linear}
\left( 
\begin{array}{c}
\fet \pi \\ \sigma 
\end{array} 
\right) \stackrel{SO(4)}{\longrightarrow}
\left( 
\begin{array}{c}
\fet \pi' \\ \sigma ' 
\end{array} 
\right) = \left[ \fet 1_{\rm 4 \times 4} + \sum_{i=1}^3 \theta_i^V V_i +  \sum_{i=1}^3 \theta_i^A A_i \right]
\left( 
\begin{array}{c}
\fet \pi \\ \sigma 
\end{array} 
\right)\,,
\eeq
where 
\beq
 \sum_{i=1}^3 \theta_i^V V_i = \left(
\begin{array}{cccc}
0 & -\theta^V_3 & \theta^V_2 & 0 \\
\theta^V_3 & 0 & -\theta_1^V & 0 \\
-\theta^V_2 & \theta_1^V & 0 & 0 \\
0 & 0 & 0 & 0
\end{array}
\right)\,, \quad \quad \quad
 \sum_{i=1}^3 \theta_i^A A_i = \left(
\begin{array}{cccc}
0 & 0 & 0 & \theta^A_1 \\
0 & 0 & 0 & \theta^A_2 \\
0 & 0 & 0 & \theta^A_3 \\
-\theta^A_1 & -\theta_2^A & -\theta_3^A & 0 
\end{array}
\right)\,.
\eeq
Notice that the set of rotations generated by $V_i$ builds a subgroup of $SO(4)$, namely 
the group of three-dimensional rotations $SO(3) \subset SO(4)$ which is locally isomorphic 
to $SU(2)$.     
The four real quantities $( \fet \pi , \, \sigma )$ define the smallest nontrivial chiral 
multiplet and represent the field content of the well-known linear sigma model. To switch from 
the above linear realization (i.e.~representation) of $SO(4)$ to the nonlinear one, we    
observe that, in fact, only three of the four components of $( \fet \pi , \, \sigma )$ are 
independent with respect to four-dimensional rotations. These three independent components 
correspond to coordinates on a four-dimensional sphere since $\fet \pi$ and $\sigma$ 
are subject to the constraint  
\beq
\label{constraint}
\fet \pi^2 + \sigma^2 = F^2\,,
\eeq
where $F$ is a constant of dimension mass. Making use of this equation to eliminate $\sigma$ in Eq.~(\ref{linear}) 
we end up with the following transformation properties of $\fet \pi$ under $SO(4)$:
\beqa   
\fet \pi &\stackrel{\theta^V}{\longrightarrow}& \fet \pi ' = \fet \pi + \fet \theta^V \times \fet \pi\,, \nn
\fet \pi &\stackrel{\theta^A}{\longrightarrow}& \fet \pi ' = \fet \pi + \fet \theta^A 
\sqrt{ F^2 - \fet \pi^2}\,,
\eeqa
where $\fet \theta^{V,A} \equiv \{ \theta^{V,A}_i \}$ with $i = 1,2,3$. 
The nonlinear terms (in $\fet \pi$) on the right-hand side of the 
second equation give rise to the nonlinear realization of $SO(4)$. 
This is exactly what we wanted to achieve: the chiral group $SU(2)_L \times SU(2)_R \simeq SO(4)$ 
is realized nonlinearly on the triplet of pions which, however, transform linearly under isospin 
$SU(2)_V \simeq SO(3)$ rotations parametrized through the angles  $\{ \fet \theta_V \}$. 

As a last remark note that the four-dimensional rotations of $( \fet \pi , \, \sigma )$ 
can be conveniently written using the $2 \times 2 $ matrix notation by introducing the unitary 
matrix\footnote{For $U$ to be unitary, $\sigma$ and $\fet \pi$ have to fulfill Eq.~(\ref{constraint}).}  
\beq
\label{matrU}
U = \frac{1}{F} \left(  \sigma  \fet 1_{\rm 2\times 2} + i \fet \pi \cdot \fet \tau \right)\,,
\eeq
and demanding the transformation properties of $U$ under chiral rotations to be:
\beq
\label{trafoU}
U \longrightarrow U' = L U R^\dagger\,.
\eeq
Here,  $L$ and $R$ are $SU(2)_L \times SU(2)_R$ matrices defined in Eq.~(\ref{trafo_quarks}). 

\begin{minipage}{\textwidth}
\vskip 0 true cm
\rule{\textwidth}{.2pt}
{\it
Exercise:  verify that infinitesimal transformations of   
$( \fet \pi , \, \sigma )$  induced by Eq.~(\ref{trafoU}) 
with $\fet \theta^V =(\fet \theta_R + \fet \theta_L )/2$ and $\fet \theta^A =(\fet \theta_R - \fet \theta_L )/2$ 
have indeed the same form as the ones given in Eq.~(\ref{linear}). 
} \\
\vskip -0.8 true cm
\rule{\textwidth}{.2pt}
\end{minipage}

\medskip
Clearly, the transition to the nonlinear realization is achieved by 
\beq
U = \frac{1}{F} \left(  \sigma \, \fet 1_{\rm 2 \times 2} + i \fet \pi \cdot \fet \tau \right) \; \longrightarrow
\; U = \frac{1}{F} \left( \sqrt{F^2 - \fet \pi^2}  \, \fet 1_{\rm 2 \times 2} + i \fet \pi \cdot \fet \tau \right) \,,
\eeq 
leaving pions as the only remaining degrees of freedom.
Notice that the ground state of the theory is characterized by a vanishing vacuum expectation values 
of $\fet \pi$ and corresponds to a particular point on the considered four-dimensional sphere (one of two 
crossing points between the sphere and the $\sigma$-axis). In accordance with the spontaneous breaking 
of chiral symmetry,  it not $SO(4)$- but only  $SO(3) \simeq SU(2)_V$-invariant.

It is now a simple exercise to construct the most general chiral-invariant Lagrangian for 
pions in terms of the matrix $U$. The building blocks are given by $U$, $U^\dagger$ and derivatives 
of these quantities. Notice that since I consider here only global chiral rotations, i.e.~$L$ and $R$ 
do not depend on space-time, the quantities like e.g.~$\partial_\mu \partial_\nu U$ transform in 
the same way as $U$ itself, i.e.~according to Eq.~(\ref{trafoU}).    
Chiral invariant terms in the effective Lagrangian terms can be constructed by taking a trace over products 
of $U$, $U^\dagger$ and their derivatives. Lorentz invariance implies that the number of derivatives 
must be even, so that the effective Lagrangian can be written as
\beq
\mathcal{L}_\pi = \mathcal{L}_\pi^{(2)} +  \mathcal{L}_\pi^{(4)} + \ldots\,.
\eeq
Notice that $\mathcal{L}_\pi^{(0)}$ is simply a constant since $U U^\dagger = \fet 1_{2\times 2}$.
The lowest-order Lagrangian involves just a single term 
\beq
\label{lagr1}
\mathcal{L}_\pi^{(2)} = \frac{F^2}{4} \langle \partial_\mu U \partial^\mu U^\dagger  \rangle \,,
\eeq
where $\langle \ldots \rangle$ denotes the trace in the flavor space. 
Terms involving $\partial_\mu \partial^\mu U$ or $\partial_\mu \partial^\mu U^\dagger$ are 
not independent and can be brought to the form of Eq.~(\ref{lagr1}) by using partial integration.  
The constant $F^2/4$
ensures that the Lagrangian has a proper dimension (both $F$ and $\pi$ have a dimension 
of mass) and is chosen in such a way that Eq.~(\ref{lagr1}) matches the usual 
free Lagrangian for massless scalar field when written in terms of pions:
\beq
\label{lagr2}
\mathcal{L}_\pi^{(2)} = \frac{1}{2} \partial_\mu \fet \pi \cdot \partial^\mu \fet \pi  + \frac{1}{2 F^2}
\left( \partial_\mu \fet \pi \cdot \fet \pi  \right)^2 + \mathcal{O} ( \fet \pi^6 )\,.
\eeq 
The a-priori unknown constants that accompany the terms in the effective Lagrangian are 
commonly called the low-energy constants (LECs), cf.~the multipoles in Eq.~(\ref{multipole}). 
The LEC $F$ can be identified with the 
pion decay constant in our idealized world (with quark masses being set to zero) \cite{Gasser:1983yg}. In the real 
world, it is measured to be $F_\pi = 92.4$ MeV.  Notice further that the pions are massless in 
the idealized world. Higher-order Lagrangians $\mathcal{L}_\pi^{(4)}$, 
$\mathcal{L}_\pi^{(6)}$, $\ldots$, can be constructed 
along the same lines by using partial integration and equations of motion to eliminate 
the redundant terms. 

Let us now pause to summarize what has been achieved so far. We have explicitly constructed 
a particular nonlinear realization of $SU(2)_L \times SU(2)_R$ in terms of  
pion fields which, as desired,  build a representation of $SU(2)_V \subset SU(2)_L \times SU(2)_R$
and learned how to write down the most general possible effective Lagrangian. The constructed 
nonlinear realization of the chiral group is, however, not unique. Different realizations emerge
by choosing different parametrizations of the matrix $U$ in Eq.~(\ref{trafoU}) in terms of pion fields such as, 
for example, the exponential parametrization $U = \exp (i \fet \pi \cdot \fet \tau/F )$. 
Generally, only the first three terms in the expansion of $U$ in powers of the pion field are 
fixed by unitarity, 
\beq
\label{Uexplicit}
U( \fet \pi )  = \fet 1_{2\times 2} + i\frac{\fet \tau \cdot \fet \pi}{F} - \frac{\fet \pi^2}{2 F^2} - i \alpha 
\frac{\fet \pi^2 \fet \tau \cdot \fet \pi}{F^3} + (8 \alpha - 1) \frac{\fet \pi^4}{8 F^4} + \mathcal{O} (\fet \pi^5)\,.
\eeq
Here, $\alpha$ is an arbitrary constant which reflects the freedom in parametrizing the matrix $U$. 
This raises the concern that observables that 
one intends to compute in the EFT are possibly affected by this non-uniqueness which would be a disaster. 
Fortunately, this is not the case. As shown by Coleman, Callan, Wess and Zumino \cite{Coleman:1969sm,Callan:1969sn}, 
all realizations of the chiral 
group are equivalent to each other modulo nonlinear field redefinitions 
\beq
\pi_i \to  \pi_i ' =  \pi_i \, F \left[ \fet \pi \right] \quad \mbox{with} \quad 
F \left[ \fet 0 \right] = 1\,.
\eeq
According to Haag's theorem \cite{Haag:1958vt}, such nonlinear field redefinitions do not affect 
S-matrix elements. 

So far, I considered the chiral limit corresponding to the idealized world with the masses of the 
up- and down-quarks being set to zero. This is fine as a first approximation but, of course, one would like to 
systematically improve on it by taking into account corrections 
due to nonvanishing quark masses. For that, we have to include in the 
effective Lagrangian all possible terms which break chiral symmetry in exactly the same way as does the 
quark mass term in $\mathcal{L}_{QCD}$. Consider, for example, the 
quark mass term with $m_u = m_d = m_q \neq 0$
which breaks chiral but preserves isospin symmetry. Recalling the geometrical interpretation 
with the four-dimensional rotation group and coordinates $(\fet \pi , \, \sigma)$, the quark mass term can be viewed  
as a vector that points along the $(\fet 0 , \, \sigma )$-direction. Its effects can be systematically taken 
into account by including in the effective Lagrangian not only $SO(4)$-scalars but also the corresponding 
components of all possible $SO(4)$-tensors and multiplying the resulting terms by the appropriate powers of 
$m_q$, see \cite{VanKolck:1993ee} for more details and the explicit construction along these lines. 
A simpler (but equivalent) method makes use of the following trick. Consider the massless QCD Lagrangian 
in the presence of an external hermitian scalar field $s$ interacting with the quarks 
via the term $- \bar q s q$. The resulting Lagrangian is chiral invariant provided the  
scalar source $s$ transforms under chiral rotations according to:
\beq
\label{fieldS}
s \to s' =L s R^\dagger = R s L^\dagger \,, 
\eeq
where the second equality follows from the hermiticity of $s$. To recover QCD from the new theory, 
the external field needs to be set to the value $s = \mathcal{M}$. To account for the explicit chiral symmetry 
breaking, we first write down the effective Lagrangian for the new theory by listing all possible 
chiral invariant terms constructed from $s$ and (derivatives of) $U$ and $U^\dagger$ and then set 
$s = \mathcal{M}$. Since the quark masses are treated as a small perturbation, the 
leading symmetry-breaking terms should contain a minimal possible number of derivatives and 
just one insertion of $s$. Given that $U \to U^\dagger$ under parity transformation, there exist only 
one symmetry-breaking term without derivatives:
\beq
\label{4.55}
\mathcal{L}_{\rm SB} = \frac{F^2 B}{2} \langle s U + s U^\dagger \rangle  \bigg|_{s=\mathcal{M}}
= F^2 B ( m_u + m_d) - \frac{B}{2} (m_u + m_d) \fet{\pi}^2 + \mathcal{O}
(\fet{\pi}^4) \;,
\eeq  
where $B$ is a LEC.  The first term is a constant and does not contribute to the $S$-matrix.
The second one gives rise to the pion mass $-(1/2) M^2 \fet{\pi}^2$
with $M^2 = (m_u + m_d ) B$. Note that to leading order in $m_{u,d}$, one 
has equal masses for all pions $\pi^+$, $\pi^-$ and $\pi^0$. Further, the LEC 
$B$ can be shown to be related to the quark condensate according to 
$\langle 0 | \bar{u} u | 0 \rangle= \langle 0 | \bar{d} d | 0 \rangle= 
- F_\pi^2 B (1 + \mathcal{O} (\mathcal{M}))$ \cite{Gasser:1983yg}. Modulo corrections
of higher order in the quark masses, the 
experimentally measured pion mass $M_\pi$ coincides with $M$: $M_\pi^2 = M^2 + \mathcal{O}(m_q^2)$. 

How important are chiral-symmetry breaking terms as compared to chiral-invariant ones?
When constructing the most general chiral-invariant effective Lagrangian, various terms 
were classified according to the number of derivatives.
When calculating observables, the derivatives acting on pion fields 
generate powers of external momenta which are assumed to be low. The EFT considered so far is 
typically applied to processes characterized by pion momenta of the order of the pion 
mass.\footnote{For an example of EFT in a different kinematical regime with nonrelativistic 
mesons see \cite{Colangelo:2006va}.} It is, therefore, natural to count the pion mass in the effective Lagrangian 
on the same footing as a derivative. We thus end up with the following 
lowest-order Lagrangian:
\beq
\label{lagrpifin}
\mathcal{L}_\pi^{(2)} = \frac{F^2}{4} \langle \partial_\mu U \partial^\mu U^\dagger  + 2 B  
(\mathcal{M} U + \mathcal{M} U^\dagger )\rangle \,.
\eeq  
For the sake of completeness, the next-higher order Lagrangian reads \cite{Gasser:1983yg}:
\beqa
\mathcal{L}_\pi^{(4)} &=& \frac{l_1}{4} \langle \partial_\mu U \partial^\mu U^\dagger \rangle^2 + 
\frac{l_2}{4} \langle \partial_\mu U \partial_\nu U^\dagger \rangle \langle \partial^\mu U \partial^\nu U^\dagger \rangle 
+ \frac{l_3}{16} \langle 2 B \mathcal{M} (U + U^\dagger ) \rangle^2 + \ldots \nn
&-& \frac{l_7}{16} \langle  2 B \mathcal{M} (U - U^\dagger ) \rangle^2\,, 
\eeqa
where the ellipses refer to terms that involve external sources and the $l_i$ are the corresponding LECs. 

\subsection{Power counting}
\begin{figure*}
\vspace{0.3cm}
\centerline{
\includegraphics[width=0.5\textwidth]{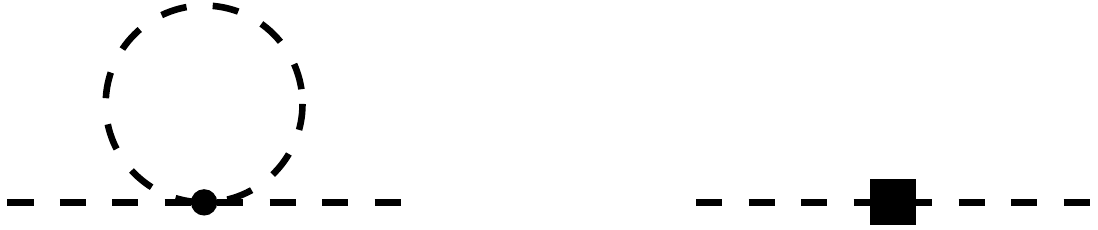}
}
\vspace{-0.2cm}
\caption[fig4aa]{\label{fig1} Tadpole contribution from $\mathcal{L}_\pi^{(2)}$ (left) and tree 
contribution from  $\mathcal{L}_\pi^{(4)}$ (right) to the pion self energy. 
}
\vspace{0.2cm}
\end{figure*}

Having constructed the most general effective Lagrangian for pions in harmony with the chiral
symmetry of QCD, we now need to figure out how to compute observables. At first sight, the effective 
Lagrangian seems to be of less practical value due to an infinite number of the unknown LECs. 
Even worse, all interaction terms entering $\mathcal{L}_\pi$ are non-renormalizable in the usual  
sense\footnote{This implies that the structure of local ultraviolet divergences generated by loop diagrams 
with vertices from  $\mathcal{L}_\pi^{(2)}$ is different from $\mathcal{L}_\pi^{(2)}$.} 
contrary to field theories such as e.g.~QED and QCD. What at first sight appears to cause a problem, 
namely non-renormalizability of the theory, in fact, turns out to be a crucial feature for the whole approach 
to be useful. As demonstrated in the seminal paper by Weinberg \cite{Weinberg:1978kz}, the effective Lagrangian $\mathcal{L}_\pi$ 
can be used to compute low-energy observables (such as e.g.~scattering amplitudes) in a systematically improvable 
way via an expansion in powers of $Q/\Lambda_\chi$, where $Q$ represents the soft scale associated with external 
momenta or the pion mass $M_\pi$ while $\Lambda_\chi$, the so-called chiral-symmetry-breaking scale, is the 
hard scale that drives the LECs in $\mathcal{L}_\pi$. This expansion is referred to as the chiral expansion, 
and the whole approach carries the name of chiral perturbation theory. 
Consider an arbitrary multi-pion scattering process 
with all initial and final pion momenta of the order of $M_\pi$. In order to decide on the importance 
of a particular Feynman diagram, we have to determine the power of the soft scale associated with it.  
For that we first need to clarify an important issue related to the counting  of virtual momenta 
which are being integrated over in the loop integrals. What scale do we associate with such virtual momenta?
When calculating Feynman diagrams in ChPT, one generally encounters two kinds of loop integrals. 
First, there are cases in which the integrand dies out fast enough when the loop momenta go to infinity so 
that the corresponding integrals are well-defined. Since the hard scale only enters the LECs 
and thus factorizes out, the integrands involve only soft scales (external momenta and the pion mass) 
and the loop momenta. Given that the integration is carried over the whole 
range of momenta, the resulting mass dimension of the integral is obviously driven by 
the soft scales. Thus, in this case we can safely count all virtual momenta as the soft scale. 
The second kind of integrals involves ultraviolet divergences and requires regularization
and renormalization. Choosing renormalization conditions in a suitable way, one can ensure that 
virtual momenta are (effectively) of the order of the soft scale. This is achieved automatically  
if one uses a mass-independent regularization such as e.g.~dimensional regularization (DR). Consider, for example,
the integral 
\beq
I = \int \frac{d^4 l}{(2 \pi)^4} \frac{i}{l^2 - M^2 + i \epsilon}\,,
\eeq
that enters the pion self energy due to the tadpole diagram shown in Fig.~\ref{fig1}. 
Evaluating this quadratically divergent integral in dimensional regularization one obtains
\beq
I \to I^{\rm reg} =   \mu^{4 - d} \int \frac{d^d l}{(2 \pi)^d} \frac{i}{l^2 - M^2} = 
\frac{M^2}{16 \pi^2} \, \ln \left( \frac{M^2}{\mu^2} \right)  
+ 2 M^2 L ( \mu )
+ \mathcal{O} (d - 4) \,,
\eeq
where $\gamma_E$ is the Euler constant and $\mu$ is the scale introduced by dimensional 
regularization and the quantity $L (\mu ) $ is given by
\beq
\label{lmu}
L (\mu )  = \frac{\mu ^{d-4}}{16 \pi^2} \left\{ \frac{1}{d-4} - \frac{1}{2} (\ln (4 \pi ) + \Gamma ' (1) + 1 ) \right\}\,, 
\mbox{\hskip 1 true cm} 
\Gamma ' (1) = -0.577215\ldots \,.
\eeq
The second term on the right-hand side of the above expression diverges
in the limit $d \to 4$ but can be absorbed into an appropriate redefinition of the LECs in $\mathcal{L}_\pi^{(4)}$
(renormalization). Notice that, as desired, the mass dimension of the finite term is driven by the 
soft scale $M$. I further emphasize that the scale $\mu$ introduced by dimensional 
regularization has to be chosen of the order $\mu \sim M_\pi$ in order to prevent the 
appearance of large logarithms in DR expressions. Last but not least, note that one can, in principle, use
different regularization methods such as e.g.~cutoff regularization provided they 
respect chiral symmetry.\footnote{When cutoff regularization is used, a special care is required regarding 
the treatment of non-covariant pieces in the pion propagator, see.~\cite{Gerstein:1971fm}.}
Contrary to dimensionally-regularized expressions, cutoff-regularized integrals  
do not scale properly i.e.~their mass dimension is not generated exclusively by the 
soft scales. The renormalized expressions emerging after absorbing the positive powers and 
logarithms of the cutoff into an appropriate redefinition of the LECs do, however,  
feature the expected scaling behavior. 
 
I am now in the position to discuss the chiral power counting, i.e.~the expression 
that determines the power $\nu$ of the expansion parameter $Q/\Lambda_\chi$ for a given Feynman 
diagram. 
This can be achieved by carefully counting the powers of small momenta associated with  
derivatives entering the vertices in $\mathcal{L}_\pi$, pion propagators, 
integrations over the loop momenta and the  $\delta$-functions. 
Using certain topological identities, one obtains the following expression for the 
chiral dimension of a connected Feynman diagram:
\beq
\label{pow_orig0}
\nu = 2  + 2 L + \sum_i V_i \Delta_i \,, \quad \quad
\Delta_i = d_i  - 2\,,
\eeq
where $L$ refers to the number of loops. 
This result has been first obtained by Weinberg in \cite{Weinberg:1978kz}. 
Notice that in order for perturbation theory to work, $\mathcal{L}_\pi$  must contain no 
interactions with $\Delta_i \leq 0$ since, otherwise, adding new vertices would not increase or even lower the chiral dimension 
$\nu$.  This feature is guaranteed by the spontaneously broken chiral symmetry of QCD which 
ensures that only non-renormalizable interactions with at least two derivatives 
or powers of the pion mass appear in  $\mathcal{L}_\pi$. In particular, chiral symmetry 
forbids the renormalizable derivative-less interaction of the type $\fet \pi^4$. 

\begin{figure*}
\vspace{0.3cm}
\centerline{
\includegraphics[width=0.9\textwidth]{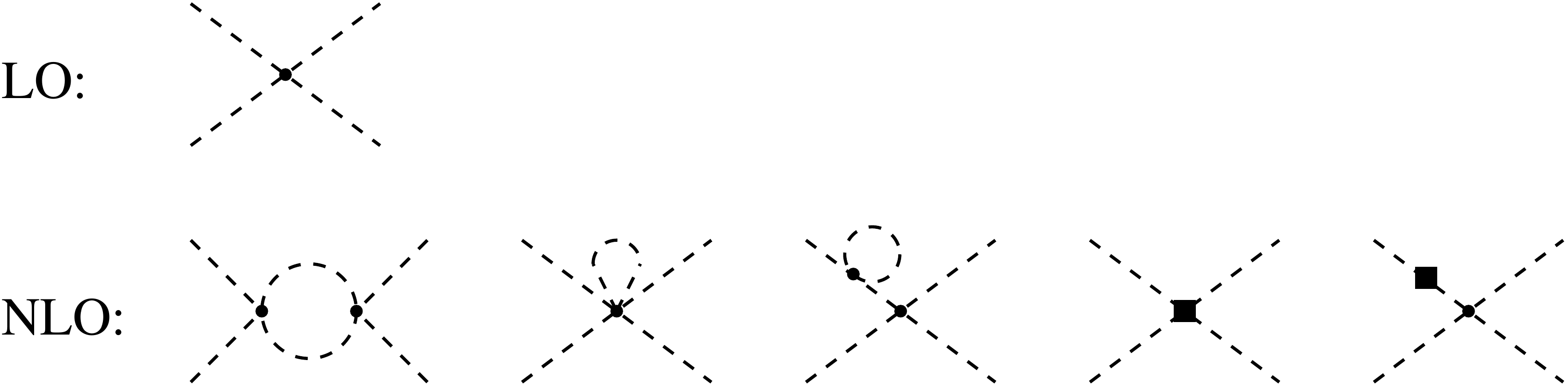}
}
\vspace{-0.2cm}
\caption[fig4aa]{\label{fig2} Diagrams contributing to pion-pion scattering in at 
leading- and next-to-leading order in ChPT. Solid dots and filled rectangles represent vertices 
from $\mathcal{L}_\pi^{(2)}$ and $\mathcal{L}_\pi^{(4)}$, respectively.  
}
\vspace{0.2cm}
\end{figure*}

Consider now pion-pion scattering as an illustrative example. Eq.~(\ref{pow_orig0})
tells us that the leading contribution to the scattering amplitude is generated 
by tree diagrams ($L=0$) constructed from the lowest-order vertices with $\Delta_i  =0$ 
i.e.~the ones from $\mathcal{L}_\pi^{(2)}$,
see Fig.~\ref{fig2}. The amplitude scales as $Q^2$. 
The corrections result from one-loop graphs involving all vertices from  $\mathcal{L}_\pi^{(2)}$
as well as tree graphs with a single insertion from $\mathcal{L}_\pi^{(4)}$, see Fig.~\ref{fig2}. 
They appear at order $Q^4$ and are suppressed by two powers of momenta or one power of the quark masses compared to the 
leading-order contribution. It is easy to verify that all diagrams in the bottom line of  this figure
scale, indeed, as $Q^4$. For example, for the first diagram, four powers of momenta arise from  
the vertices and another four powers of momenta emerge from the loop integration. One should further take 
into account four powers of momenta generated  in the denominator by the pion propagators. Thus, the total 
power of the soft scale is indeed four.  All ultraviolet divergences entering the loop integrals are local
and absorbable into redefinition of the LECs in   
$\mathcal{L}_\pi^{(4)}$ (when using dimensional regularization),  
as it represents the most general, approximately chiral invariant, 
local interaction of Goldstone bosons at order $Q^4$. 
The divergent parts of the LECs $l_i$ have been worked out in \cite{Gasser:1983yg} using the heat-kernel method.
The finite parts of the $l_i$'s are not fixed by chiral symmetry and have to be determined 
from the data or lattice QCD calculations. 
In the Goldstone boson sector, even a number of two-loop 
calculations (i.e.~at order $Q^6$) have already been performed, see Ref.~\cite{Bijnens:2006zp} for a review article. 
In particular, an impressive theoretical prediction has been made for 
the isoscalar S-wave $\pi \pi$ scattering length $a_0^0$ by Colangelo et al.~\cite{Colangelo:2000jc} who 
combined the two-loop calculation \cite{Bijnens:1997vq} with dispersion relations to predict 
$a_0^0 = 0.220 \pm 0.005$. 
To compare, the leading-order calculation by Weinberg yielded $a_0^0 = 0.16$  \cite{Weinberg:1966kf} while 
the next-to-leading value obtained by Gasser and Leutwyler is $a_0^0 = 0.20$ \cite{Gasser:1983yg}.
The results of the recent E865 experiment at Brookhaven \cite{Pislak:2003sv} 
and the NA48/2 experiment 
at CERN  \cite{Batley:2007zz} combined with the older measurement by the Geneva-Saclay collaboration  
beautifully confirmed the prediction of the two-loop analysis of  
Ref.~\cite{Colangelo:2000jc} yielding the value 
$a_0^0 = 0.217 \pm 0.008\mbox{ (exp)} \pm 0.006\mbox{ (th)}$  \cite{Colangelo:2008sm}.  
This combined result accounts for isospin breaking corrections, see Ref.~\cite{Colangelo:2008sm}
for more details.  

The procedure outlined above can, in principle, be extended to arbitrarily high orders in the 
low-energy expansion. Clearly, the accuracy of the calculations depends crucially on the value 
of the hard scale $\Lambda_\chi$ which   sets the (maximal) radius of convergence of the 
chiral expansion. The $\rho$-meson is the first meson of the non-Goldstone type 
and shows up as a resonance in p-wave $\pi \pi$ scattering. Such resonances represent truly non-perturbative
phenomena that cannot be described in standard ChPT\footnote{For an extension of ChPT to the resonance 
region, the so-called unitarized ChPT, see \cite{Pelaez:2003ip} and references therein.}. Consequently, their appearance 
signals the breakdown of the chiral expansion. This leads to the estimation 
$\Lambda_\chi \sim M_\rho \simeq 770$ MeV. A related observation that matches naturally 
the above estimation was made by Manohar and Georgi who pointed out that $\Lambda_\chi$
cannot be larger than $4 \pi F_\pi \simeq 1200$ MeV since this number sets the scale  
that controls the running of the renormalized LECs when shifting the renormalization point.   

Last but not least, I would like to summarize and underline the special role and importance 
of the chiral symmetry for the whole approach. 
First of all, it implies severe constraints on the interactions in the effective 
Lagrangian and relates the strengths of various multi-pion vertices. For example, the leading-order Lagrangian 
$\mathcal{L}_\pi^{(2)}$ in Eq.~(\ref{lagrpifin}) gives rise to infinitely many vertices, when 
expanded in powers of pion fields, whose strengths are determined by just two (!) LECs $F$ and $B$.   
$\mathcal{L}_\pi^{(2)}$ allows to compute the leading contribution to scattering amplitudes for multi-pion processes 
and to relate the strengths of the corresponding matrix elements, thus featuring a remarkable predictive 
power. Moreover, as we saw through the explicit construction, the spontaneously
broken chiral symmetry of QCD prevents the appearance of derivative-less interactions between pions
in the effective Lagrangian. The only derivative-less interactions in $\mathcal{L}_\pi$ are due 
to explicit chiral symmetry breaking in  $\mathcal{L}_{QCD}$ and are suppressed by powers of the 
light quark masses. When calculating S-matrix elements, the derivatives entering the vertices generate powers 
of external momenta. Consequently, the interaction between pions becomes weak at vanishingly low 
energies and would even completely disappear if chiral symmetry were exact. This turns out to be  
a general feature of Goldstone bosons and is not restricted to the $SU(2)_L \times SU(2)_R$ group. 
This allows to compute low-energy hadronic observables in a systematic way via the chiral expansion, 
i.e.~the dual expansion in powers of momenta and quark masses about the kinematical point 
corresponding to the free theory (assuming that the actual quark masses in the real world 
are low enough for such an
expansion to converge).

\subsection{Inclusion of nucleons}
So far we only discussed interactions between Goldstone bosons. 
We now extend these considerations to include nucleons. More precisely, we are interested 
in describing reactions involving pions with external momenta of the order of $M_\pi$ 
and (essentially) non-relativistic nucleons whose three-momenta 
are of the order of $M_\pi$. Similarly to the triplet 
of pion fields, the isospin doublet of the nucleon fields should transform 
nonlinearly under the chiral $SU(2)_L \times SU(2)_R$ but linearly under the vector subgroup 
$SU(2)_V$. The unitary matrix $U$ introduced in Eq.~(\ref{matrU}) is less 
useful when constructing the Lagrangian involving the nucleons. It is more convenient to 
introduce its square root $u$, $U = u^2$. The transformation properties of $u$ under chiral 
rotations can be read off from Eq.~(\ref{trafoU}):
\beq
\label{defu}
u \to u' = \sqrt{L u R^\dagger} \equiv L  u h^{-1} = h u R^\dagger \,,
\eeq
where I have introduced the unitary matrix $h = h (L, R, U)$ given by 
$h = \sqrt{L U R^\dagger}^{-1} L \sqrt{U}$ which is sometimes referred to as a compensator field. 
The last equality in Eq.~(\ref{defu}) 
follows from $U' = u' u' = Luh^{-1} u'= L u u R^\dagger$. Notice that 
since pions transform linearly under isospin rotations corresponding to $L = R = V$ with 
$U \to  U' = V U V^\dagger$ and, accordingly,  $u \to  u' = V u V^\dagger$, the compensator 
field  in this particular case  becomes $U$-independent and coincides with $V$. 

\begin{minipage}{\textwidth}
\vskip 0 true cm
\rule{\textwidth}{.2pt}
{\it
Exercise:  calculate the explicit form of the compensator field $h(L, R, \fet \pi)$ 
for infinitesimal chiral transformations using 
Eq.~(\ref{Uexplicit}) and keeping only terms that are at most 
linear in the pion fields. Verify that $h$ indeed reduces to the isospin transformation for  $L = R = V$.
} \\
\vskip -0.8 true cm
\rule{\textwidth}{.2pt}
\end{minipage}

\medskip
It can be shown that $\{ U, \, N \}$ define a nonlinear realization of the chiral group
if one demands that 
\beq
N \to N' = h N\,.  
\eeq
I do not give here the proof of this statement and refer the interesting reader 
to Ref.~\cite{Coleman:1969sm,Callan:1969sn}. Moreover, this nonlinear realization 
obviously fulfills the desired feature that pions and nucleons transform linearly under 
isospin rotations. Similarly to the purely Goldstone boson case, one can show that 
all other possibilities to introduce the nucleon fields are identical with the above realization 
modulo nonlinear field redefinitions. The most general chiral invariant Lagrangian 
for pions and nucleons can be constructed from \emph{covariantly} transforming building 
blocks, i.e.~$O_i \to O_i ' = h O_i h^{-1}$, by writing down all possible terms of 
the form $\bar N O_1 \ldots O_n N$. The covariant (first) derivative of the pion field 
is given by 
\beq
u_\mu \equiv i u^\dagger (\partial_\mu U) u^\dagger = - 
\frac{\fet \tau \cdot \partial_\mu \fet \pi}{F} + \mathcal{O} (\fet \pi^3) \to u_\mu ' =  h u_\mu h^{-1}\,,
\eeq
and is sometimes referred to as chiral vielbein. The derivative of the nucleon field, 
$\partial_\mu N$, does not transform covariantly, i.e.~$\partial_\mu N \to (\partial_\mu N)' \neq
h \partial_\mu N$ since the compensator field $h$ does, in general, depend on space-time 
(through its dependence on $U$). The covariant derivative of the nucleon field $D_\mu N$, 
$D_\mu N \to (D_\mu N)' = h D_\mu N$, is given by 
\beq
D_\mu N \equiv  (\partial_\mu  + \Gamma_\mu ) N\,, \quad
\mbox{with} \quad
\Gamma_\mu \equiv \frac{1}{2} \left( u^\dagger \partial_\mu u + u \partial_\mu u^\dagger \right) =
\frac{i}{4 F^2} \fet \tau \cdot \fet \pi \times \partial_\mu \fet \pi + \mathcal{O} (\fet \pi^4) 
\,.
\eeq
The so-called connection $\Gamma_\mu$ can be used to construct higher covariant derivatives 
of the pion field, for example:
\beq
u_{\mu \nu} \equiv \partial_\mu u_\nu + [ \Gamma_\mu , \, u_\nu ]\,.
\eeq 

\begin{figure*}
\vspace{0.3cm}
\centerline{
\includegraphics[width=0.3\textwidth]{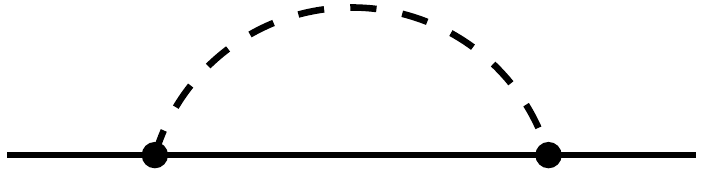}
}
\vspace{-0.2cm}
\caption[fig4aa]{\label{fig_nucl} Leading pion loop contribution to the nucleon self energy. 
Solid line represents the nucleon. 
}
\vspace{0.2cm}
\end{figure*}

To first order in
the  derivatives, the most general pion-nucleon 
Lagrangian takes the form \cite{Gasser:1987rb}
\beq
\label{LeffN}
\mathcal{L}_{\pi N}^{(1)} = \bar N \left( i \gamma^\mu D_\mu - m  + \frac{g_A}{2} \gamma^\mu \gamma_5 u_\mu \right) N\,,
\eeq
where $m$ and $g_A$ are the bare nucleon mass and the axial-vector coupling constant and 
the superscript of $\mathcal{L}_{\pi N}$ denotes the power of the soft scale $Q$.
Contrary to the pion mass, the nucleon mass 
does not vanish in the chiral limit and introduces an additional hard scale in the problem. 
Consequently, terms proportional to $D_0$ and $m$ in Eq.~(\ref{LeffN})
are individually large. 
It can, however, be shown that $(i \gamma^\mu D_\mu - m )N \sim \mathcal{O} (Q )$ \cite{Krause:1990xc}. 
The appearance of the additional hard scale associated with the nucleon mass invalidates the 
power counting for dimensionally regularized expressions since the contributions from 
loop integrals involving nucleon propagators are not automatically suppressed. 
To see this consider the correction to the nucleon mass $m_N$ due to the pion loop shown in Fig.~\ref{fig_nucl}. 
Assuming that the nucleon and pion propagators scale as $1/Q$ and $1/Q^2$, respectively, and taking into account 
$Q^4$ from the loop integration and $Q^2$ from the derivatives entering the $g_A$-vertices, the pion loop 
contribution to the nucleon self energy $\Sigma ( p )$ is expected to be of the order $\sim Q^3$. 
Consequently, the corresponding nucleon mass shift $\delta m_N = \Sigma (m_N)$ 
is expected to be $\propto M_\pi^3$ (since no other soft scale is left). 
Explicit calculation, however, shows that the resulting nucleon mass shift does not vanish in the 
chiral limit \cite{Gasser:1987rb}:
\beq
\label{mNrelativ}
\delta m_N \big|_{\rm loop , \, rel} \stackrel{\mathcal{M} \to 0}{=}
 - \frac{3 g_A^2 m^3}{F^2} \left( L(\mu ) + \frac{1}{32 \pi^2} \ln \frac{m^2}{\mu^2} \right) + \mathcal{O} (d-4 )\,,
\eeq
where 
the quantity $L(\mu )$ is defined in Eq.~(\ref{lmu}). 
The result in Eq.~(\ref{mNrelativ}) implies that the nucleon mass receives a contribution 
which is formally of the order $\sim m \, (m/4 \pi F )^2$ and is not suppressed compared to $m$. 
The bare nucleon mass $m$ that enters the lowest-order Lagrangian $\mathcal{L}_{\pi N}^{(1)}$
gets renormalized. This is in contrast  
to the purely mesonic sector where loop contributions are always suppressed by powers of the soft 
scale and the parameters $F$ and $B$ in the lowest-order Lagrangian $L_\pi^{(2)}$ remain unchanged 
by higher-order corrections (if mass-independent regularization is used). I emphasize, however, 
that even though DR expressions do not automatically obey the dimensional power counting with nucleons 
being treated relativistically, the proper scaling in agreement with naive dimensional analysis 
can be restored via appropriately chosen renormalization 
conditions \cite{Fuchs:2003qc}. Stated differently, one can (and should in order for the EFT to be useful) 
choose renormalization conditions in such a way, 
that all momenta flowing through diagrams are effectively of the order of $Q$. 
Another, simpler way to ensure the proper power counting exploits   
the so-called heavy-baryon formalism \cite{Jenkins:1990jv,Bernard:1992qa} which is closely related to the 
nonrelativistic expansion due to Foldy and Wouthuysen \cite{Foldy:1950aa} and is also widely used in 
heavy-quark effective field theories. The idea is to decompose 
the nucleon four-momentum $p^\mu$ according to
\beq
\label{HBmomentum}
p_\mu = m v_\mu + k_\mu \,,
\eeq
with $v_\mu$ the four-velocity of the nucleon satisfying $v^2 = 1$ and $k_\mu$ its small residual momentum,
$v \cdot k \ll m$. One can thus decompose the nucleon field $N$ in to the velocity eigenstates 
\beq
N_v = e^{i m v \cdot x} P_v^+ N\,, \mbox{\hskip 1.5 true cm}
h_v = e^{i m v \cdot x} P_v^- N\,,
\eeq
where $P_v^\pm = (1 \pm \gamma_\mu v^\mu )/2$ denote the corresponding projection operators.
In the nucleon rest-frame with  
$v_\mu = (1, 0, 0, 0 )$, the quantities $N_v$ and $h_v$ coincide with the 
familiar large and small components of the
free positive-energy Dirac field (modulo the modified time dependence).  
One, therefore, usually refers to $N_v$ and $h_v$ as to the large and small 
components of $N$. The relativistic Lagrangian $\mathcal{L}_{\pi N}^{(1)}$ in Eq.~(\ref{LeffN}) 
can be expressed in terms of $N_v$ and $h_v$ as:
\beq
\mathcal{L}_{\pi N}^{(1)} = \bar N_v  \mathcal{A} N_v  + \bar h_v \mathcal{B} N_v + \bar N_v  \gamma_0 \mathcal{B}^\dagger 
\gamma_0 h_v - \bar h_v \mathcal{C} h_v \,,
\eeq
where 
\beq
\mathcal{A} = i (v \cdot D ) + g_A (S \cdot u )\,,\quad 
\mathcal{B} = - \gamma_5 \left[ 2 i (S \cdot D) + \frac{g_A}{2} (v \cdot u ) \right]\,, \quad
\mathcal{C} = 2 m + i (v \cdot D ) + g_A (S \cdot u )\,,
\eeq
and $S_\mu = i \gamma_5 \sigma_{\mu \nu} v^\nu$ is the nucleon spin operator. 
One can now use the equations of motion for the large and small component fields to 
completely eliminate $h_v$ from the Lagrangian. Utilizing the more elegant path integral formulation \cite{Mannel:1991mc}, 
the heavy degrees of freedom can be integrated out performing the Gaussian integration over the 
(appropriately shifted) variables $h_v$, $\bar h_v$. This leads to the effective Lagrangian of the form \cite{Bernard:1992qa}
\beq
\label{Lfin}
\mathcal{L}_{\pi N} = \bar N_v  \left[ \mathcal{A} + (\gamma_0 \mathcal{B}^\dagger \gamma_0 ) 
\mathcal{C}^{-1} \mathcal{B} \right] N_v
= \bar N_v  \left[ i (v \cdot D ) + g_A (S \cdot u ) \right] N_v  + \mathcal{O} \left(\frac{1}{m} \right)\,.
\eeq
Notice that the (large) nucleon mass term has disappeared from the Lagrangian, 
and the dependence on $m$ in $\mathcal{L}_{\pi N}^{\rm eff}$
resides entirely in new vertices suppressed by powers of $1/m$. The heavy-baryon propagator 
of the nucleon is simply $1/(v \cdot k + i \epsilon )$ and can be obtained from the $1/m$ expansion 
of the Dirac propagator using Eq.~(\ref{HBmomentum}) and assuming $v \cdot k \ll m$:
\beq
\label{Dirac}
\frac{p \palka + m}{p^2 - m^2 + i \epsilon} = \frac{\Lambda_+}{v \cdot k + i \epsilon} + 
\mathcal{O} \left( m^{-1} \right)\,,
\eeq
where $\Lambda_+ = (p \palka  + m)/(2m)$ is a projection operator on the states of positive energy.   
The advantage of the heavy-baryon formulation (HBChPT) compared to
the relativistic one can be illustrated using the previous example of
the leading one-loop correction to the nucleon mass 
\beq
\label{mNHB}
\delta m_N \big|_{\rm loop, \, HB} = - \frac{3 g_A^2 M_\pi^3}{32 \pi F^2} \,.
\eeq
Contrary to the relativistic CHPT result in Eq.~(\ref{mNrelativ}), the loop correction in HBChPT 
is finite (in DR) and vanishes in the chiral limit. The parameters in the lowest-order 
Lagrangian do not get renormalized due to higher-order corrections which are suppressed by powers of $Q/\Lambda_\chi$. 
Notice further that Eq.~(\ref{mNHB}) represents the leading contribution to the nucleon mass which 
is nonanalytic in quark masses. It agrees with the result obtained by Gasser et al.~based on  the relativistic 
Lagrangian in Eq.~(\ref{LeffN}) \cite{Gasser:1987rb}. 
In general, the power $\nu$ of a soft scale $Q$ for connected contributions to 
the scattering amplitude can be read off from the extension of Eq.~(\ref{pow_orig0}) to the single-nucleon 
sector which has the form:
\beq
\label{pow1N}
\nu = 1 + 2 L + \sum_i V_i \Delta_i\,,
\quad \quad \mbox{with}
\quad \quad \Delta_i = -2 + \frac{1}{2} n_i + d_i\,,
\eeq
with $n_i$ being the number of nucleon field operators at a vertex $i$ with the chiral dimension $\Delta_i$.  
Notice that no closed fermion loops appear in the heavy-baryon approach, so that exactly one nucleon line 
connecting the initial and final states runs through all diagrams in the single-baryon sector.

The heavy-baryon formulation outlined above can be straightforwardly extended to higher 
orders in the chiral expansion. At lowest orders in the derivative expansion, 
the effective Lagrangian $\mathcal{L}^{\Delta_i}$
for pions and nucleons takes the form \cite{Fettes:1998ud}:
 \beqa
\label{lagr0}
\mathcal{L}_{\pi N}^{(0)} &=& \bar{N} \left[
 i \, v\cdot D + g_A \, u \cdot S \right] N\,,\nn
\mathcal{L}^{(1)}_{\pi N} & = & \bar{N} \left[  c_1 \, \langle \chi_+ \rangle + c_2 \, (v \cdot
       u)^2 + c_3 \, u \cdot u 
+ c_4 \, [ S^\mu, S^\nu ] u_\mu u_\nu    + c_5 \langle \hat \chi_+ \rangle \right]
  N \,, \nn
\mathcal{L}_{\pi N}^{(2)} &=& \bar N \left[ \frac{1}{2 m} (v \cdot D)^2 
- \frac{1}{ 2 m}
D \cdot D + d_{16} S \cdot u \langle \chi_+ \rangle 
+ i d_{18} S^\mu [ D_\mu , \, \chi_-] + \ldots
 \right] N \,, \nn
\mathcal{L}_{\pi NN}^{(0)} &=& {} -\frac{1}{2} C_S ( \bar N N)   ( \bar N N )  +
2 C_T  ( \bar N S N ) 
\cdot ( \bar N S N ) \,, \nn
{\cal L}^{(1)}_{\pi NN} & = & \frac{D}{2} (\bar{N} N) (\bar{N} S \cdot u N)\,, \nn
\mathcal{L}_{\pi NN}^{(2)} &=& {} - \tilde C_1 \left[ ( \bar N D N) \cdot  ( \bar N D N) 
+ ( ( D \bar N) N) \cdot  ((D  \bar N) N) \right]\nn
&-&  2 (\tilde C_1  + \tilde C_2 )  ( \bar N D N) \cdot  ( (D \bar N ) N) 
-  \tilde C_2   ( \bar N N) \cdot  \left[ (D^2 \bar N ) N + \bar N D^2 N \right] + \ldots
\,, \nn
\mathcal{L}_{\pi NNN}^{(1)} & = &{} - \frac{E}{2}  (\bar{N} N) (\bar{N} \fet \tau N)
\cdot (\bar{N} \fet \tau N)\,.
\eeqa
Here, the ellipses refer to terms which do not contribute to the nuclear
forces up to next-to-next-to-leading order (N$^2$LO) except for
$\mathcal{L}_{\pi NN}^{(2)}$ where I have shown only a few terms in order to keep
the presentation compact. Further, here and in what follows I omit the subscript $v$ of the 
nucleon field operators. 
The quantity $\chi_+ = u^\dagger \chi u^\dagger + u
\chi^\dagger u$ with $\chi = 2 B  \mathcal{M}$ involves the explicit chiral
symmetry breaking due to the finite light quark masses and $\tilde O \equiv O - \langle O \rangle /2$.  
Finally, $c_i$, $d_i$, $C_i$, $\tilde C_i$, $D$ and $E$ denote the corresponding LECs.   

The presented elementary introduction into ChPT aims at providing the 
main conceptual ideas of this framework and is neither complete nor comprehensive. 
Excellent lecture notes on the discussed and related subjects  
\cite{Leutwyler:1994fi,Meissner:1997ws,Ecker:1998ai,Pich:1998xt,Gasser:2003cg,Kubis:2007iy} 
are highly recommended for 
further reading. A very comprehensive, textbook-like 
lecture notes can be found in Ref.~\cite{Scherer:2002tk}. Current frontiers and challenges in these 
fields are addressed in recent review articles \cite{Bernard:1995dp,Bernard:2007zu}, see also Ref.~\cite{CD09}.

\section{EFT for two nucleons}
\def\theequation{\arabic{section}.\arabic{equation}}
\label{sec4}

\subsection{ChPT and nucleon-nucleon scattering}
\label{sec:nuclearEFT}

As outlined in the previous section, ChPT can be straightforwardly extended to the single-nucleon sector
(apart from the complication related to the treatment of the nucleon mass). 
A generalization to processes involving two and more nucleons is much more difficult.  
Contrary to the interaction between Goldstone bosons, nucleons do interact with each other 
in the limit of vanishingly small momenta and quark masses. Chiral symmetry does not constrain 
few-nucleon operators in the effective Lagrangian which contains derivative-less terms, 
see Eq.~(\ref{lagr0}). In fact, the interaction between the nucleons at low energy is even strong 
enough to bind them together. Shallow bound states such as the deuteron, triton etc.~represent 
non-perturbative phenomena that cannot be described in perturbation theory. 

\begin{figure*}
\vspace{0.3cm}
\centerline{
\includegraphics[width=0.85\textwidth]{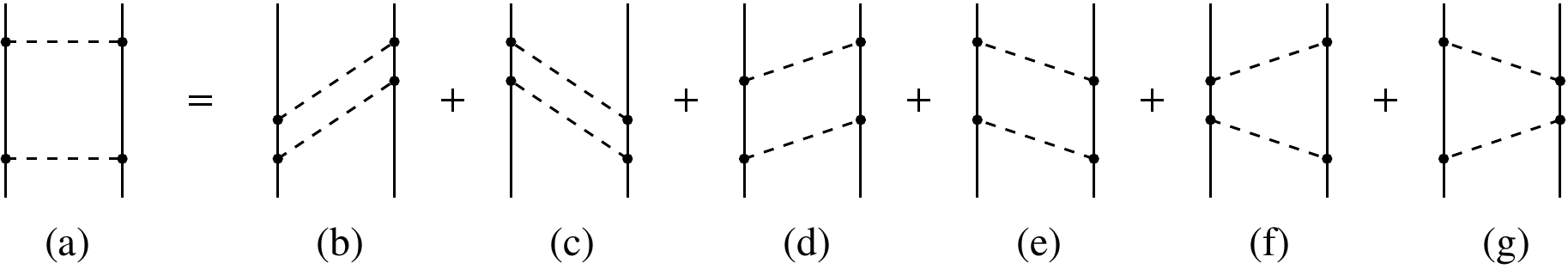}
}
\vspace{-0.2cm}
\caption[fig4aa]{\label{fig4aa} Two-pion exchange: Feynman diagram (a) and the corresponding time-ordered graphs (b-g). 
Solid (dashed) lines correspond to nucleons (pions). }
\vspace{0.2cm}
\end{figure*}

On the other hand, just following the naive dimensional analyses as we did in the previous section, 
the power counting can be straightforwardly generalized to connected Feynman diagrams involving $N$ nucleons 
leading to 
\beq
\label{powNN}
\nu = 2 - N + 2 L  + \sum_i V_i \Delta_i \,.
\eeq
This implies the usual suppression for loop diagrams and 
thus suggests that the interaction is weak. This conclusion is certainly not correct.  
So, what goes wrong? The reason why the naive dimensional analysis yields a wrong 
result is due to the appearance of infrared divergences (in the HBChPT) 
in diagrams which contain purely nucleonic intermediate states \cite{Weinberg:1990rz,Weinberg:1991um}. 
Consider the two-pion- ($2\pi$-) exchange box Feynman diagram shown in Fig.~\ref{fig4aa} (the diagram on the left-hand side).
In the nucleon rest frame with $v_\mu = (1, 0, 0, 0 )$, the four-momenta of the incoming nucleons 
are $(\vec k^2/(2 m) + \mathcal{O} (m^{-3}), \, \vec k )$ and 
$( \vec k^2/(2 m) + \mathcal{O} (m^{-3}), \,- \vec k )$. In the infrared regime with $\vec k =0$,  
the contribution of the box diagram takes the form
\beq
\int \frac{d^4 l }{(2 \pi)^4} \frac{P (l)}{( l^0 + i \epsilon ) ( - l^0 + i \epsilon ) ( l^2 - M_\pi^2 + i \epsilon )^2}\,,
\eeq 
where $l$ is the loop momentum and 
$P(l)$ is a polynomial whose explicit form is determined by the pion-nucleon vertex.  
The integral over $l^0$ possesses the so-called pinch singularity due to the poles at $l^0 = \pm i \epsilon$. 
Notice that such pinch singularities only show up in the case of at least two nucleons since 
for a single nucleon the contour of integration can be distorted to avoid the singularity. 
The singularities that appear in the box diagram are not ``real'' but an artefact of the heavy-baryon 
approximation for the nucleon propagators (static nucleons) that is not valid for such diagrams. 

\begin{minipage}{\textwidth}
\vskip 0 true cm
\rule{\textwidth}{.2pt}
{\it
Exercise:  verify this statement by 
using the standard Dirac propagators for the nucleon field, see Eq.~(\ref{Dirac}),
and making the nonrelativistic expansion \underline{after} carrying out the integration over $l^0$. 
} \\
\vskip -0.8 true cm
\rule{\textwidth}{.2pt}
\end{minipage}

\medskip
An alternative and, perhaps, more instructive way to explore the origin of the infrared enhancement is 
by using the so-called ``old-fashioned'' time-ordered perturbation theory instead of the covariant one. 
In time-ordered perturbation theory, the $T$-matrix is given by  
\beq
\label{old-fashioned}
T_{\alpha \beta} = (H_I)_{\alpha \beta} + \sum_a \frac{(H_I)_{\alpha a}  (H_I)_{a \beta}}{E_\beta - E_a + i \epsilon}
+ \sum_{a b} \frac{(H_I)_{\alpha a}  (H_I)_{ab}   (H_I)_{b \beta}}{(E_\beta - E_a + i \epsilon)
(E_\beta - E_{b} + i \epsilon)} + \ldots\,,
\eeq
where $H_I$ is the interaction Hamiltonian corresponding to the effective Lagrangian for pions and nucleons.
This expression should be familiar from Quantum Mechanics. Its derivation and application to 
quantum field theory can be found e.g.~in \cite{Schweber:1966aa}. 
Here, I use Latin letters for intermediate states,
which, in general, may contain any number of pions, in order to distinguish them from purely nucleonic states
denoted by Greek letters. I remind the reader that no nucleon-antinucleon pairs can be created or destroyed 
if nucleons are treated nonrelativistically. Consequently, all states contain the same number of nucleons. 
It is useful to represent various contributions to the scattering amplitude in terms of time-ordered diagrams. 
For example, the Feynman box diagram for NN scattering via  $2\pi$-exchange can be expressed 
as a sum of six time-ordered graphs, see Fig.~\ref{fig4aa}, 
which correspond to the following term in Eq.~(\ref{old-fashioned}): 
\beq
\label{TPEold-fashioned}
\sum_{abc} \frac{(H_{\pi NN})_{\alpha a}  (H_{\pi NN})_{ab}   
(H_{\pi NN})_{bc} (H_{\pi NN})_{c \beta}}{(E_\beta - E_a + i \epsilon)
(E_\beta - E_b + i \epsilon) (E_\beta - E_c + i \epsilon)} \,,
\eeq
where $H_{\pi NN}$ denotes the $\pi NN$ vertex. Actually, this expression can be obtained   
from carrying out the $l^0$-integration in the corresponding Feynman diagram (using Dirac propagators for the nucleons). 
It is easy to see that the contributions of diagrams (d-g) are 
\emph{enhanced} due to the presence of the small (of the order $Q^2/m$) energy denominator associated with the 
purely nucleonic intermediate state $| b \rangle $ which in the center-of-mass system (CMS) takes the form:
\beq
\frac{1}{E_\beta - E_b + i \epsilon} = \frac{1}{\vec p_\beta^{\, 2} /m  - \vec p_b^{\, 2} /m  + i \epsilon}\,.
\eeq
Notice that the energy denominators corresponding to the $\pi NN$ states  
$| a \rangle$ and $| c \rangle$ contain the pion energy $\omega_k \equiv \sqrt{\vec k \, ^2 + M_\pi^2}$ and 
are of the order $M_\pi \sim Q$ in agreement with the dimensional analysis. 
According to Weinberg, the failure of perturbation theory in the few-nucleon sector is caused
by the enhanced contributions of reducible diagrams, i.e.~those ones which contain purely 
nucleonic intermediate states. It should, however, be emphasized that the 
infrared enhancement is not sufficient to justify the need of non-perturbative 
resummation of the amplitude if one counts $m = \mathcal{O} (\Lambda_\chi )$. According to Eq.~(\ref{powNN})
and taking into account the infrared enhancement $\sim m/Q$ due to the purely nucleonic intermediate states, 
loop contributions are still suppressed by  $\sim Q m /\Lambda_\chi^2 \sim  Q /\Lambda$ for $m \sim \Lambda_\chi$. 
To overcome this conceptual difficulty, Weinberg proposed to treat the nucleon mass as 
a separate hard scale according to the rule \cite{Weinberg:1990rz,Weinberg:1991um}: 
\beq
\label{counting_m}
m \sim \frac{\Lambda_\chi^2}{Q} \gg \Lambda_\chi\,.
\eeq
The resulting power counting 
is referred to as the Weinberg power counting. 
I will also discuss some alternative scenarios. 

The infrared enhancement of the few-nucleon diagrams can be naturally taken into account 
by re-arranging the expansion in Eq.~(\ref{old-fashioned}) and casting it into  
the form of the Lippmann-Schwinger (LS) equation
\beq
\label{LSeqO}
T_{\alpha \beta}
= (V_{\rm eff})_{\alpha \beta} + \sum_\gamma \frac{(V_{\rm eff})_{\alpha \gamma}  
T_{\gamma \beta}}{E_\beta - E_\gamma + i \epsilon}\,,
\eeq
with the effective potential $(V_{\rm eff})_{\alpha \beta}$ defined as a sum of all 
possible irreducible diagrams (i.e. the ones 
which do not contain purely nucleonic intermediate states):
\beq
\label{ep}
(V_{\rm eff})_{\alpha \beta} = (H_I)_{\alpha \beta} + \sum_{a} \frac{(H_I)_{\alpha  a}  (H_I)_{a \beta}}
{E_\beta - E_{a} + i \epsilon}
+ \sum_{a b} \frac{(H_I)_{\alpha a}  (H_I)_{ a b}  
(H_I)_{ b \beta}}{(E_\beta - E_{ a} + i \epsilon)
(E_\beta - E_{ b} + i \epsilon)} + \ldots\,.
\eeq
Here, the states $| a \rangle$,  $| b \rangle$ contain at least one pion. 
The effective potential in Eq.~(\ref{ep}) does not contain small energy denominators and 
can be worked out within the low-momentum expansion
following the usual procedure of ChPT.  
After the potential is obtained at a given order in the chiral expansion, few-nucleon 
observables can be computed by solving the LS equation (\ref{LSeqO}), 
which leads to a nonperturbative resummation of the contributions resulting from reducible diagrams. 
The resulting two-step approach will be referred to as ChEFT in order to distinguish it 
from ChPT in the Goldstone boson and single-nucleon sectors.

\subsection{Analytic properties of the non-relativistic scattering amplitude}
\label{sec:analyt}

Before discussing various scenarios of organizing EFT for two nucleons, it is useful 
to recall general constraints imposed  on the partial wave scattering amplitude 
by analyticity.  
Consider two non-relativistic nucleons interacting via a potential
$V$. The corresponding $S$-matrix for an uncoupled channel with the orbital angular momentum $l$ 
is parametrized in terms of a single phase shift $\delta_l$ and can be written in terms of $T$-matrix as
\beq
\label{Tnorm}
S_l = e^{2 i \delta_l (k) } = 1 - i \left( \frac{k m}{8 \pi^2} \right) T_l (k)\,,
\eeq 
with $k$ denoting the CMS scattering momentum. 
The $T$-matrix can then be 
expressed in terms of the so-called effective range function $F_l (k) \equiv k^{2l+1} {\rm cot} \delta_l (k)$ via
\beq
\label{tmat}
T_l (k) = -\frac{16 \pi^2}{m} \frac{k^{2l}}{F_l (k) - i k^{2l+1}} \,. 
\eeq
In the complex energy plane, the scattering amplitude and thus also the $T$-matrix 
possess a so-called unitarity cut, a kinematic singularity due to two-body unitarity.  
The unitarity cut starts from the branch point at the threshold ($E=0$) and goes to 
positive infinity.  The dynamic singularities are associated with the interaction mechanism 
and are located at the negative real axis. 
For example, in the case of Yukawa potential $\sim \exp (-M r )/r$ corresponding to an exchange of 
a meson of mass $M$, the amplitude has a left-hand cut starting at $k^2 = -M^2/4$.  
Bound and virtual states reside as poles at the negative real axis ($k = i | k |$ and $k = -i | k |$ for 
bound- and virtual-state poles, respectively) while resonances show up as poles at complex energies.  
\begin{figure*}
\vspace{0.3cm}
\centerline{
\includegraphics[width=0.31\textwidth]{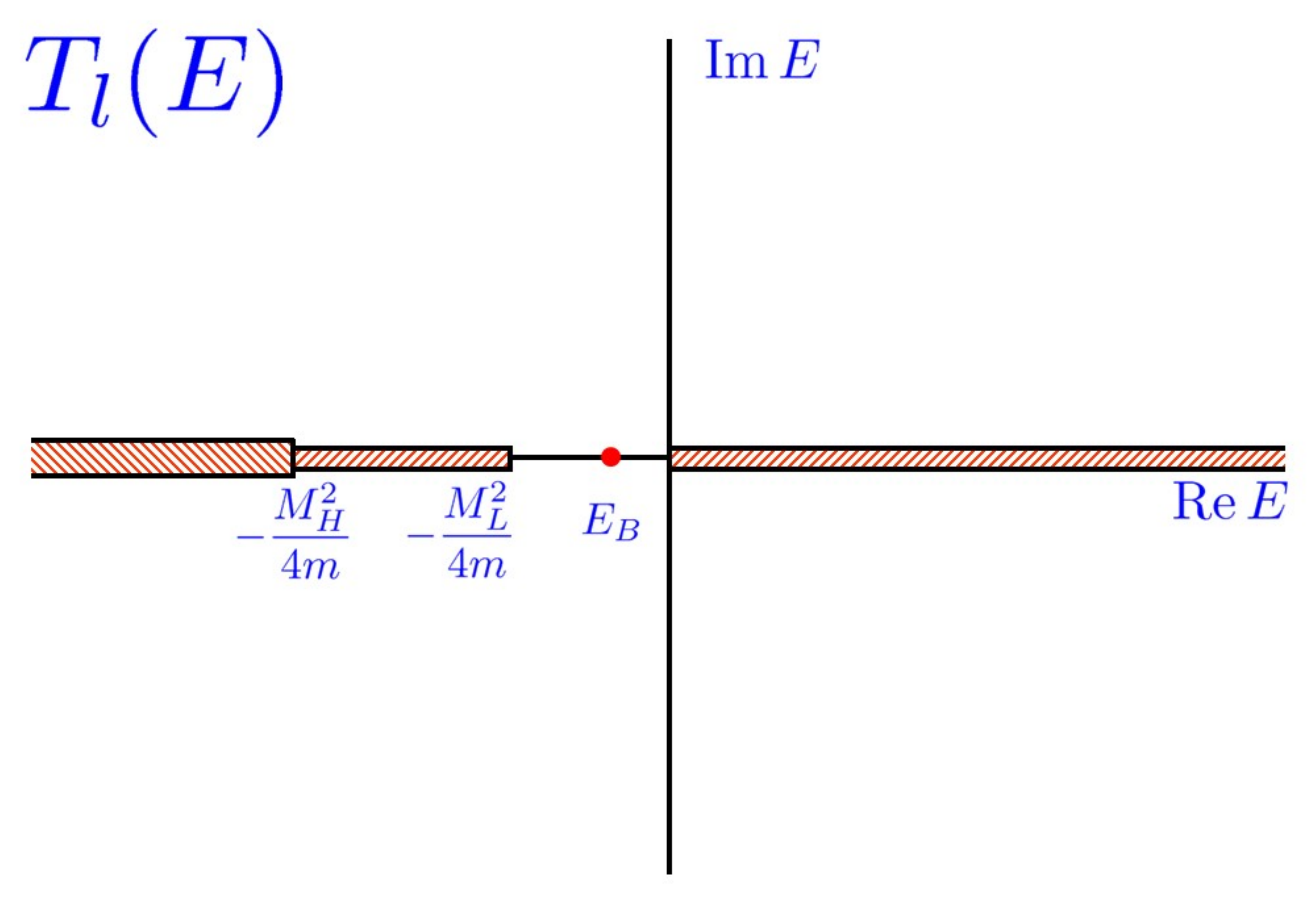}
\hskip 0.035\textwidth
\includegraphics[width=0.31\textwidth]{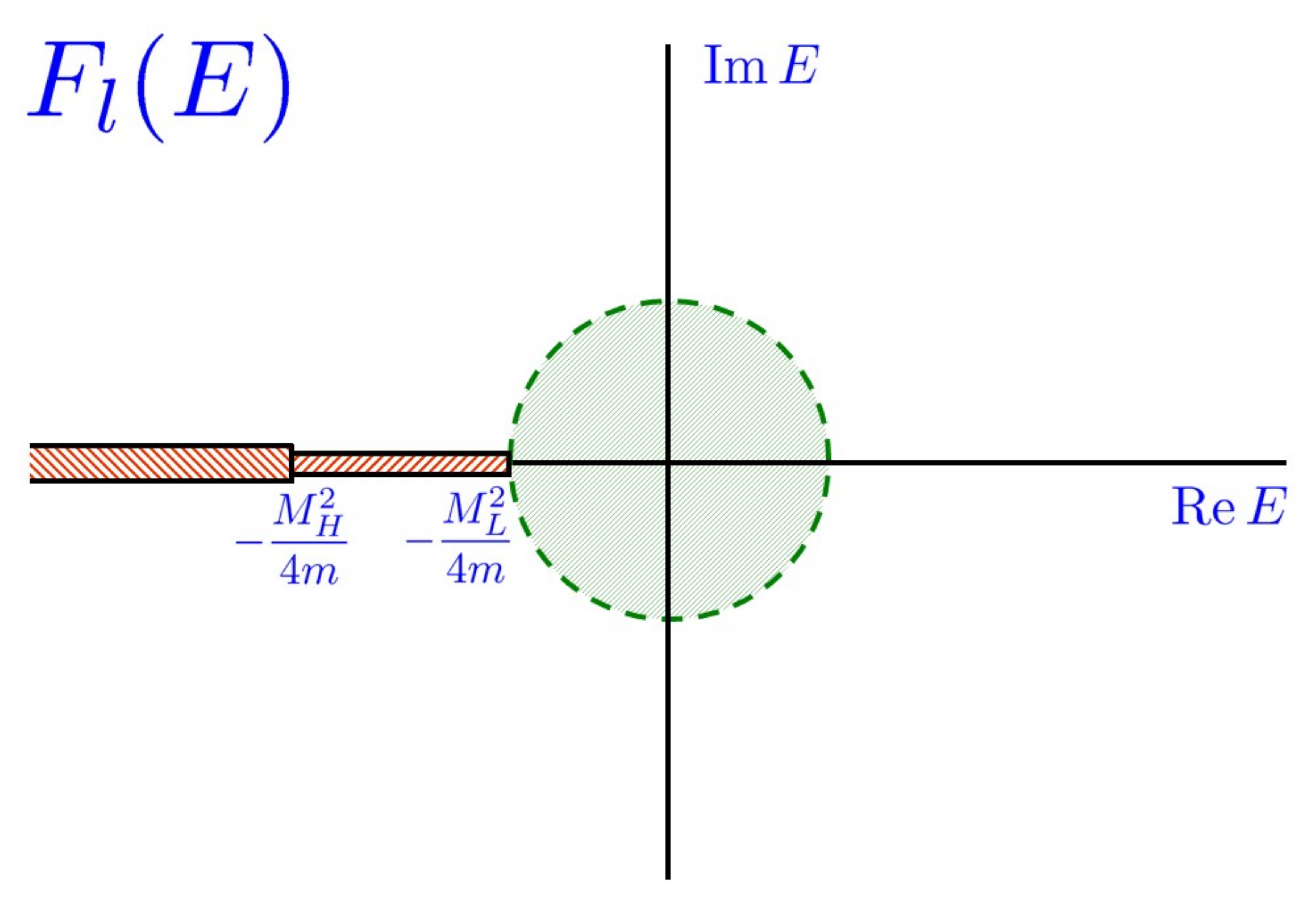}
\hskip 0.035\textwidth
\includegraphics[width=0.31\textwidth]{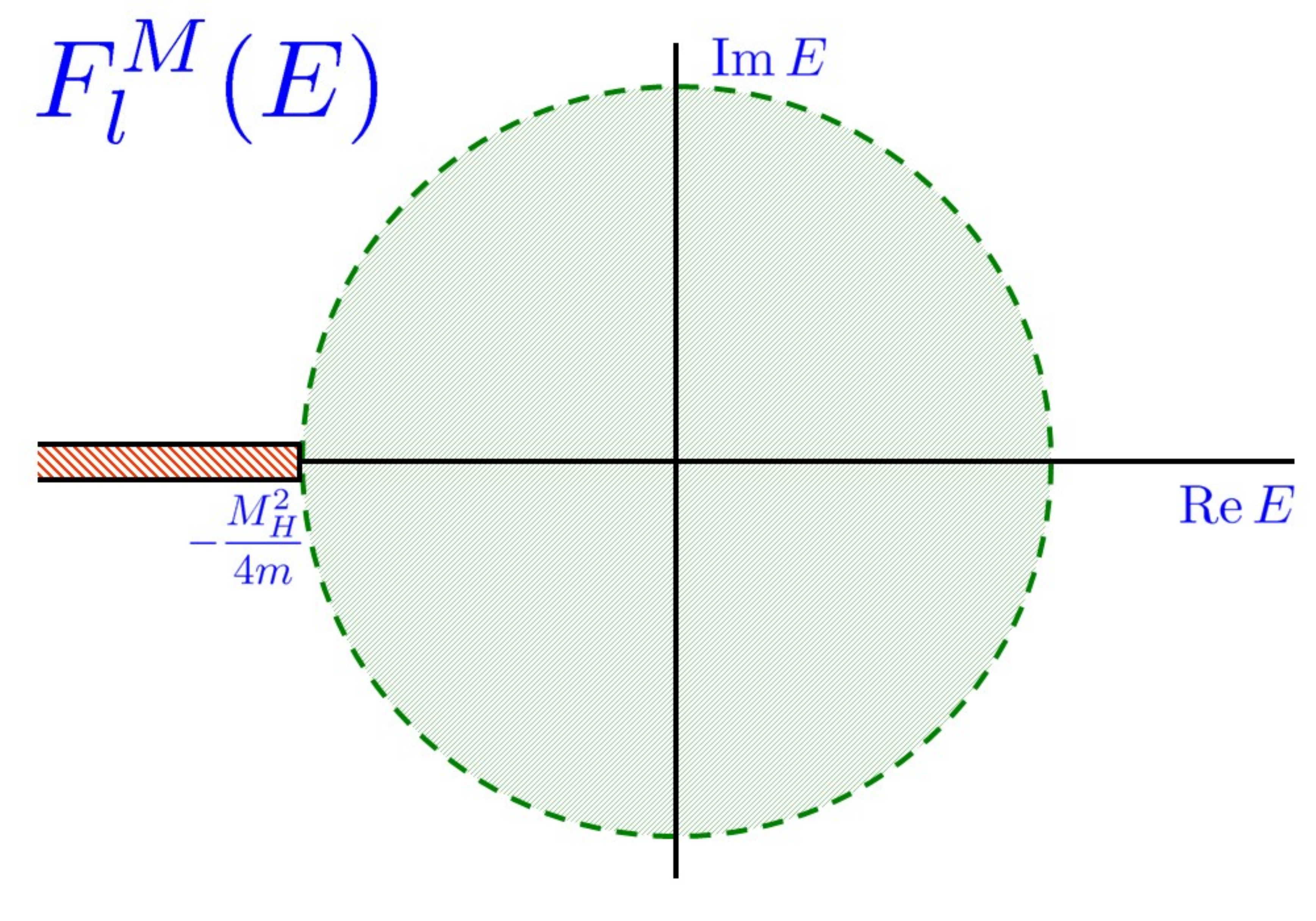}
}
\vspace{-0.2cm}
\caption{\label{analyticity} Some singularities of the partial wave $T$-matrix (left panel), effective range function $F_l (E)$ (middle panel) 
and the modified effective range function $F_l^M (E)$ (right panel) . The shaded areas show the (maximal) range of applicability of the ERE 
and MERE.   
}
\vspace{0.2cm}
\end{figure*}

\begin{minipage}{\textwidth}
\vskip 0 true cm
\rule{\textwidth}{.2pt}
{\it
Exercise: verify the appearance of the left-hand cut in the scattering amplitude 
for the one-pion exchange potential
\begin{displaymath}
V(\vec p \, ', \, \vec p \, ) \propto \frac{\vec \sigma_1 \cdot \vec q \, \vec \sigma_2 \cdot \vec q}{\vec q\, ^2 + M_\pi^2}\,,
\quad \mbox{ with } \quad
\vec q \equiv \vec p \, ' - \vec p\,,
\end{displaymath}
using the  first Born approximation. } \\
\vskip -0.8 true cm
\rule{\textwidth}{.2pt}
\end{minipage}

\medskip
Contrary to the scattering amplitude, the effective range function does not possess the kinematic 
unitarity cut and can be shown to be a
real meromorphic function of $k^2$ near the origin for non-singular
potentials of a finite range \cite{Blatt:49,Bethe:49}. It can, therefore, be
Taylor-expanded about the origin leading to the well-known effective range expansion (ERE)
\be
\label{ere}
F_l (k^2  ) =  - \frac{1}{a} + \frac{1}{2}r k^2 + v_2 k^4 + v_3 k^6 +
\ldots\,,
\ee
with $a$, $r$ and $v_i$ being  the scattering length, effective range and
the so-called shape parameters. Generally, the maximal radius of convergence of the ERE 
is limited by the lowest-lying left-hand dynamic singularity associated with the
potential. For Yukawa-type potentials with the range $r \sim M^{-1}$, 
the (maximal) radius of convergence of the ERE is given
by $k^2 < M^2/4$. For (strong) nucleon-nucleon interaction with the one-pion- ($1\pi$-) exchange potential 
constituting the longest-range contribution, the ERE is expected to 
converge for energies up to $ E_{\rm lab} \sim M_\pi^2/(2 m_N)  =
10.5$ MeV. 
Notice that apart from the singularities associated with the structure of the potential, 
$F_l (k^2)$ may also contain discrete poles whose positions are determined by the strength 
of the interaction. The appearance of such poles near the origin would spoil 
the convergence of the ERE.  

The framework of ERE can be generalized to the case in which the potential is
given by a sum of a long-range ($r_l \sim
m_l^{-1}$) and short-range ($r_s \sim m_s^{-1} \ll m_l^{-1}$) potentials $V_L$ and $V_S$,
respectively. Following van Haeringen and Kok \cite{vanHaeringen:1981pb}, one
can define the \emph{modified} effective range function $F_l^M$ via
\be
\label{mere}
F_l^M (k^2) \equiv M_l^L (k) + \frac{k^{2l+1}}{|f_l^L (k)|} \cot [\delta_l
(k) - \delta_l^L (k)]\,,
\ee
The 
Jost function $f_l^L (k)$ is defined according to $f_l^L (k) \equiv f_l^L (k, r) \big|_{r = 0}$
with $f_l^L (k, r)$ being the Jost solution of the Schr\"odinger equation corresponding to the potential $V_L$, 
i.e.~the particular solution that fulfills
\beq
\lim_{r \to \infty} e^{- i k r} f_l (k, \, r) = 1\,. 
\eeq
Further, $\delta_l^L (k)$ denotes to the phase shift associated with the potential $V_L$
and the the quantity $M_l^L (k)$ can be computed from  $f_l^L (k, r)$ as follows:
\beq
M_l^L (k) = \left( - \frac{i k}{2} \right)^l \frac{1}{l!} \, \lim_{r \to 0} \left[\frac{d^{2l+1}}{dr^{2l+1}}
\, r^l\frac{f_l (k, \, r) }{f_l (k)} \right] \,.
\eeq
I denote here with the superscript ``L'' all quantities that can be computed solely from the 
long-range part of the potential. The modified effective range function $F_l^M (k^2)$ 
defined in this way does not contain the left-hand singularity 
associated with the long-range potential and
reduces, per construction, to the ordinary effective range function 
$F_l (k^2)$ for $V_L=0$. 
It is a real meromorphic function in a much larger region given by  $r_s^{-1}$ as  compared 
to $F_l (k^2)$.\footnote{Note that the existence of $M_l^L (k)$ implies
  certain  constraints on
  the small-$r$ behavior of $V_L(r)$.} 
If the long-range interaction is due to a Coulomb potential, $V_L(r)  = \alpha/r$, the 
Jost solution and, consequently, the function $M_l^L (k)$  
can be calculated analytically. For example, for $l=0$ and the repulsive Coulomb potential, 
the MERE takes the following well-known form:
\beq
F_C  (k^2)  = C_0^2 (\eta ) \, k \, \cot[\delta
(k) - \delta^C (k)] + 2 k \, \eta  \, h (\eta  )\,,
\eeq
where the Coulomb phase shift is $\delta^C \equiv \arg \, \Gamma (1 + i \eta )$ 
and the quantity $\eta$ is given by
\beq
\eta = \frac{m}{2 k} \alpha  \,.
\eeq
Further,  the functions $C_0^2 (\eta )$ (the Sommerfeld factor) and $h (\eta )$ read
\beq
C_0^2 (\eta ) = \frac{2 \pi \eta }{e^{2 \pi \eta } - 1} \,,  \quad  \quad  \mbox{and} \quad  \quad 
h (\eta ) = {\rm Re} \Big[ \Psi ( i \eta  ) \Big] - \ln (\eta  ) \,.
\eeq
Here, $\Psi (z) \equiv \Gamma ' (z)/\Gamma (z)$ denotes the digamma function. 
For more details on the analytic properties of the scattering amplitude and related topics 
I refer the reader to the review article \cite{Badalian:1981xj}. 

After these preparations, we are now in the position to discuss the implications of the 
long-range interaction on the energy dependence of the phase shift.  
It is natural to assume that the coefficients in the ERE and MERE 
(except for the scattering length)
are driven by the scales $m_l$ and $m_s$ associated with the lowest left-hand singularities, 
see \cite{Steele:1998un} for a related discussion. The knowledge of the long-range 
interaction $V_L$ allows to compute the quantities $f_l^L (k)$,  $M_l^L (k)$ and $\delta_l^L (k)$ 
entering the right-hand side of Eq.~(\ref{mere})
and thus to express $\delta_l (k)$ and the ordinary effective range function $F_l(k^2)$
in terms of the modified one,  $F_l^M (k^2)$. 
The MERE for $F_l^M
(k^2)$ then yields an expansion of the subthreshold parameters entering Eq.~(\ref{ere})
in powers of $m_l/m_s$. In particular, using the first few terms in the MERE
as input allows to make predictions for \emph{all} coefficients in the ERE. 
The appearance of the correlations between the
subthreshold parameters in the above-mentioned sense which I will 
refer to as low-energy theorems (LETs) is the only signature of the long-range interaction 
at low energy (in the two-nucleon system). The LETs allow to test whether the long-range interactions 
are incorporated properly in nuclear chiral EFT and thus 
provide an important consistency check.

\subsection{EFT for two nucleons at very low energy}
\label{sec_piless}

Before discussing chiral EFT for two nucleons, let us consider, as a warm-up exercise, 
a simpler EFT for very low energies with $Q \ll M_\pi$.  
Then, no pions need to be taken into account explicitly, 
and the only relevant degrees of freedom are the nucleons themselves. The corresponding 
EFT with the hard scale $\Lambda \sim M_\pi$ 
is usually referred to as  pionless EFT. The most general effective Lagrangian 
consistent with Galilean invariance, baryon number conservation and the isospin symmetry 
takes in the absence of external sources the following form:
\beq
\label{Lagr_nopi}
\mathcal{L} = N^\dagger \left( i \partial_0 + \frac{\vec \nabla ^2}{2 m} \right) N
- \frac{1}{2} C_S \, ( N^\dagger N ) ( N^\dagger N )  - \frac{1}{2}  C_T \, 
( N^\dagger \vec \sigma N ) \cdot ( N^\dagger \vec \sigma N ) + \ldots\,,
\eeq
where $C_{S,T}$ are LECs and the ellipses denote operators with derivatives. 
Isospin-breaking and relativistic corrections to Eq.~(\ref{Lagr_nopi}) can be 
included perturbatively. 

What can be expected from the pionless EFT as compared to the ERE? 
In the absence of external sources and restricting ourselves to the two-nucleon system, 
both approaches provide an expansion of NN low-energy observables 
in powers of $k/M_\pi$, have the same validity range and incorporate the same physical 
principles. Pionless EFT can, therefore, not be expected to do any better than 
ERE. Our goal will be thus to design the EFT in such a way that it matches  
the ERE for the scattering amplitude
\beq
\label{tmat1}
T =  - \frac{16 \pi^2}{m} \, \frac{1}{\left(- \frac{1}{a} + \frac{1}{2} r_0 k^2 
+ v_2 k^4 + v_3 k^6 + \ldots \right)  - i k}\,. 
\eeq
Here, I restrict myself to S-waves only. 
While the coefficients in the effective range expansion are, in general, driven by the range of the potential 
and thus expected to scale with the appropriate powers of $M_\pi$,    
the scattering length can, in principle, take any value. 
In particular, it diverges in the presence of a bound or virtual state at threshold. 
It is, therefore, useful to distinguish between a natural case with $|a | \sim M_\pi^{-1}$ 
and an unnatural case with $|a | \gg  M_\pi^{-1}$.  In the natural case,
the $T$-matrix  in Eq.~(\ref{tmat1}) can be expanded in powers of $k$ as:
\beq
\label{Tnatural}
T = T^{(0)} + T^{(1)} + T^{(2)} + \ldots  = \frac{16 \pi^2 a}{m} \left[ 1 - i a k + \left(\frac{a r_0}{2} - 
a^2 \right) k^2 + \ldots \right]  \,,
\eeq
where the superscripts of $T$ denote the power of the soft scale $Q$.  
A natural value of the scattering length implies the absence of bound and virtual states 
close to threshold. 
The $T$-matrix can then be evaluated perturbatively in the EFT provided one uses an appropriate 
renormalization scheme (i.e.~the one that does not introduce an additional large scale). When solving the 
LS equation with point-like contact interactions, 
one encounters divergent loop integrals of the kind 
\beq
I_n  =  -{m \over (2\pi )^3}\int {d^3l}\, l^{n-3} \,,\; \mbox{
  with } \; n=1,3,5,\ldots\,, \quad \quad
I(k)  =  {m\over (2\pi )^3}\int d^3l \, \frac{1}{k^2-l^2+i\eta} \,.
\eeq
The integrals can be evaluated using a cutoff regularization:  
\begin{eqnarray}
I_n  &\to & I_n^\Lambda  =  -{m \over (2\pi )^3}\int {d^3l}\, l^{n-3}\,\theta \left(\Lambda-l
\right) = -\frac{m\,\Lambda^n}{2n\pi^2} \,, \nn
I (k)  &\to & I^\Lambda(k)  =  {m\over (2\pi )^3}\int {d^3l \,\theta \left(\Lambda-l
  \right)\over
k^2-l^2+i\eta} = I_1^\Lambda -\frac{i\,m\, k}{4\pi} -
\frac{m k}{4\pi^2} \,\ln \frac{\Lambda-k}{\Lambda+k}
 \,, \label{int_def}
\end{eqnarray}
where the last equation is valid for $k < \Lambda$.
To renormalize the scattering amplitude, I divide loop integrals into the divergent and
finite parts and take the limit $\Lambda \to \infty$\footnote{While extremely convenient 
in the case under consideration, taking the limit $\Lambda \to \infty$ is, strictly speaking,
not necessary in an EFT. It is sufficient to ensure that the error from keeping the cutoff
finite is within the accuracy of the EFT expansion. In the considered case, taking  
$\Lambda \sim M_\pi$ would do equally good job in describing the scattering amplitude. 
}:
\begin{eqnarray}
I_n & \equiv & \lim_{\Lambda \to \infty} I_n^\Lambda = \lim_{\Lambda \to
  \infty} \left(I_n^\Lambda + \frac{m \mu_n^n}{2 n \pi^2}\right)
-\frac{m \mu_n^n}{2 n \pi^2} \equiv \Delta_n(\mu_n)+I_n^R(\mu_n)\,, \nonumber\\
I(k) & \equiv & \lim_{\Lambda \to \infty}  I^\Lambda (k) = \lim_{\Lambda \to \infty}
\left(I_1^\Lambda + \frac{m \mu}{2 \pi^2}\right)
+\left[-\frac{m \mu}{2 \pi^2}  -\frac{i\,m\, k}{4\pi}  \right] \equiv
\Delta(\mu)+I^R(\mu,k)\,. \label{splitting2}
\end{eqnarray}
Here, $\Delta_n(\mu_n)$ and $\Delta(\mu)$ denote the divergent parts
of the loop integrals while $I_n^R(\mu_n)$ and $I^R(\mu,k)$ are finite. 
The procedure is analogous to the standard treatment of divergences arising from pion loops 
in ChPT, see section \ref{sec3}. The splitting of
loop integrals in Eq.~(\ref{splitting2}) is not unique. The freedom in the
choice of renormalization
conditions is parameterized by $\mu$ and $\mu_n$. The divergent parts
$\Delta_n(\mu_n)$ and $\Delta(\mu)$ are to be canceled by contributions of
counterterms. The renormalized expression for the amplitude, therefore, emerge from 
dropping the divergent parts in Eq.~(\ref{splitting2}) and replacing the 
bare LECs by the renormalized ones $C_i \to C_i^r (\{ \mu ,\,\mu_n \} )$. 
The proper choice of renormalization conditions requires choosing  
$\mu, \, \mu_n \sim Q \ll M_\pi$. Dimensional regularization  with the minimal subtraction
can be viewed as a special case corresponding to $\mu = \mu_n =0$. Another special case 
is given by DR with the power divergence subtraction (PDS) 
\cite{Kaplan:1998tg,Kaplan:1998we}. In this scheme, the power law divergences, 
which are normally discarded in DR, are 
explicitly accounted for by subtracting from dimensionally regulated loop 
integrals not only $1/(d-4)$-poles but also the $1/(d-3)$-poles. 
Its formulation used in Refs.~\cite{Kaplan:1998tg,Kaplan:1998we} 
corresponds to the choice $\mu_n = 0$ and $\mu \to \mu \pi/2$ in Eq.~(\ref{splitting2}). 

\begin{figure*}
\vspace{0.3cm}
\centerline{
\includegraphics[width=0.5\textwidth]{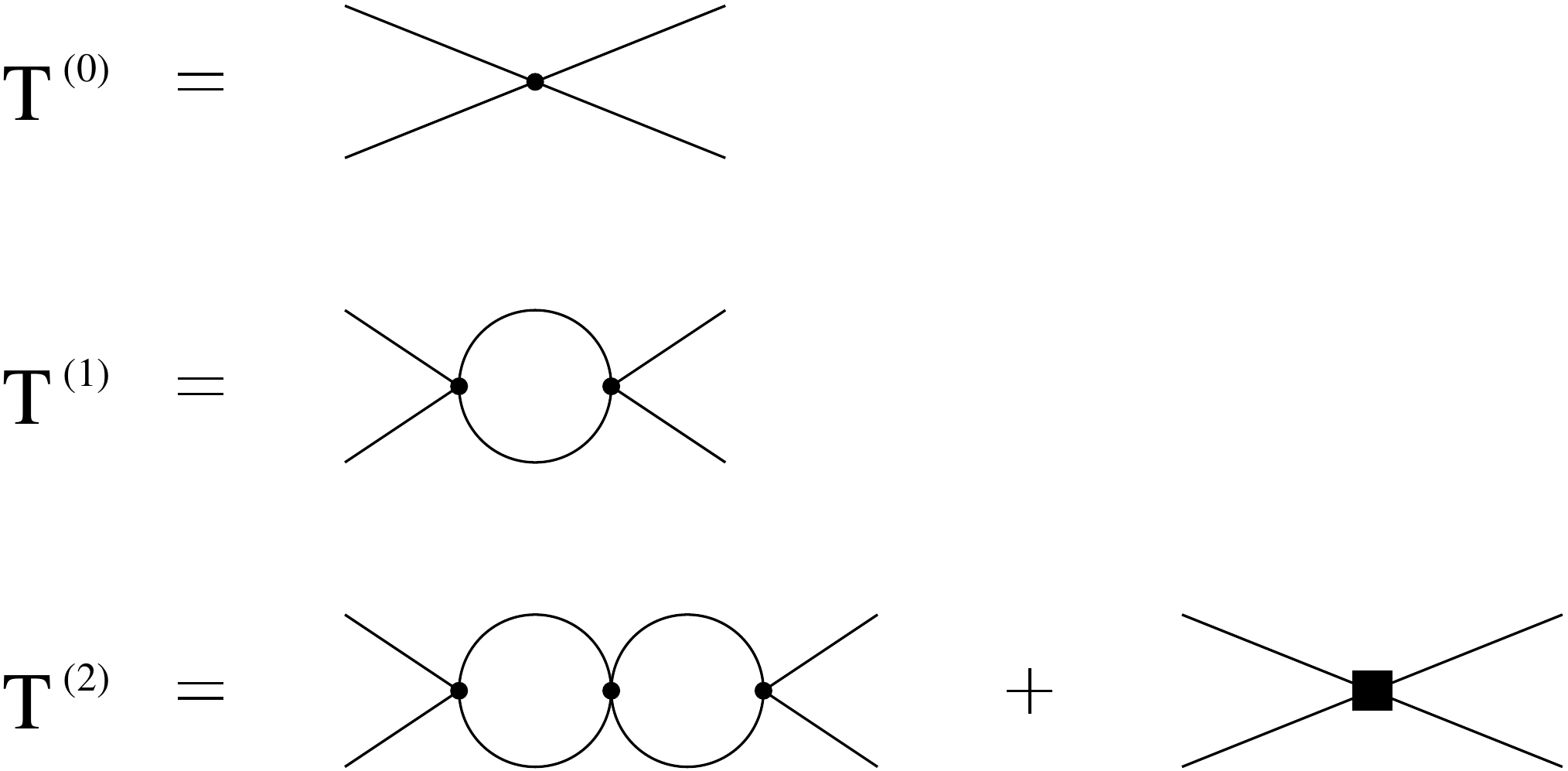}
}
\vspace{0.2cm}
\caption[fig4aa]{\label{natural} Leading, subleading and sub-subleading contributions to the S-wave $T$--matrix 
in the case of a natural scattering length. Solid dots (filled rectangles) refer to contact vertices without 
(with two) derivatives. Lines represent the nucleon propagators.   
}
\vspace{0.2cm}
\end{figure*}

The dimensional analysis for the renormalized scattering amplitude  
implies that the leading and subleading terms $T^{(0)}$ 
and $T^{(1)}$ are given by the tree- and one-loop graphs constructed with the lowest-order vertices 
from Eq.~(\ref{Lagr_nopi}), see Fig.~\ref{natural}.
$T^{(2)}$ receives a contribution from the two-loop graph 
with the lowest-order 
vertices and the tree graph with a subleading vertex \cite{Kaplan:1998tg,Kaplan:1998we}. 
Higher-order corrections can be evaluated straightforwardly.  
Matching the resulting $T$-matrix to the ERE in Eq.~(\ref{Tnatural}) 
order by order in the low-momentum expansion allows to 
fix the LECs $C_i^r$. At next-to-next-to-leading order (N$^2$LO), for example, one finds:
\beq
\label{Cnatural}
C_0^r = \frac{4 \pi a}{m} \Big[ 1 + \mathcal{O} (a \mu ) \Big] \,, 
\mbox{\hskip 2 true cm}
C_2^r = \frac{2 \pi a^2}{m} \,r_0\,,
\eeq
where the LECs $C_0$ and $C_2$ are defined via the tree-level $T$-matrix: 
$T_{\rm tree} = 4 \pi (C_0 + C_2 \, k^2 + \ldots )$. The LEC $C_0$ is related to $C_{S,T}$ in Eq.~(\ref{Lagr_nopi})
as $C_0 = C_S - 3 C_T$.
Here and in the remaining part of this section, the expressions are given in DR with PDS. 

\begin{figure*}
\vspace{0.3cm}
\centerline{
\includegraphics[width=0.9\textwidth]{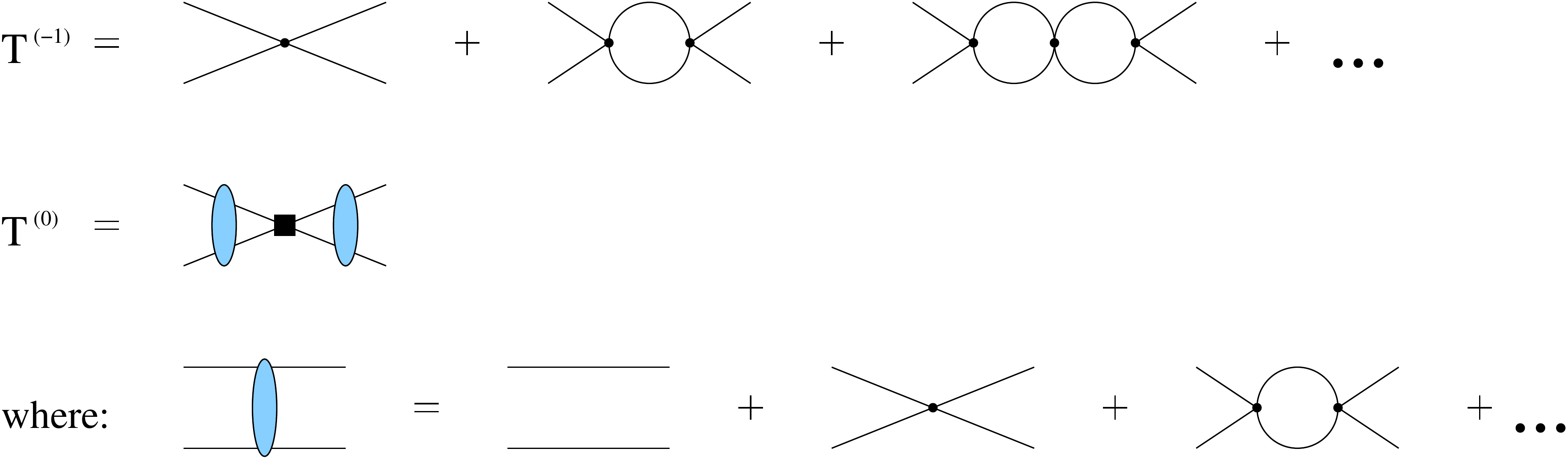}
}
\vspace{0.2cm}
\caption{\label{unnatural} Leading and subleading contributions to the S-wave $T$-matrix 
in the case of an unnaturally large scattering length. For notation, see Fig.~\ref{natural}.    
}
\vspace{0.2cm}
\end{figure*}

For the physically interesting case of neutron-proton scattering, 
the two S-wave scattering lengths appear to be large:
\beq
\label{scattl}
a_{^1S_0} = -23.714 \mbox{ fm} \sim - 16.6 \, M_\pi^{-1}\,, 
\mbox{\hskip 2 true cm}
a_{^3S_1} = 5.42 \mbox{ fm} \sim 3.8 \, M_\pi^{-1}\,.
\eeq
Instead of using the low-momentum representation in Eq.~(\ref{Tnatural}) which is valid only for $k < 1/a$,
it is advantageous to expand the $T$-matrix in powers of $k$ keeping 
$a k \sim 1$ \cite{Kaplan:1998tg,Kaplan:1998we}:
\beqa
\label{Tunnatural}
T &=& T^{(-1)} + T^{(0)} + T^{(1)} + \ldots  \\
&=& \frac{16 \pi^2}{m} \, \frac{1}{(a^{-1} + i k)} \, 
\left[ 1 +  \frac{r_0}{2 (a^{-1} + i k)} k^2 + \left(  
\frac{r_0^2}{4 (a^{-1} + i k)^2} + \frac{v_2}{(a^{-1} + i k)} \right) k^4 + \ldots \right]\,.
\nonumber
\eeqa
The EFT expansion can be cast into the form of Eq.~(\ref{Tunnatural}) if one assumes that the 
LECs $C_i^r$ scale according to $C_{2n} \sim 1/Q^{n+1}$. The leading contribution $T^{(-1)}$ 
then results from summing up an infinite chain of bubble diagrams with the 
lowest-order vertices, see Fig.~\ref{unnatural}. All diagrams constructed only from $C_0^r$ scale as $1/Q$.  
For example, for the one-loop graph one has $Q^3$ from the integration, $1/Q^2$ from the 
nucleon propagator and another $1/Q^2$ from the LECs $C_0^r$.
The corrections are given by perturbative insertions of higher-order interactions dressed 
to all orders by the leading vertices. Matching the resulting $T$-matrix with the one in 
Eq.~(\ref{Tunnatural}) one finds at NLO:
\beq
\label{matchingKSW}
C_0^r = \frac{4 \pi}{m} \; \frac{1}{a^{-1} - \mu } \,, 
\mbox{\hskip 2 true cm}
C_2^r = \frac{4 \pi}{m} \; \frac{1}{(a^{-1} - \mu )^2} \frac{r_0}{2}\,.
\eeq

\begin{minipage}{\textwidth}
\vskip 0 true cm
\rule{\textwidth}{.2pt}
{\it
Exercise: calculate the S-wave scattering amplitude up to NLO for the case of 
unnaturally large scattering length and verify the expressions for the LECs 
given in Eq.~(\ref{matchingKSW}). 
} \\
\vskip -0.8 true cm
\rule{\textwidth}{.2pt}
\end{minipage}

\medskip
The real power of pionless EFT comes into play when one goes beyond the two-nucleon system by 
considering e.g.~low-energy reactions involving external electroweak sources and/or three- and 
more nucleon systems. A discussion of these topics goes beyond the scope of these lectures. 
I refer an interested reader to the recent review articles  
\cite{Beane:2000fx,Bedaque:2002mn,Braaten:2004rn,Platter:2009pi}.

\subsection{Chiral EFT for two nucleons with perturbative pions}
\label{KSW}

\begin{figure*}
\vspace{0.3cm}
\centerline{
\includegraphics[width=0.82\textwidth]{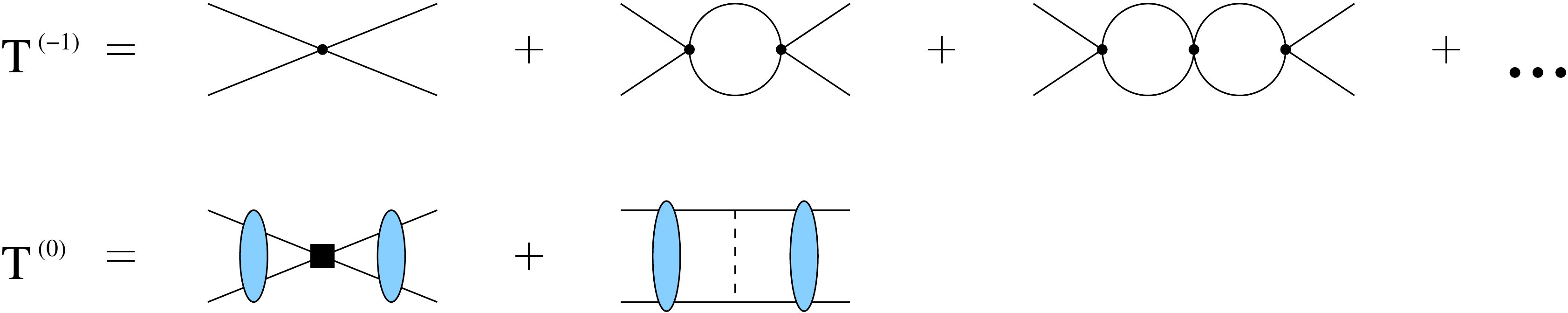}
}
\vspace{0.2cm}
\caption{\label{pionsKSW} Leading and subleading contributions to the S-wave $T$-matrix 
in the case of unnaturally large scattering length in the KSW approach with explicit pions. 
Filled rectangle denotes contact interactions with two derivatives or a single insertion of $M_\pi^2$.
For remaining notation, see Fig.~\ref{natural}.    
}
\vspace{0.2cm}
\end{figure*}

We have seen in the previous section how the EFT without explicit pions can be organized to describe
strongly interacting nucleons at low energy. The limitation in energy of this approach, cf.~the discussion 
in section \ref{sec:analyt}, appears to be 
too strong for most nuclear physics applications. To 
go to higher energies it is necessary to include pions as explicit degrees of freedom. 
I have already outlined in section  \ref{sec:nuclearEFT} one possible way to extend ChPT to the 
few-nucleon sector following Weinberg's original proposal \cite{Weinberg:1990rz,Weinberg:1991um}. 
In this approach, the nonperturbative dynamics is generated by iterating the lowest-order 
two-nucleon potential $V^{(0)}_{2N}$ which subsumes irreducible (i.e.~non-iterative) contributions 
from tree diagrams with the leading vertices (i.e.~$\Delta_i =0$), see Eq.~(\ref{powNN}).  
The only possible contributions are due to derivative-less contact interaction and the static 
$1\pi$-exchange, so that the resulting potential reads:
\beq
\label{VLO}
V_{\rm 2N}^{(0)} = -\frac{g_A^2}{4 F_\pi^2} \frac{\vec
  \sigma_1 \cdot \vec q \, \vec \sigma_2 \cdot \vec q}{\vec q \, ^2 +M_\pi^2} \fet \tau_1
\cdot \fet \tau_2  + C_S + C_T
\vec \sigma_1 \cdot \vec \sigma_2\,.
\eeq 
Here, $\vec \sigma_i$ ($\fet \tau_i$) are the Pauli spin (isospin) matrices of the nucleon $i$, $\vec q
= \vec p \, ' - \vec p$ is the nucleon momentum transfer and $\vec{p}$
($\vec{p}~'$) refers to initial (final) nucleon momenta in the CMS. 
As pointed out in section  \ref{sec:nuclearEFT}, the justification for resumming $V_{\rm 2N}^{(0)}$ 
to all orders in the LS equation is achieved in the Weinberg approach via a fine 
tuning of the nucleon mass, see Eq.~(\ref{counting_m}). With this counting rule, it follows immediately  
that every iteration of $V_{\rm 2N}^{(0)}$ in Eq.~(\ref{LSeqO}) generates a contribution of the order 
$Q^0/\Lambda_\chi$. 
On the other hand, in the pionless EFT with unnaturally large scattering length outlined in section 
\ref{sec_piless}, the non-perturbative resummation of the amplitude was enforced by fine-tuning the 
LECs accompanying the contact interactions while treating the nucleon mass on the same footing as 
all other hard scales. While these two scenarios are essentially equivalent in the pionless case, 
they lead to an important difference in organizing EFT with explicit pions.    
The approach due to Kaplan, Savage and Wise (KSW) \cite{Kaplan:1998tg,Kaplan:1998we} 
represents a straightforward generalization of the pionless EFT 
to \emph{perturbatively} include diagrams with exchange of one or more pions. 
The scaling of the contact interactions is assumed to be the same as
in pionless EFT (provided one uses  DR with PDS or an equivalent scheme to regularize divergent 
loop integrals). Notice that in contrast to pionless EFT, the pion mass is treated as a soft scale with  
$Q \sim M_\pi \sim a^{-1}$. The only new ingredients in the calculation of the amplitude 
up to next-to-leading order in the KSW expansion are given by the dressed $1\pi$-exchange 
potential and derivative-less interaction $\propto M_\pi^2$, see Fig.~\ref{pionsKSW}.  
$2\pi$-exchange is suppressed 
and starts to contribute at N$^2$LO. At each order in the perturbative expansion, the amplitude is made 
independent on the renormalization scale by an appropriate running of the LECs.\footnote{Strictly speaking, 
an exact scale independence of the NLO amplitude in the KSW approach with explicit pions is achieved at the 
cost of resumming a certain class of higher-order terms, see the discussion in Ref.~\cite{Epelbaum:2009sd}.} 
Compact analytic expressions for the scattering amplitude represent another nice feature of the KSW approach. 

As explained in section \ref{sec:analyt}, the appearance of a long-range interaction implies 
strong constraints on the energy dependence of the amplitude and imposes certain correlations between the 
coefficients in the ERE (LETs). EFT with explicit pions aims at a correct description of 
non-analyticities in the scattering amplitude associated with exchange of pions which in this 
framework represent truly long-range phenomena. 
Thus, the correct treatment of the long-range interaction  by including pions perturbatively can 
be ultimately judged by testing the corresponding low-energy theorems. This idea was picked up 
by Cohen and Hansen \cite{Cohen:1998jr,Cohen:1999ia}. 
Fixing the LECs accompanying the contact interactions from the scattering length and effective range, 
they obtained the following predictions for the first three S-wave shape parameters at NLO in the KSW 
scheme \cite{Cohen:1998jr,Cohen:1999ia}:
\beqa
v_2 &=& \frac{g_A^2 m}{16 \pi F_\pi^2} \left( -\frac{16}{3 a^2 M_\pi^4} + \frac{32}{5 a M_\pi^3} 
- \frac{2}{M_\pi^2} \right) \,,  \nn
v_3 &=& \frac{g_A^2 m}{16 \pi F_\pi^2} \left( \frac{16}{a^2 M_\pi^6} - \frac{128}{7 a M_\pi^5} 
+ \frac{16}{3 M_\pi^4} \right) \,, \nn
v_4 &=& \frac{g_A^2 m}{16 \pi F_\pi^2} \left( -\frac{256}{5 a^2 M_\pi^8} + \frac{512}{9 a M_\pi^7} 
- \frac{16}{M_\pi^6} \right) \,.
\eeqa
Plugging in the numbers for the nucleon and pion masses, $g_A \simeq 1.27$, $F_\pi = 92.4$ MeV
and the corresponding scattering lengths, see Eq.~(\ref{scattl}), Cohen and Hansen obtained 
the results quoted in Table~\ref{tab1}.\footnote{A slightly different values compared to the ones 
quoted in this table are extracted from the Nijmegen PWA in Ref.~\cite{PavonValderrama:2005ku}.}
A clear failure for the LETs in both channels serves as an indication that the long-range physics 
is not properly taken into account if pions are treated perturbatively. The convergence 
of the KSW expansion was further tested in Ref.~\cite{Fleming:1999ee} where the amplitude is calculated 
up to N$^2$LO. While the results for the $^1S_0$ and some other partial waves including spin-singlet channels 
were found to be in a reasonable agreement with the Nijmegen PWA, large corrections show up in spin-triplet 
channels already at momenta of the order of $\sim 100$ MeV and lead to strong disagreements with the data. 
The perturbative inclusion of the pion-exchange contributions does not allow to increase the region of 
validity of the EFT compared to the pionless theory, 
see however, Ref.~\cite{Beane:2008bt} for a new formulation
which is claimed to yield a convergent expansion. The failure of the KSW approach in the spin-triplet channels 
was attributed in \cite{Fleming:1999ee} to the iteration of the tensor part of the $1\pi$-exchange potential.  
This appears to be in line with phenomenological successes of Weinberg's approach
which treats pion exchange contributions  nonperturbatively. The most
advanced analyses of the NN system at next-to-next-to-next-to-leading order (N$^3$LO) 
in Weinberg's power counting scheme demonstrate the ability to accurately
describe NN scattering data up to center-of-mass momenta  at least of the order $\sim 2
M_\pi$ \cite{Entem:2003ft,Epelbaum:2004fk}. 
\begin{table}[tb]
\begin{center}
\begin{tabular}{|c|ccc|ccc|}
\hline
&&&&&& \\[-10pt]
  &  \multicolumn{3}{|c|}{$^1S_0$ partial wave}  &  \multicolumn{3}{|c|}{$^3S_1$ partial wave} \\
\hline
&&&&&& \\[-10pt]
  &  $v_2\; $ [fm$^3$] &  $v_3\; $ [fm$^5$] &  $v_4\; $ [fm$^7$] &  $v_2\; $ [fm$^3$] &  $v_3\; $ [fm$^5$] &  $v_4\; $ [fm$^7$] \\
\hline
&&&&&& \\[-10pt]
LETs & $-3.3$ & $17.8$ & $-108$ & $-0.95$ & $4.6$ & $-25$ \\
Nijmegen PWA & $-0.48$ & $3.8$ & $-17$ & $0.4$ & $0.67$ & $-4$ \\
\hline
\end{tabular}
\caption{A comparison of the predicted S-wave shape parameters from Ref.~\cite{Cohen:1998jr} with coefficients 
extracted from the Nijmegen PWA.}
\label{tab1}
\end{center}
\end{table}
 
\subsection{Towards including pions nonperturbatively: playing with toy models}
\label{toy}

While the power counting approach due to Weinberg allows for a nonperturbative 
resummation of the $1\pi$-exchange potential, there is a price to pay. 
Contrary to the KSW approach, the leading NN potential is non-renormalizable in the traditional sense,
i.e.~iterations of the LS equation generate divergent terms with structures 
that are not included in the original potential. 
Moreover, resummation of the  $1\pi$-exchange potential in the LS equation can 
only be carried out numerically. This prevents the use of regularization prescriptions 
such as e.g.~DR
which avoid the appearance of a hard scale and maintain the manifest power counting 
for regularized loop contributions making renormalization considerably more subtle. Most of the 
available calculations employ various forms of cutoff regularization with the cutoff 
being kept finite. The purpose of this section is to provide an in-depth discussion 
on regularization and renormalization in this context 
by considering a simple and exactly solvable quantum mechanical 
model for two nucleons interacting via the long- and short-range forces. This may be
regarded as a toy model for chiral EFT in the two-nucleon sector. In this model, resummation 
of the long-range interaction can be carried out analytically. This allows to employ
and compare the subtractive renormalization that maintains the manifest power counting and the cutoff 
formulation of the effective theory. I also explore the consequences of taking very
large values of the cutoff in this model. The presentation follows closely the one of 
Ref.~\cite{Epelbaum:2009sd}.

\subsubsection{The model}

Consider two spin-less nucleons interacting
via the two-range separable potential
\be
\label{Vunderlying}
V (p ',\, p) =  v_{l} \, F_{l}(p')\,   F_{l}(p)+
  v_{s}\, F_{s}(p')\,   F_{s}(p)\,, \quad
F_l (p) \equiv \frac{\sqrt{p^2 + m_s^2}}{p^2 + m_l^2}\,,  \quad
F_s (p) \equiv \frac{1}{\sqrt{p^2 + m_s^2}}\,,
\ee
where the masses $m_l$ and $m_s$ fulfill the condition $m_l \ll
m_s$. Further, the dimensionless quantities
$v_l$ and $v_s$ denote the strengths of the long- and short-range
interactions, respectively. The choice of the explicit form of $F_{l,s}(p)$
is motivated by the simplicity of calculations. 

The potential in Eq.~(\ref{Vunderlying}) 
does not depend on the angle between $\vec p$ and $\vec p \, '$ and, therefore, only gives rise 
to S-wave scattering. The projection onto the S-wave in this case is trivial and simply yields 
the factor of $4 \pi$ from the integration over the angles. 
For an interaction of a separable type, the off-shell
T-matrix and, consequently, also the coefficients in the ERE, see Eqs.~(\ref{tmat}) and (\ref{ere}),  
can be calculated analytically by solving the corresponding LS equation
\be
\label{LS}
T (p' ,\, p; \, k ) =  V (p' ,\, p) + 4 \pi  \int \frac{l^2 dl}{(2 \pi)^3} V
(p' ,\, l)
\frac{m}{k^2-l^2 + i \epsilon} T (l ,\, p; \, k )\,,
\ee
where $m$ is the nucleon mass and $k$ corresponds to the on-shell momentum
which is related to the two-nucleon center-of-mass energy via $E_{\rm CMS} =
k^2/m$. Note that we have absorbed the factor $4 \pi$ into the 
normalization of the T-matrix which is, therefore, different from the one in Eq.~(\ref{LSeq}). 
In particular, the relation between the S- and T-matrices is given by 
$S (p) = 1 - i  p   m   T (p, \, p; \, p)/(2 \pi )$. 

As explained in section \ref{sec:analyt}, the coefficients in the ERE generally scale
with the mass corresponding to the long-range interaction that gives rise to
the first left-hand singularities in the T-matrix. Thus, in the considered two-range model, 
the coefficients in the ERE can be expanded in powers of $m_l/m_s$ leading to the ``chiral'' expansion:
\bea
\label{EREexpanded}
a &=&  \frac{1}{m_l} \bigg( \alpha_a^{(0)}  + \alpha_a^{(1)} \frac{m_l}{m_s} +
\alpha_a^{(2)} \frac{m_l^2}{m_s^2} + \ldots  \bigg) \,, \nn
r &=& \frac{1}{m_l} \bigg( \alpha_r^{(0)}  + \alpha_r^{(1)} \frac{m_l}{m_s} +
\alpha_r^{(2)} \frac{m_l^2}{m_s^2} + \ldots  \bigg) \,, \nn
v_i &=& \frac{1}{m_l^{2 i -1}} \bigg( \alpha_{v_i}^{(0)} + \alpha_{v_i}^{(1)}
\frac{m_l}{m_s} +
\alpha_{v_i}^{(2)} \frac{m_l^2}{m_s^2} + \ldots  \bigg) \,,
\eea
where $\alpha_a^{(m)}$,  $\alpha_r^{(m)}$ and  $\alpha_{v_i}^{(m)}$ are
dimensionless constants whose values are determined by the form of the
interaction potential. I fine tune the strengths of the long- and short-range
interactions in our model in such a way that they generate scattering lengths of a natural
size. More
precisely, I require that the scattering length takes the value $a =
\alpha_l/m_l$
($a = \alpha_s/m_s$) with a dimensionless constant $| \alpha_l | \sim 1$ ($|
\alpha_s  | \sim 1$)
when the short-range (long-range) interaction is switched off.  This allows to express the corresponding 
strengths $v_l$ and $v_s$ in terms of $\alpha_l$ and $\alpha_s$ as follows:
\be
\label{strengths}
v_l = -\frac{8 \pi  m_l^3 \alpha _l}{m \left(\alpha _l m_s^2+m_l^2 \alpha _l-2
    m_s^2\right)}\,,
\quad
v_s = -\frac{4 \pi  m_s \alpha _s}{m \left(\alpha _s-1\right)}\,.
\ee
One then finds the following expressions for the first three terms in the
``chiral'' expansion of the scattering length 
\be
\label{LET1}
\alpha_a^{(0)} =  \alpha_l \,, \quad  \quad
\alpha_a^{(1)} =  (\alpha_l -1 )^2 \alpha_s \,, \quad \quad
\alpha_a^{(2)} =  (\alpha_l -1 )^2 \alpha_l \alpha_s^2 \,,
\ee
and effective range 
\bea
\label{LET2}
\alpha_r^{(0)} &=&  \frac{3 \alpha_l - 4}{\alpha_l} \,, \nn
\alpha_r^{(1)} &=&  \frac{2 \left(\alpha _l-1\right) \left(3 \alpha
    _l-4\right) \alpha _s}{\alpha _l^2} \,, \nn
\alpha_r^{(2)} &=& \frac{\left(\alpha _l-1\right) \left(3 \alpha _l-4\right)
  \left(5 \alpha _l-3\right) \alpha _s^2+\left(2-\alpha _l\right) \alpha
   _l^2}{\alpha _l^3} \,.
\eea
Notice that in the model considered the leading terms in the $m_l/m_s$-expansion of the ERE
coefficients are completely fixed by the long-range interaction. The scenario
realized corresponds to a strong (at momenta $k \sim
m_l$) long-range interaction which needs to be treated non-perturbatively and
a weak short-range interaction which can be accounted for in perturbation theory.  

\begin{minipage}{\textwidth}
\vskip 0 true cm
\rule{\textwidth}{.2pt}
{\it
Exercise: calculate the $T$-matrix by solving the LS equation (\ref{LS}) for the potential given 
in Eq.~(\ref{Vunderlying}). Verify the ``chiral'' expansion for the scattering length and effective range. 
Work out the first terms in the ``chiral'' expansion of the shape parameters $v_2$ and $v_3$. 
The results can be found in Ref.~\cite{Epelbaum:2009sd}. 
} \\
\vskip -0.8 true cm
\rule{\textwidth}{.2pt}
\end{minipage}

\medskip

\subsubsection{KSW-like approach}

\begin{figure}[tb]
  \begin{center}
\includegraphics[width=0.99\textwidth]{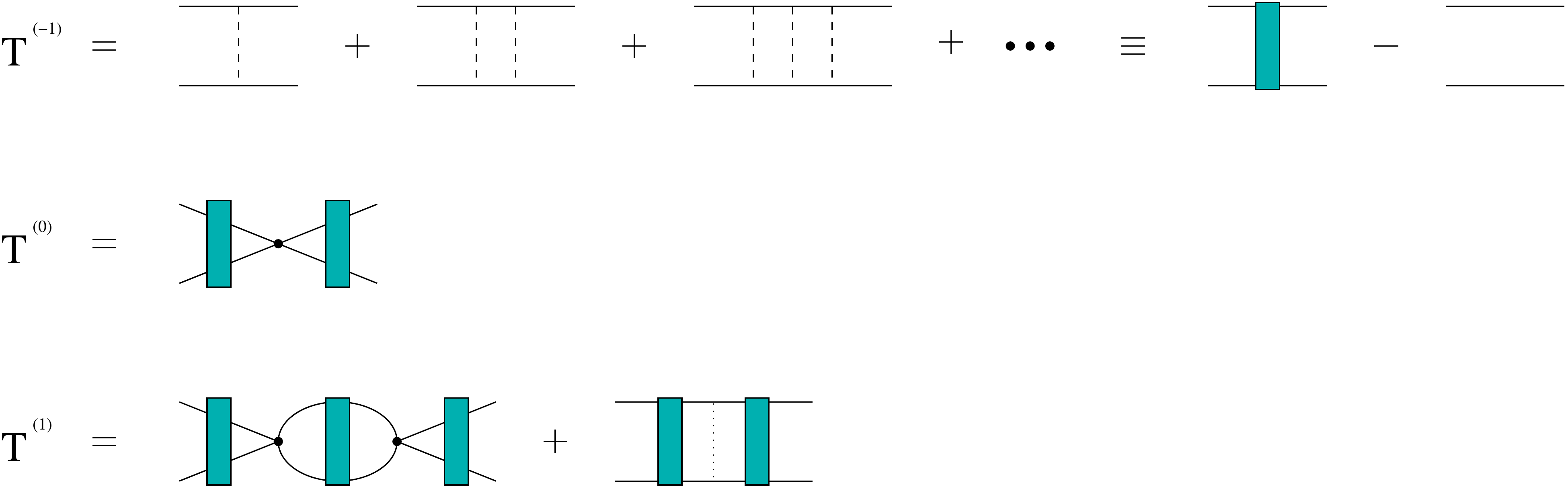}
\vskip 0.5 true cm
    \caption{Leading, next-to-leading and next-to-next-to-leading order
      contributions to the scattering amplitude in the KSW-like approach. The
      solid lines denote nucleons while the dashed ones represent an insertion
      of the lowest-order (i.e.~$\mathcal{O} (Q^{-1})$) long-range
      interaction. Solid dots (dotted lines) 
      denote an insertion of the lowest-order contact interaction $\propto
      C_0$ (subleading order-$\mathcal{O} (Q)$ contribution to the long-range interaction).
    \label{fig1_1}
 }
  \end{center}
\end{figure}

At momenta of the order $k \sim m_l \ll m_s$, the details of the short-range
interaction cannot be resolved. An EFT description emerges by keeping the
long-range interaction and replacing the short-range one by a series of
contact terms $V_{\rm short} (p', \, p ) = C_0 + C_2 (p^2 + {p'} ^2) +
\ldots$. Iterating such an effective potential in the LS equation leads to ultraviolet 
divergences which need to be regularized and renormalized. The renormalization prescription 
plays  an important role in organizing the EFT expansion. I first consider the most
convenient and elegant formulation by employing the subtractive renormalization
which respects dimensional power counting at the level of diagrams. In this sense, 
the considered formulation is conceptually similar to the KSW framework with perturbative 
pions discussed above and will be referred to as the KSW-like 
approach. The available soft
and hard scales in the problem are given by $Q = \{k, \, \mu, \, m_l \}$ and
$\lambda = \{m_s , \, m \}$, respectively. Here $\mu \sim m_l$ denotes the subtraction point.
(There is just a single linearly divergent integral at the order in the low-momentum expansion I will consider.)
The contributions to the amplitude up to N$^2$LO in the
$Q/\lambda$-expansion are visualized in Fig.~\ref{fig1_1} and can be easily verified
using naive dimensional analysis. 
In particular, the leading term arises at order $Q^{-1}$ is
generated by the leading term in the $Q/\lambda$-expansion of the long-range
interaction
\bea
\label{long}
V_{\rm long} (p', \, p ) &=&  v_{l} \, F_{l}(p')\,   F_{l}(p) \\
&\simeq&-
\frac{8 \pi  m_l^3 \alpha _l}{m \left(\alpha _l-2\right) ( p^2 + m_l^2)( {p
    '}^2 + m_l^2)}
\left[ 1 -\frac{\alpha _l m_l^2 }{\left(\alpha _l-2\right) m_s^2}
+ \frac{p^2}{2 m_s^2} + \frac{{p '}^2}{2 m_s^2} + \mathcal{O} \left(
  \frac{Q^4}{\lambda^4}
\right)\right]\,, \nonumber
\eea
which scales as $Q^{-1}$ and, therefore,
needs to be summed up to an infinite order, see Fig.~\ref{fig1_1}.
Notice that the natural size of
the short-range effects in our model suggests the scaling of the short-range
interactions in agreement with the naive dimensional analysis, i.e.~$C_{2n}
\sim Q^0$. 
This leads to the following expression for the on-the-energy shell T-matrix:
\be
T^{(-1)}=-\frac{8 \pi  m_l^3 \alpha _l}{m \left(k-i m_l\right){}^2 \left[k^2
    \left(\alpha _l-2\right)+2 i k m_l \left(\alpha _l-2\right)+2
    m_l^2\right]} \,,
\ee
from which one deduces
\be
k \cot \delta = - \frac{4 \pi}{m} \frac{1}{T^{(-1)}} + i k 
= {} -\frac{m_l}{\alpha _l} + \frac{ \left(3
    \alpha _l-4\right)}{2 m_l \alpha _l} k^2  +
\frac{\left(\alpha _l-2\right)}{2 m_l^3 \alpha _l} k^4 \,.
\ee
Not surprisingly, one observes that the leading terms in the expansion of the ERE
coefficients in Eq.~(\ref{EREexpanded}) are correctly reproduced.

The first correction at order $Q^0$ is given by
the leading-order contact interaction dressed with the iterated leading
long-range  interaction as visualized in
Fig.~\ref{fig1_1}. One finds
\be
T^{(0)} = \frac{C_0 \left(k+i m_l\right){}^2 \left[k^2 \left(\alpha
      _l-2\right)+2 m_l^2  \left(\alpha _l-1\right)\right]{}^2}{\left(k-i
    m_l\right){}^2 \left[k^2 \left(\alpha
   _l-2\right)+2 i k m_l \left(\alpha _l-2\right)+2 m_l^2\right]{}^2}.
\ee
Notice that all integrals entering $T^{(-1)}$ and $T^{(0)}$ are finite.
The effective range function $k \cot \delta$ at NLO
can be computed via
\be
k \cot \delta = - \frac{4 \pi}{m}  \frac{1}{T^{(-1)}} \bigg( 1 -
\frac{T^{(0)}}{T^{(-1)}}  \bigg) +  i k \,.
\ee
The ``chiral'' expansion of the coefficients in the ERE results from
expanding the right-hand side in this
equation in powers of $k^2$ and, subsequently, in powers of $m_l$. The
LEC $C_0$ can be determined from matching to $\alpha_a^{(1)}$ in
Eq.~(\ref{LET1}) which yields
\be
\label{C0_LO}
C_0 = \frac{4 \pi \alpha_s}{m m_s}\,.
\ee
This leads to the following predictions for the effective range:
\be
r=\frac{1}{m_l} \Bigg[\frac{3 \alpha_l-4}{\alpha _l}
+\frac{2 \left( \alpha_l-1\right) \left(3 \alpha _l-4\right)
\alpha _s}{\alpha _l^2 m_s} m_l \Bigg]\,.
\ee
One observes that  $\alpha_r^{(1)}$ is correctly reproduced at NLO. The same holds true  
for the first shape parameters, see Ref.~\cite{Epelbaum:2009sd} for explicit expressions. 
Moreover, using dimensional analysis it is easy to verify
that, in fact, $\alpha_{v_i}^{(1)}$ for
all $i$ \emph{must} be reproduced correctly at this order.
This is a manifestation of the LETs discussed in section \ref{sec:analyt}. 

At N$^2$LO, the linearly divergent integral $I(k)$ occurs, see Eq.~(\ref{int_def}), which is treated 
according to Eq.~(\ref{splitting2}). Renormalization is carried out by dropping its divergent part 
$\Delta (\mu )$ and replacing the bare LEC $C_0$ by the renormalized one $C_0 (\mu)$ in the expression for
the amplitude. A straightforward calculation using MATHEMATICA yields:
\bea
T^{(1)} &=& \frac{\left(k+i m_l\right){}^2}{4 \pi ^2 m m_s^2 \left(k-i
    m_l\right){}^2 \left[k \alpha _l \left(k+2 i m_l\right)-2 \left(k+i
   m_l\right){}^2\right]{}^2} \Bigg[
-32 \pi ^3 k^2 m_l^3 \left(\alpha _l-2\right) \alpha _l \nn
&& {}+ \left( C_0 (\mu)\right) ^2 m^2 m_s^2 \left[k^2 \left(\alpha
    _l-2\right)+2 m_l^2
    \left(\alpha _l-1\right)\right]{}^2 \\
&& {} \times \frac{
\alpha _l \left[k^2 (-2 \mu
      -i
   \pi  k)+2 k (\pi  k-2 i \mu ) m_l+2 \pi  m_l^3\right]+2 (2 \mu +i
 \pi  k) \left(k+i m_l\right){}^2}{k \alpha _l \left(k+2 i
   m_l\right)-2 \left(k+i m_l\right){}^2} \Bigg].
\nonumber
\eea
The LEC $C_0( \mu )$ can be written in terms of a perturbative expansion as
follows
\be
C_0( \mu ) = C_0^{(0)} + C_0^{(1)}( \mu ) + \ldots \,,
\ee
where the superscript refers to the power of the soft scale $Q$. The first term
does not depend on $\mu$ and equals $C_0$ in Eq.~(\ref{C0_LO}).
The $\mu$-dependence of $C_0^{(1)} (\mu )$ can be determined by solving the
renormalization group equation
\be
\frac{d}{d \mu} \bigg[ T^{(-1)} + T^{(0)} + T^{(1)} \bigg] = 0\,.
\ee
One also needs one additional input parameter, such as e.~g.~$\alpha_a^{(2)}$,
in order to fix the
integration constant.  This leads to
\be
\label{C0_NLO}
C_0^{(1)} (\mu ) = \frac{8 \mu  \alpha _s^2}{m m_s^2}\,.
\ee
It is then easy to verify that the scattering amplitude $T^{(-1)} + T^{(0)} +
T^{(1)}$ is $\mu$-independent up to terms of order $Q^2$. Further, the
effective range function is given at this order by
\be
k \cot \delta = - \frac{4 \pi}{m}  \frac{1}{T^{(-1)}} \Bigg[ 1 -
\frac{T^{(0)}}{T^{(-1)}} +
\left( \frac{T^{(0)}}{T^{(-1)}} \right)^2   -
\frac{T^{(1)}}{T^{(-1)}}
 \Bigg] +  i k \,,
\ee
which can be used to predict the ``chiral'' expansion for the coefficients in the ERE. 
Here I list only the result for the effective range which is sufficient for our purposes. 
The expressions for $v_{2,3}$ can be found in \cite{Epelbaum:2009sd}. 
\bea
\label{LETKSW}
r &=&\frac{1}{m_l} \bigg[\frac{3\alpha _l -4}{\alpha _l}
+ \frac{2 \left(\alpha _l-1\right) \left(3 \alpha _l-4\right)
  \alpha _s}{\alpha _l^2 m_s} m_l 
 +\frac{\left(\alpha _l-1\right) \left(3
      \alpha _l-4\right) \left(5 \alpha _l-3\right) \alpha _s^2+\left(2-\alpha
   _l\right) \alpha _l^2}{\alpha _l^3 m_s^2} m_l^2  \nn
&-& \frac{4 \mu m_l  \left(\alpha _l-1\right) \left(3 \alpha _l-4\right)
  \alpha _s^3 \left(\pi  m_l \left(3-5 \alpha _l\right)+4 \mu  \alpha
   _l\right)}{\pi ^2 \alpha _l^3 m_s^3}
+ \mathcal{O} \left( Q^4 \right)\bigg] \,.
\eea
As expected, the first three terms in the ``chiral'' expansion of $r$ are correctly
reproduced at N$^2$LO being protected by the LETs. The same holds true for the 
shape parameters $v_i$, see Ref.~\cite{Epelbaum:2009sd}. 
The knowledge of $\alpha_{x_j}^{(i)}$ for one particular
$x_j$ is sufficient to predict $\alpha_{x_k}^{(i)}$ for all $k \neq j$.

\subsubsection{Weinberg-like approach with a finite cutoff}

\begin{figure}[tb]
  \begin{center}
\includegraphics[width=0.75\textwidth]{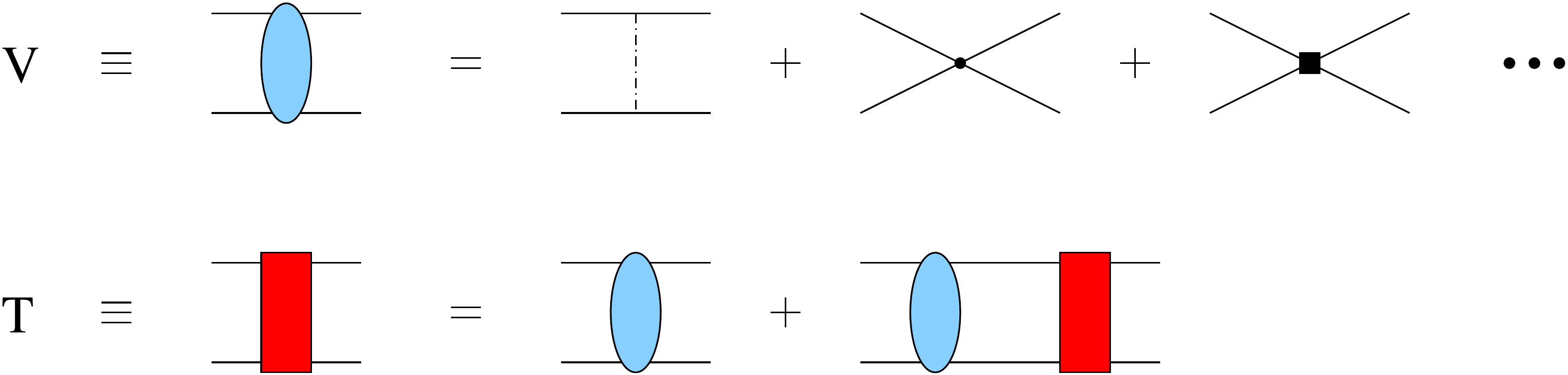}
\vskip 0.5 true cm
    \caption{Effective potential and scattering amplitude in the Weinberg-like
      approach. The dashed-dotted line refers to the full long-range
      interaction. Solid dot and filled rectangle refer to the leading and
      subleading contact interactions, respectively. For remaining notation
      see Fig.~1.
    \label{figW}
 }
  \end{center}
\end{figure}

An elegant EFT formulation like the one described above which respects the
manifest power counting at every stage of the calculation is not
available in the realistic case of nucleon-nucleon interaction.  
Here, one lacks a regularization prescription for \emph{all} divergent integrals
resulting from iterations of the potential in the LS equation
which would keep regularization artefacts small without, at the same time,
introducing a new hard scale in the problem. Contrary to the considered model, 
the $1\pi$-exchange potential is non-separable and cannot be analytically 
resummed in the LS equation. In the context of chiral EFT for
few-nucleon systems, the divergent integrals are usually dealt with by
introducing an UV cutoff $\Lambda$, which needs to be taken of the order
$\Lambda \sim m_s$ or higher in order to keep regularization artefacts small. 
Clearly, cutoff-regularized diagrams will not obey dimensional power
counting anymore. This, however, does not mean a breakdown of EFT since power 
counting is only required for the \emph{renormalized} amplitude.
I now consider the
Weinberg-like formulation in which the effective potential, given by the
long-range interaction and a series of contact terms, is iterated in
the LS equation to all orders, see the work by Lepage \cite{Lepage:1997} for a
related discussion. This is visualized in Fig.~\ref{figW}. 
I carry out \emph{renormalization} by literally following
the steps outlined in Ref.~\cite{Lepage:1997} and summarized
in Ref.~\cite{Gasser:1982ap} in the following way: \emph{''The
theory is fully specified by the values of the bare constants ...
once a suitable regularization procedure is chosen. In principle,
the renormalization program is straightforward: one calculates
quantities of physical interest in terms of the bare parameters at
given, large value of (ultraviolet cutoff) $\Lambda$. Once a
sufficient number of physical quantities have been determined as
functions of the bare parameters one inverts the result and
expresses the bare parameters in terms of physical quantities,
always working at some given, large value of $\Lambda$. Finally,
one uses these expressions to eliminate the bare parameters in all
other quantities of physical interest. Renormalizability
guarantees that this operation at the same time also eliminates
the cutoff.''} Notice that by iterating the truncated expansion for the
effective potential in the LS equation one unavoidably generates
higher-order contributions without being able to absorb all arising
divergences into redefinition of the LECs present in the considered truncated
potential. Thus, for the case at hand,
cutoff dependence in observables is expected to be eliminated only up to the
considered order in the EFT expansion. 
I further emphasize that expressing the bare parameters
(i.~e.~LECs $C_i$) in terms of physical quantities is a non-trivial step as
the resulting nonlinear
equations for $\{C_i \}$ do not necessarily possess  real solutions for all values
of $\Lambda$, especially  when it is chosen to be considerably larger than $m_s$.

To be specific, consider the effective potential at subleading order
in the Weinberg-like approach as depicted in Fig.~\ref{figW}
\be
V_{\rm eff} (p' ,\, p) =  v_{l} \, F_{l}(p')\,   F_{l}(p)
 + C_0 \,.
\ee
The off-shell T-matrix $T (p',\, p; \, k)$ can be easily calculated by
solving the $2 \times 2$ matrix equation
\be
\label{LSmatrix}
t (k) =  v_{\rm eff} +  v_{\rm eff} \, \mathcal{G}(k) \,   t(k)
\ee
where I have defined
\be
V_{\rm eff} (p' ,\, p) =\gamma^T (p') \, v_{\rm eff} \, \gamma (p ), \quad
T (p' ,\, p, \, k) =\gamma^T (p') \, t (k) \, \gamma (p )\,,
\ee
with
\be
v_{\rm eff} \equiv \left( \begin{array}{cc} v_l & 0 \\ 0 & C_0 \end{array}
\right)\,, \quad
\gamma ( p) \equiv \left( \begin{array}{c} F_l (p) \\ 1 \end{array}
\right)\,,  \quad
\mathcal{G}(k) \equiv \left( \begin{array}{cc} I_l(k) & I_{l1}^{\rm reg}(k) \\
    I_{l1}^{\rm reg} (k)
    &  I_1^{\rm reg} (k) \end{array}
\right)\,.
\ee
The integral $I_l(k)$ is given by 
\be
I_l (k) = 4 \pi m \int_0^\infty \frac{l^2 \, dl}{(2 \pi)^3}
\frac{l^2+ m_s^2}{[k^2 - l^2 + i \epsilon][l^2 + m_l^2]^2} \nn
= \frac{m \left(-2 i k m_l+m_l^2+m_s^2\right)}{8 \pi  m_l \left(k+i
    m_l\right){}^2} \,, 
\ee
and is ultraviolet-finite. The divergent integrals $I_1 (k)$ and $ I_{l1}(k)$ are regularized by 
means of a finite cutoff $\Lambda$:
\bea
\label{otherintegrals}
I_1^{\rm reg} &\equiv& 4 \pi m \int_0^\Lambda \frac{l^2 dl}{(2 \pi)^3}
\frac{1}{k^2 - l^2 + i \epsilon} = - \frac{m \Lambda}{2 \pi^2} - i \frac{m k
}{4 \pi}  + \mathcal{O} (\Lambda^{-1}) \,, \nn 
I_{l1}^{\rm reg} &\equiv & 
4 \pi m \int_0^\Lambda \frac{l^2 \, dl}{(2 \pi)^3}
\frac{\sqrt{l^2+ m_s^2}}{[k^2 - l^2 + i \epsilon][l^2 + m_l^2]} 
= \frac{m}{2 \pi^2}
\bigg[ k \frac{\sqrt{k^2 + m_s^2}}{k^2 + m_l^2} \ln \bigg(\frac{k + \sqrt{k^2
      + m_s^2}}{m_s} \bigg)  \nn
&-&
\frac{m_l m_s s}{2(k^2 + m_l^2)} + 
\ln \left(\frac{m_s}{2 \Lambda }\right) - \frac{i \pi k
  \sqrt{k^2+m_s^2}}{2 \left(k^2+m_l^2\right)} \bigg)  + \mathcal{O}
(\Lambda^{-1})\,,
\eea
where $s \equiv \left( 2 \sqrt{m_s^2-m_l^2}/m_s \right)
\, {\rm arccot}\left(m_l/\sqrt{m_s^2-m_l^2}\right)$.
Neglecting, for the sake of simplicity, the finite cutoff artefacts
represented by the $\mathcal{O} (\Lambda^{-1})$-terms in
Eq.~(\ref{otherintegrals}) and performing straightforward calculations, one obtains for the
scattering length: 
\be
\label{aWeinb1}
a_\Lambda = \frac{\pi  m_s \left\{C_0 m \left[2 \alpha _l \left(m_s \left(\Lambda
          -s  m_l\right)+2 m_l^2 \ln (m_s/2 \Lambda )
\right)+\pi  m_l m_s\right]+4 \pi ^2 \alpha _l m_s\right\}}{m_l
\left\{2 \pi  m_s^2 \left(C_0 m \Lambda +2 \pi ^2\right)-C_0 m m_l
   \alpha _l \left[s m_s-2 m_l \ln (m_s/2 \Lambda)
       \right]^2\right\}} \,.
\ee
\emph{Renormalization} is 
carried out by matching the above expression to the value of the scattering length
in the underlying model which is regarded as data, 
\be
\label{a_data}
a_{\rm underlying} = \frac{m_l \left(2 \alpha _l-1\right) \alpha _s-\alpha _l
  m_s}{m_l \left(m_l \alpha _l \alpha _s-m_s\right)}\,,
\ee
and expressing $C_0 (\Lambda ) $ in terms of $a_{\rm underlying}$.  
A straightforward calculation yields the following
\emph{renormalized} expression for the effective range:
\bea
\label{LETWeinberg}
r_\Lambda &=& \frac{1}{m_l} \bigg[\frac{3 \alpha_l - 4}{\alpha _l}
+\frac{2 \left(\alpha _l-1\right) \left(3 \alpha _l-4\right) \alpha _s}{\alpha
  _l^2 m_s} m_l + \bigg( \frac{4 \left(\alpha _l-2\right) \alpha _s }{\pi
  \alpha _l m_s^2} \left(\ln \frac{m_s}{2 \Lambda }+1\right) \nn
&+&
\frac{\left(\alpha _l-1\right) \left(3 \alpha _l-4\right) \left(5 \alpha
      _l-3\right) \alpha _s^2+\left(2-\alpha _l\right) \alpha
   _l^2}{\alpha _l^3 m_s^2} \bigg) m_l^2 + \mathcal{O} \left( m_l^3
\right)\bigg]\,.
\eea
In agreement with the LETs discussed above, one observes that the subleading terms in the
``chiral'' expansion of $r$ (and $v_i$, see \cite{Epelbaum:2009sd}) 
are correctly reproduced once $C_0$ is appropriately tuned. Notice that the smallness 
of the subleading correction to $r_\Lambda$ due to the $C_0$-term in the effective potential
as compared to the leading contribution given by the first term on the right-hand side of 
Eq.~(\ref{LETWeinberg}) is only guaranteed \emph{after} carrying out renormalization by 
properly tuning $C_0 (\Lambda )$.  The sub-subleading
and higher-order terms in the ``chiral'' expansion of $r$ and $v_i$ are not
reproduced correctly being not protected by the LETs at the considered order. 
Moreover, since the included LEC is insufficient to absorb all divergencies
arising from iterations of the LS equation, nothing prevents the appearance of
positive powers or logarithms of the cutoff $\Lambda$ in the expressions for
$\alpha_{r}^{(\geq 2)}$. The results
in Eq.~(\ref{LETWeinberg}) show that this is indeed the case. The
dependence on $\Lambda$ occurs, however, only in contributions beyond the
accuracy of calculation and, obviously, does not affect the predictive power
of the EFT as long as the cutoff is chosen to be of the order of the
characteristic hard scale in the problem, $\Lambda \sim m_s$. 

An important misconception that appears frequently in the literature is related 
to the treatment of the cutoff by employing very large values of $\Lambda$ or even 
regarding $\Lambda \to \infty$. While this is perfectly fine  
in ChPT, where observables are calculated perturbatively and \emph{all} emerging UV divergencies can
be absorbed by the corresponding counterterms at any fixed order in the chiral
expansion, this is not a valid procedure for the case at hand. Let us further elaborate 
on this issue using the above example. 
At first sight, the appearance of positive powers of $\Lambda$ and/or logarithmic terms in the
predicted ``chiral'' expansion of the subthreshold parameters, see
Eq.~(\ref{LETWeinberg}), may give a (wrong)
impression that no finite limit exists for $r_\Lambda $ and $(v_i)_\Lambda$ 
as $\Lambda \to \infty$.  Actually, taking
the limit $\Lambda \to \infty$ does not commute with the Taylor expansion of
the ERE coefficients in
powers of $m_l$. 
Substituting the value for $C_0 (\Lambda )$ resulting from matching Eq.~(\ref{aWeinb1}) to (\ref{a_data}) 
into the solution of the LS equation (\ref{LSmatrix}) and taking the limit $\Lambda
\to \infty$ yields the following finite, cutoff-independent result for the
inverse amplitude:
\bea
\label{Tperatized}
(T_{\infty})^{-1} &=& i \frac{k m}{4 \pi}  - \frac{m}{8 \pi  m_l^3
  \left(k^2+m_s^2\right)
 \left(\alpha _l m_s+m_l \left(1-2 \alpha _l\right) \alpha _s\right)} \Big(
2 m_l^4 m_s^2 \left(m_s-m_l \alpha _l \alpha _s\right) \nn
&& {} + k^2 m_l^2 \left(\left(4-3 \alpha _l\right) m_s^3+m_l^2 \alpha _l
  m_s+m_l \alpha _s \left(\left(2 \alpha _l-3\right) m_s^2+m_l^2 \left(1-2
      \alpha
   _l\right)\right)\right) \nn
&& {} + k^4 \left(-\alpha _l m_s \left(m_l^2+m_s^2\right)-m_l \alpha _s
  \left(m_l^2 \left(1-2 \alpha _l\right)+m_s^2\right)+2 m_s^3\right)  \Big)\,.
\eea
The corresponding infinite-cutoff prediction for the effective range has the form:
\be
r_{\infty} =  \frac{1}{m_l} \bigg[\frac{3 \alpha_l - 4}{\alpha _l}
+\frac{4 \left(\alpha _l-1\right){}^2 \alpha _s}{\alpha _l^2 m_s} m_l
 +
\frac{\alpha _l^3 \left(8 \alpha _s^2-1\right)+\alpha _l^2
   \left(2-20 \alpha _s^2\right)+16 \alpha _l \alpha _s^2-4 \alpha
   _s^2}{\alpha _l^3 m_s^2} m_l^2
+ \ldots \bigg],
\ee
where the ellipses refer to $\mathcal{O} \left( m_l^3 \right)$-terms.
One observes that the result after removing the cutoff fails to reproduce the
low-energy theorem by yielding a wrong value for $\alpha_{r}^{(1)}$. This also
holds true for the $\alpha_{v_i}^{(1)}$ \cite{Epelbaum:2009sd}.  Notice that, by 
construction, the scattering length is still correctly reproduced. 
The breakdown of LETs in the Weinberg-like approach in the $\Lambda \to
\infty$ limit can be traced back to spurious $\Lambda$-dependent contributions
still appearing in renormalized expressions for observables, 
see Eq.~(\ref{LETWeinberg}), which are irrelevant at the order 
the calculations are performed in the regime $\Lambda
\sim m_s$ but become numerically dominant if $\Lambda \gg m_s$.
Due to non-renormalizability of the effective potential as discussed above, 
such spurious terms do, in general, 
involve logarithms and positive powers of $\Lambda$ which, as $\Lambda$ 
gets increased beyond the hard scale $m_s$, become, at some point, comparable
in size with lower-order terms in the ``chiral'' expansion.
For example, the appearance of terms linear in $\Lambda$ would suggest the breakdown
of LETs as the cutoff approaches the scale $\Lambda \sim m_s^2/m_l$. 
The unavoidable appearance of ever higher power-law divergences when going to
higher orders in the EFT expansion implies that the cutoff should not be
increased beyond the pertinent hard scale in Weinberg-like or Lepage-like
approach to NN scattering leading to $\Lambda \sim m_s$ as the optimal
choice. It is furthermore instructive to compare the predictions for the effective range
in Eqs.~(\ref{LETKSW}) and (\ref{LETWeinberg}) corresponding to two different
renormalization schemes. One observes that taking $\Lambda \gg m_s$ in
Eq.~(\ref{LETWeinberg}) has an effect which is qualitatively similar to choosing $\mu \gg m_l$ in
Eq.~(\ref{LETKSW}) and corresponds to an improper choice of renormalization conditions in
the EFT framework.

\subsubsection{Toy model with local interactions: a numerical example}

Having clarified the important conceptual issues related to nonperturbative 
renormalization in the context of chiral EFT for two nucleons, I now turn to the  
last toy-model example and give some numerical results. 

\begin{figure}[t]
\begin{center}
\includegraphics[width=0.5\textwidth]{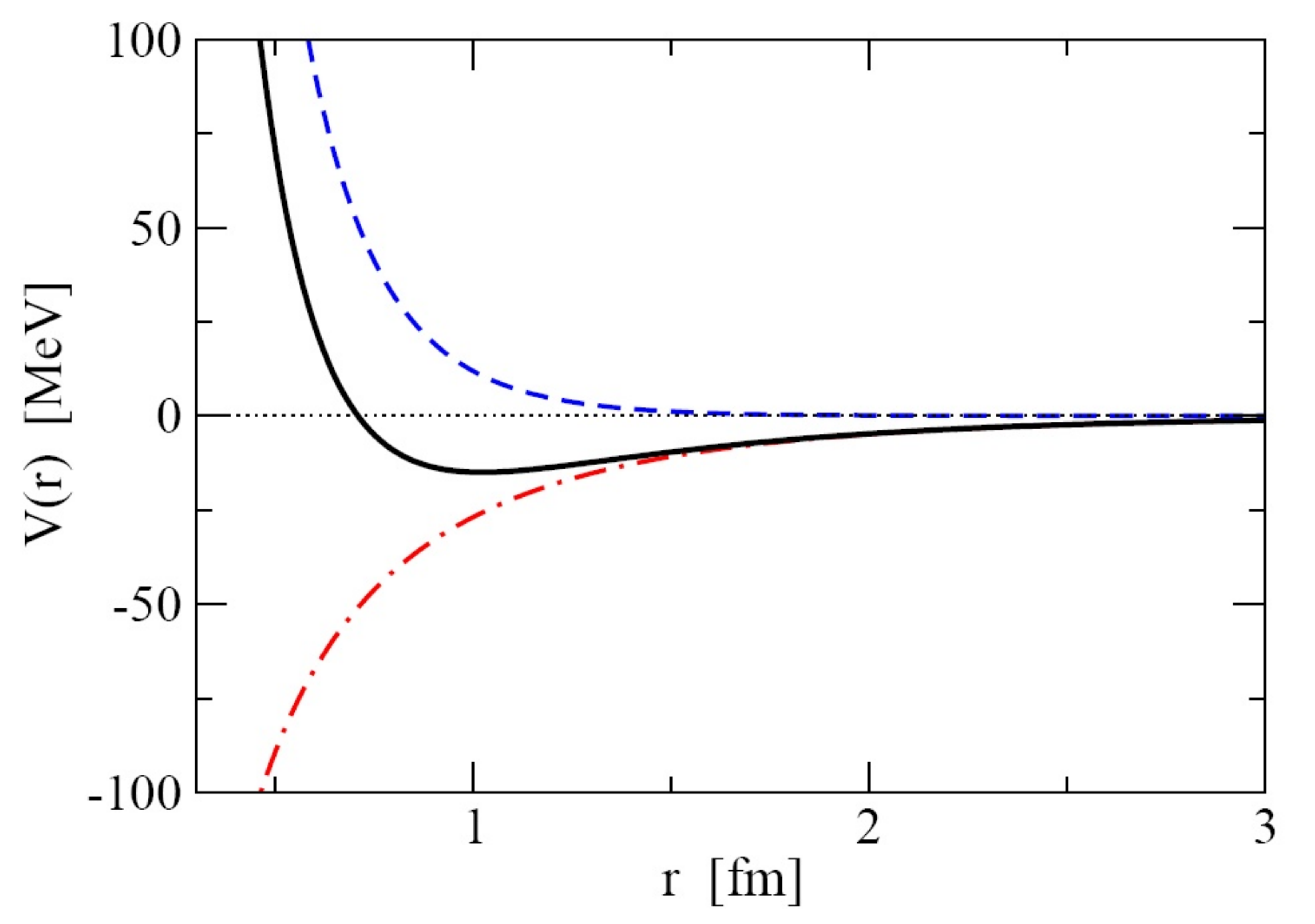}
\end{center}
\vskip -0.7 true cm 
\caption{Toy-model potential in Eq.~(\ref{pot_r}). 
The dashed (dashed-dotted) line depicts the short-range (long-range) part proportional to $\alpha_s$ ($\alpha_l$)
while the full potential is shown by the solid line.   } \label{fig_toy2}
\end{figure}

\begin{figure}[t]
\includegraphics[width=0.48\textwidth]{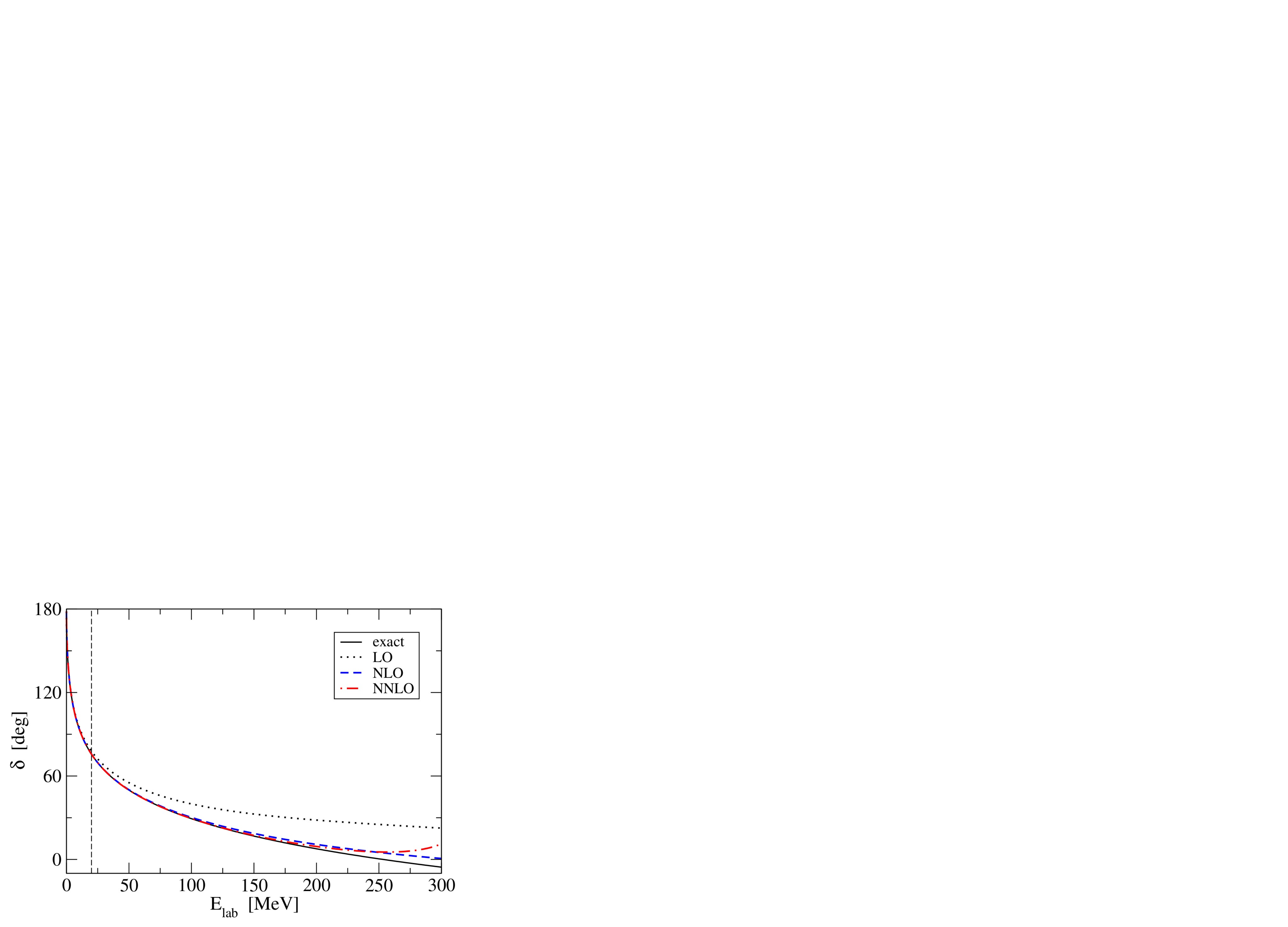}
\hfill
\includegraphics[width=0.49\textwidth]{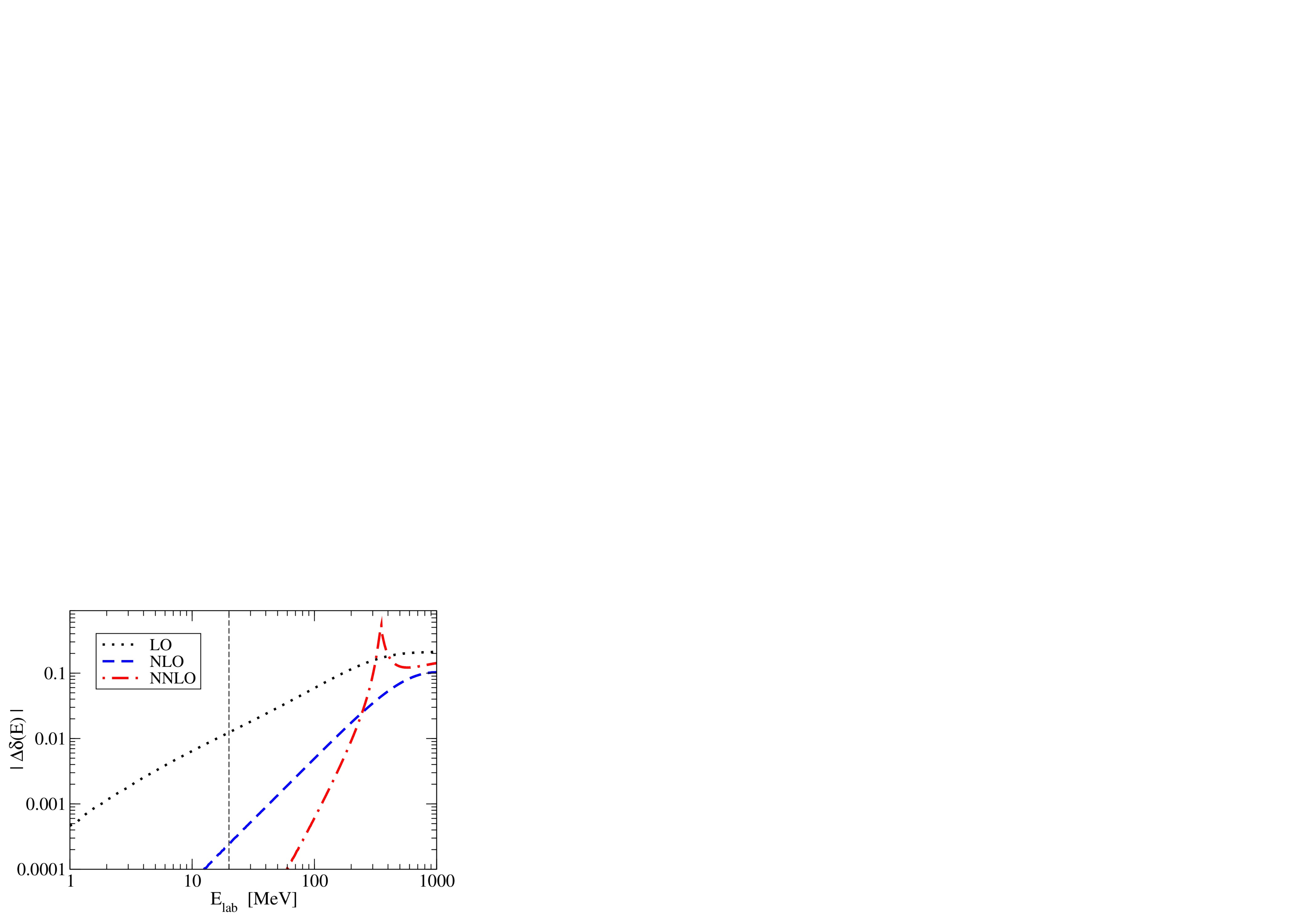}
\vskip 0.0 true cm 
\caption{Left panel: phase shifts resulting from the original and  effective potentials in Eqs.~(\ref{pot_loc_mom}) 
and (\ref{pot_eff_mom}), respectively. Right panel: Lepage plot showing the absolute error in phase shifts 
at various orders in the low-momentum expansion versus lab.~energy. 
} \label{toy_results}
\end{figure}

Consider two nucleons interacting via the local force given by a superposition of two Yukawa potentials 
corresponding to the (static) exchange of the scalar light and heavy mesons of masses 
$m_l$ and $m_s$, respectively:
\be
\label{pot_r}
V (r) = \alpha_l \frac{e^{-m_l r}}{r} +   \alpha_s \frac{e^{-m_s r}}{r} 
\ee
This type of potentials is sometimes referred to as Malfliet-Tjon potential.   
Motivated by the realistic case of the two-nucleon force, I choose 
the meson masses to be $m_l = 200$ MeV and $m_s = 750$ MeV. Further, I adjust the dimensionless 
strengths $\alpha_{l,s}$ in such a way that the potential features an S-wave  bound state 
(``deuteron'') with the binding energy $E_B = 2.2229$ MeV. A suitable combination is given by 
$\alpha_l = -1.50$ and $\alpha_s = 10.81$.
With the parameters specified in this way, the potential is depicted in Fig.~\ref{fig_toy2}. 
The corresponding momentum-space potential can be easily obtained by making the Fourier transformation:
\beq
\label{pot_loc_mom}
V (\vec p \,', \, \vec p \, ) = \frac{4 \pi \alpha_l}{\vec q \, ^2 + m_l^2} +  \frac{4 \pi \alpha_s}{\vec q \, ^2 + m_s^2}\,.
\eeq
Here, $\vec q = \vec p \, ' - \vec p$ denotes the momentum transfer. 
I treat the nucleons in this example as identical, spin-less particles. 
Thus, in the partial wave basis, the only nonvanishing matrix 
elements $\langle l ',  j, p' \, | V | l , j , p \rangle$ correspond to $l = l' = j$. 
I only consider the S-wave here.  Because of no spin dependence, the matrix element 
$V_0 (p', \, p) \equiv \langle 0, 0,  p' \, | V | 0 , 0 , p \rangle$ can be obtained by simply integrating over 
the angle $\theta$ between $\vec p \, '$ and $\vec p\,$:
\beqa
V_0 (p', \, p)  &=& 2 \pi \int_{-1}^{+1} d (\cos \theta)\, 
V (p', p, \theta ) \nn
&=&  \alpha_l \frac{4\pi^2}{p' p} \ln \left( \frac{(p' + p)^2 + m_l^2}{(p' - p)^2 + m_l^2} \right)
+ \alpha_s \frac{4\pi^2}{p' p} \ln \left( \frac{(p' + p)^2 + m_s^2}{(p' - p)^2 + m_s^2} \right) \nn
&\equiv& V_0^l (p', \, p)  + V_0^s (p', \, p) \,.
\eeqa
Contrary to the previously considered case of a separable interaction, the LS equation 
\beq
T_0 (p ,\, p'; \, k ) =  V_0 (p ,\, p') + \int \frac{l^2 dl}{(2 \pi )^3} V_0
(p ,\, l)
\frac{m}{k^2-l^2 + i \epsilon} T_0 (l ,\, p'; \, k )\,,
\eeq
cannot be solved analytically for the Malfliet-Tjon-type potentials. It can, however, be 
solved numerically using the standard methods, see e.g.~\cite{Gloeckle:1983aa}. With the parameters 
specified above, one obtains the phase shift which is shown by the solid line in the left panel 
of Fig.~\ref{toy_results}. It is fairly similar to the neutron-proton $^3$S$_1$ phase shift, 
cf.~the left panel of Fig.~\ref{phases_nijm}. 

I now develop an effective potential that describes the same physics as the underlying 
one at momenta of the order of $Q \sim m_l$. Up to N$^2$LO, it takes the form:\footnote{One can write 
down another contact interaction with four derivatives whose matrix elements in the on-shell kinematics, 
i.e. with $p = p'$, cannot be disentangled from the $C_4$-term.}
\beq
\label{pot_eff_mom}
V^{\rm eff}_0  (p', \, p)  = V_0^l (p', \, p) + \left[ C_0 + C_2 (p'^2 + p^2) + C_4  p'^2   p^2 \right]
f_\Lambda (p', p)
\eeq 
The regulator function prevents the appearance of ultraviolet divergences in the LS equation
and is chosen of the form:
\beq
f_\Lambda (p', p) = \exp \left(- \frac{p'^2 + p^2}{\Lambda^2} \right) \,.
\eeq
I set the cutoff $\Lambda = 500$ MeV, solve the LS equation with the effective potential 
$V^{\rm eff}_0 (p', \, p)$ and adjust the LECs to reproduce the coefficients in the ERE as follows:
\begin{itemize}
\item
LO: $C_2 = C_4 =0$, $C_0$ is tuned to reproduce $a$;
\item
NLO: $C_4 =0$, $C_{0,2}$ are tuned to reproduce $\{a, \, r \}$;
\item
N$^2$LO: $C_{0,2,4}$ are tuned to reproduce $\{a, \, r, \, v_2 \}$.
\end{itemize}
With the LECs being fixed as described above, the predictions for the S-wave phase shift at various 
orders in the low-momentum expansion are summarized in the left panel of Fig.~\ref{toy_results}. 
The so-called Lepage plot in the right panel of this figure shows absolute errors 
in the phase shift, $\Delta \delta (E_{\rm lab} ) \equiv  \delta_{\rm underlying} -   \delta_{\rm eff} $, 
versus energy. It is plotted in radians. One reads off from this plot that the laboratory energy, at 
which the expansion breaks down, is of the order of $E_{\rm lab} = 2 k^2/m \sim 250$ MeV. This corresponds 
to the momentum scale of the order of $\tilde \Lambda \sim 350$ MeV, in a good agreement with the 
expected breakdown scale of the modified effective range expansion of the order of $m_s/2$, see section \ref{sec:analyt}. 
Notice that the effective theory is, as desired,  able to go beyond the ERE, whose range of 
convergence is indicated by the vertical lines in  Fig.~\ref{toy_results}. The ``deuteron'' binding energy 
is found to be reproduced correctly with 5 significant digits at N$^2$LO:
\beq
E_B^{LO} + \delta E_B^{NLO} + \delta E_B^{N^2LO} = 2.1594 + 0.638 - 0.0003 = 2.2229 \mbox{ MeV}\,.
\eeq
For further illustrative quantum mechanical examples and a discussion on renormalization in the context 
of the Schr\"odinger equation, the reader is referred to the excellent lecture notes by Lepage
\cite{Lepage:1997}.

\section{Nuclear forces from chiral EFT}
\def\theequation{\arabic{section}.\arabic{equation}}
\label{sec5}

In this section I  outline and exemplify some methods which can be used to derive nuclear 
forces from chiral EFT. 

\subsection{Derivation of nuclear potentials from field theory}

The derivation of a potential from field theory is an extensively studied problem in nuclear physics. 
Historically, the important conceptual achievements in this field have been done in the fifties 
of the last century
in the context of the so-called meson field theory. The problem can be formulated in 
the following way: given a field theoretical Lagrangian for interacting mesons 
and nucleons, how can one reduce the (infinite dimensional) equation of motion for 
mesons and nucleons to an effective Schr\"odinger equation for nucleonic 
degrees of freedom, which can be solved by standard methods?
It goes beyond the scope of this work to address the whole variety of
different techniques which have been developed to construct effective interactions, 
see Ref.~\cite{Phillips:1959aa}
for a comprehensive review. I will now briefly outline a few methods which have been used in the 
context of chiral EFT. Similar methods are frequently used in computational nuclear 
physics in order to reduce a problem to a smaller model space which can be treated numerically.  

I begin with the approach developed by Tamm \cite{Tamm:1945qv} and Dancoff \cite{Dancoff:1950ud} which in the following 
will be referred to as the Tamm-Dancoff (TD) method. Consider the time-independent Schr\"odinger equation 
\beq
\label{schroed1}
(H_0 + H_I) | \Psi \rangle = E | \Psi \rangle\,,
\eeq
where $|\Psi \rangle$ denotes an eigenstate of the Hamiltonian $H$ 
with the eigenvalue $E$. 
One can divide the full Fock space in to the nucleonic subspace $|\phi \rangle$ 
and the complementary one $|\psi \rangle$  and rewrite the Schr\"odinger equation 
(\ref{schroed1}) as 
\begin{equation}
\label{schroed2}
\left( \begin{array}{cc} \eta H \eta & \eta H \lambda \\ 
\lambda H \eta & \lambda  H 
\lambda \end{array} \right) \left( \begin{array}{c} | \phi \rangle \\ 
| \psi \rangle \end{array} \right)
= E  \left( \begin{array}{c} | \phi \rangle \\ 
| \psi \rangle \end{array} \right)~,
\quad \,
\end{equation}
where I introduced the projection operators $\eta$ and 
$\lambda$ such that $|\phi \rangle = \eta | \Psi \rangle$,
$| \psi \rangle = \lambda | \Psi \rangle$.
Expressing the state $| \psi \rangle$ from the second line 
of the matrix equation (\ref{schroed2}) as 
\begin{equation}
\label{5.3}
| \psi \rangle = \frac{1}{ E - \lambda H \lambda}  H  | \phi \rangle~,
\end{equation}
and substituting this in to
the first line,  one obtains the Schroedinger-like equation for the projected 
state $| \phi \rangle$:
\begin{equation}
\label{TDschroed}
\left( H_0 + V_{{\rm eff}}^{\rm TD} ( E ) \right) | \phi \rangle  = E | \phi \rangle \,,
\end{equation}
with an effective potential $V_{\rm eff} (E)$ given by
\begin{equation}
\label{TDpot}
V_{\rm eff}^{\rm TD} (E)= \eta H_I \eta + \eta H_I \lambda 
\frac{1}{E - \lambda H \lambda} \lambda H_I \eta  \,\, .
\end{equation}
This definition of the effective potential corresponds exactly to the one given in section \ref{sec:nuclearEFT}  
in the context of ``old-fashioned'' time-ordered perturbation theory. 
To evaluate $V_{\rm eff}^{\rm TD} (E)$ one usually relies on perturbation theory. 
For example, for the Yukawa theory with a single $\pi NN$ vertex $H_I = g H_1$, $V_{\rm eff}^{\rm TD} (E)$ 
up to the fourth order in the coupling constant $g$ is given by
\beq
\label{TDg4}
V_{\rm eff}^{\rm TD} (E) = - \eta ' \bigg[ g^2 H_1 \frac{\lambda^1}{H_0 - E} H_1  +  g^4 
H_1 \frac{\lambda^1}{H_0 - E} H_1 \frac{\lambda^2}{H_0 - E}  H_1 \frac{\lambda^1}{H_0 - E} H_1  + \mathcal{O} (g^6) \bigg] \eta \,,
\eeq
where the superscripts of $\lambda$ refer to the number of mesons in the corresponding state. 
It is important to realize that the effective potential $V_{{\rm eff}} ( E )$ in this scheme depends explicitly on the energy,
which makes it inconvenient for practical applications (especially for calculations beyond the two-nucleon system). 
In addition, the projected nucleon states $| \phi \rangle$ have a different 
normalization compared to the states $| \Psi \rangle$ we have started from 
(which are assumed to span a complete and orthonormal set in the whole Fock space)
\beq
\langle \phi_i | \phi_j \rangle =  
\langle \Psi_i | \Psi_j \rangle - \langle \psi_i | \psi_j \rangle=
\delta_{ij} - \langle \phi_i | H_I \lambda 
\left( \frac{1}{E - \lambda H \lambda} \right)^2 
\lambda H_I | \phi_j \rangle ~,
\eeq
since the components $\psi_i$ do, in general, not vanish.

The above mentioned deficiencies are naturally avoided in the method of unitary transformation \cite{Okubo:1954aa,Fukuda:1954aa}.
In this approach, the decoupling of the $\eta$- and $\lambda$-subspaces of the Fock space is achieved via a 
unitary transformation $U$   
\beq
\label{decoupling}
\tilde H \equiv U^\dagger H U = \left( \begin{array}{cc} \eta \tilde H \eta  & 0 \\ 0 & 
\lambda \tilde H \lambda \end{array} \right)\,.
\eeq
Following Okubo \cite{Okubo:1954aa}, the unitary operator $U$ can be parametrized as
\begin{equation}
\label{5.9}
U = \left( \begin{array}{cc} \eta (1 +  A^\dagger  A )^{- 1/2} & - 
 A^\dagger ( 1 +  A A^\dagger )^{- 1/2} \\
 A ( 1 +  A^\dagger  A )^{- 1/2} & 
\lambda (1 +  A  A^\dagger )^{- 1/2} \end{array} \right)~,
\end{equation}
with the operator $A= \lambda  A \eta$. The operator $A$ has to satisfy 
the decoupling equation 
\begin{equation}
\label{5.10}
\lambda \left( H - \left[ A, \; H \right] - A H A \right) \eta = 0
\end{equation}
in order for the transformed Hamiltonian $\tilde H$ to be of block-diagonal form. 
The effective $\eta$-space potential $\tilde V_{\rm eff}^{\rm UT}$ can be expressed in terms of the operator $A$ as: 
\beq
\label{effpot}
\tilde{V}_{\rm eff}^{\rm UT} =  \eta (\tilde H  - H_0 ) = \eta \bigg[ (1 + A^\dagger A)^{-1/2} (H + A^\dagger H + H A + A^\dagger H A )  
(1 + A^\dagger A)^{-1/2} - H_0 \bigg] \eta~.
\eeq
The solution of the decoupling equation and the calculation of the effective potential 
according to Eq.~(\ref{effpot}) can be carried out perturbatively in the weak-coupling case. 
For the previously considered case of the Yukawa theory, the decoupling equation can be solved 
recursively by making the ansatz
\beq
A = \sum_{n=1}^\infty g^n A^{(n)}. 
\eeq
The resulting effective potential $V_{\rm eff}^{\rm UT}$  takes the form:
\beqa
\label{UTg4}
V_{\rm eff}^{\rm UT} &=&  - g^2 \, \eta ' \Bigg[ \frac{1}{2} H_1 \frac{\lambda^1}{H_0 - E_\eta} H_1 + \mbox{h.~c.} \Bigg] \eta 
  - g^4 \, \eta ' \Bigg[ \frac{1}{2} H_{1} \frac{\lambda^1}{(H_0 - E_{\eta})}  H_{1} \,  \frac{\lambda^2}{(H_0 - E_{\eta})} 
\,  H_{1} \frac{\lambda^1}{(H_0 - E_{\eta} )}  H_{1}  \nn 
&& {} -
\frac{1}{2} H_{1} \frac{\lambda^1}{(H_0 - E_{\eta '})}  H_{1} \, \tilde \eta 
\,  H_{1} \frac{\lambda^1}{(H_0 - E_{\tilde \eta} )( H_0 - E_{\eta '} )}  H_{1}  \nn
&& {} 
+\frac{1}{8} H_{1} \frac{\lambda^1}{(H_0 - E_{\eta '})}  H_{1} \, \tilde \eta 
\,  H_{1} \frac{\lambda^1}{(H_0 - E_{\tilde \eta} )( H_0 - E_{\eta} )}  H_{1}  \nn
&&  {} - \frac{1}{8} H_{1} \frac{\lambda^1}{(H_0 - E_{\eta '}) ( H_0 - E_{\tilde \eta} )}  
H_{1} \, \tilde \eta 
\,  H_{1} \frac{\lambda^1}{(H_0 - E_{\tilde \eta} )}  H_{1}  + \mbox{h.~c.}
\Bigg] \eta  + \mathcal{O}(g^6)\,.
\eeqa
Here, $\eta$, $\eta '$ and $\tilde \eta$ denote projection operators onto the purely nucleonic states. 
Different notation is only used to indicate what state the energies in the denominators correspond to.
In contrast to $V_{\rm eff}^{\rm TD}$, $V_{\rm eff}^{\rm UT}$ does not depend on 
the energy $E$ which enters the Schr\"odinger equation. 
Another difference to the Tamm-Dancoff method is given by the presence of terms 
with the projection operator $\tilde \eta$ which give rise to purely nucleonic intermediate states.
These terms are responsible for the proper normalization of the few-nucleon states.  
In spite of the presence of the purely nucleonic intermediate states, such terms are not 
generated through the iteration of the dynamical equation and are truly irreducible.  
Since \emph{all} energy denominators entering $V_{\rm eff}^{\rm UT}$ correspond to intermediate states 
with at least one pion, there is no enhancement by large factors of $m/Q$ that occurs for reducible 
contributions. 

\begin{figure*}
\vspace{0.3cm}
\centerline{
\includegraphics[width=0.1\textwidth]{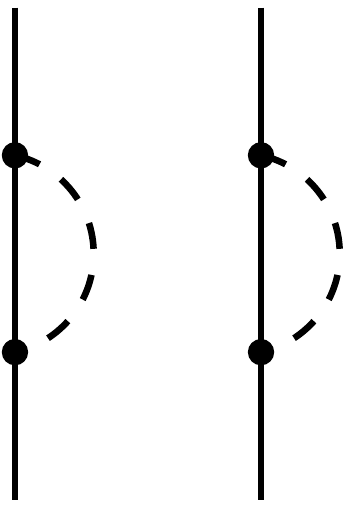}
}
\vspace{-0.2cm}
\caption[fig4aa]{\label{figEx} An example of a disconnected diagram that does not contribute to the NN 
potential in the method of unitary transformation. }
\vspace{0.2cm}
\end{figure*}

\begin{minipage}{\textwidth}
\vskip 0 true cm
\rule{\textwidth}{.2pt}
{\it
Exercise:  \\
1. Calculate $A^{1}$, $A^{(2)}$ and $A^{(3)}$ by solving the decoupling equation for the considered case of 
Yukawa theory and verify the expression for  $V_{\rm eff}^{\rm UT}$  in Eq.~(\ref{UTg4}). \\
2. Consider the disconnected Feynman diagram in Fig.~\ref{figEx} and draw all possible time-ordered diagrams. 
Using Eqs.~(\ref{TDg4}) and (\ref{UTg4}) show that, in contrast to the TD approach,  
these diagrams do not contribute to the nucleon-nucleon potential in the method of unitary transformation. 
Use the static approximation for the nucleons in order to simplify the calculations (i.e. set:
$E = E_\eta = E_{\eta '} = E_{\tilde \eta} =0$).
} \\
\vskip -0.8 true cm
\rule{\textwidth}{.2pt}
\end{minipage}

\medskip
The two methods of deriving effective nuclear potentials are quite general 
and can, in principle, be applied to any field theoretical meson-nucleon Lagrangian. 
In the weak-coupling case, the potential can be obtained straightforwardly via the expansion 
in powers of the corresponding coupling constant(s). 
For practical applications, it is helpful to use time-ordered diagrams to visualize 
the contributions to the potential, see Fig.~\ref{fig4aa}. 
In ``old-fashioned'' perturbation theory or, equivalently, the Tamm-Dancoff approach, 
only irreducible diagrams contribute to the potential. 
In the method of unitary transformation one can draw both irreducible and reducible graphs 
whose meaning, however, differs from that of diagrams emerging in time-ordered perturbation theory.  
The coefficients in front of various operators and the energy denominators can, in general,  
not be guessed by looking at a given diagram and have to be determined by 
solving the decoupling equation (\ref{5.10}) for the operator $A$ and using Eq.~(\ref{effpot}). 

Application of the above methods to  the effective chiral Lagrangian requires
the expansion in powers of the coupling constants to be replaced by the chiral expansion in 
powers of $Q/\Lambda_\chi$. This issue will be dealt with in the next section.

\subsection{Method of unitary transformation}

To apply the method of unitary transformation to derive nuclear forces in chiral EFT 
it is useful to rewrite the power counting discussed in section \ref{sec:nuclearEFT} into 
a different form which is more suitable to carry out algebraic manipulations described 
above. 

\begin{figure}[t]
\includegraphics[width=0.99\textwidth]{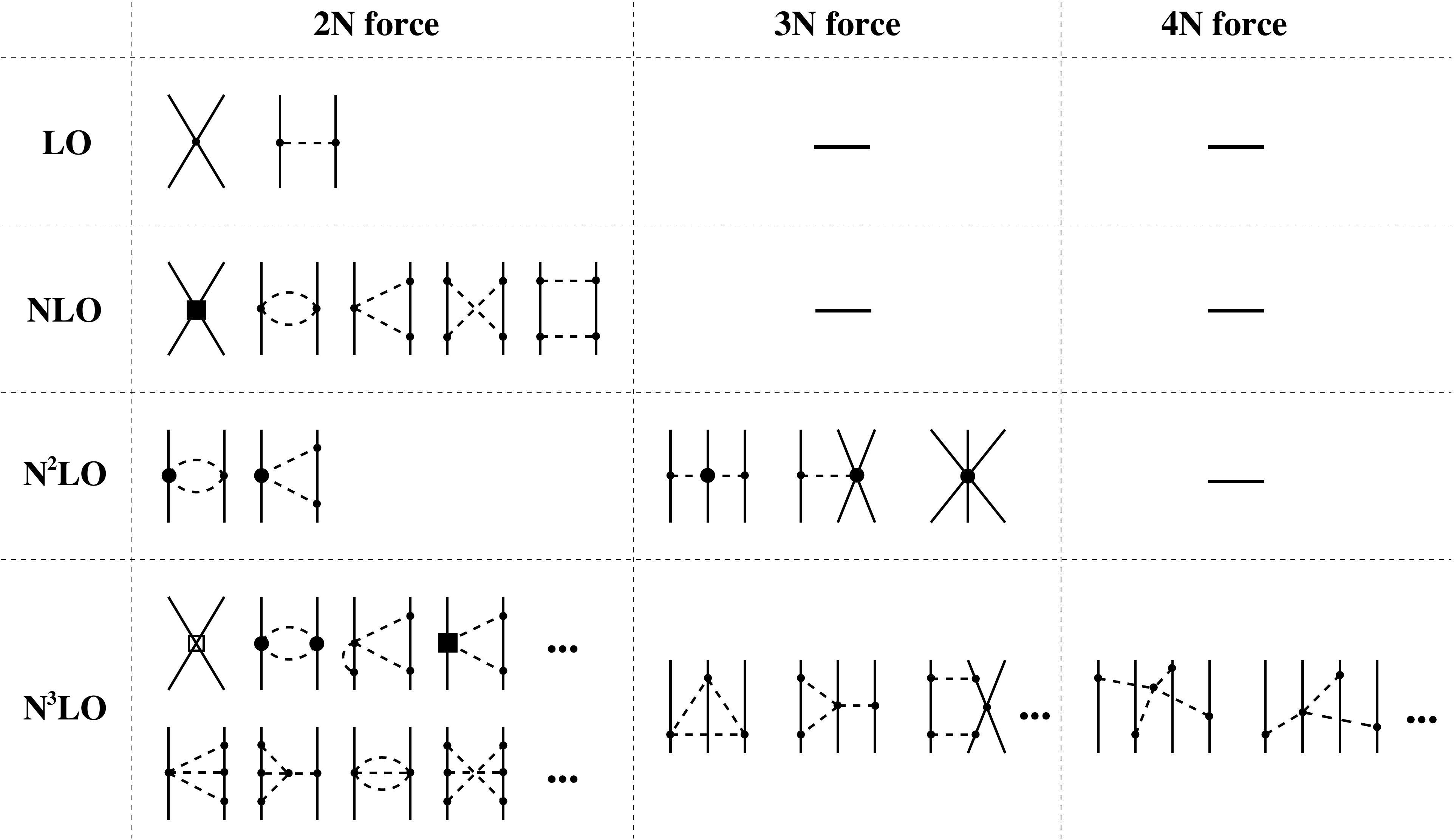}
\caption{Diagrams that give rise to nuclear forces in ChEFT based on Weinberg's power
  counting. Solid and dashed lines denote nucleons
  and pions, respectively. Solid dotes, filled circles and filled squares
  and crossed squares refer to vertices with $\Delta_i =0, \, 1, \, 2$ and $4$, respectively.} 
\label{hierarchy}
\end{figure}

We begin with Weinberg's original power counting expression for $N$-nucleon 
diagrams involving $C$ separately connected pieces:   
\beq
\label{pow_orig}
\nu = 4 - N + 2 (L - C) + \sum_i V_i \Delta_i \,, \quad \quad
\Delta_i = d_i + \frac{1}{2} n_i - 2\,.
\eeq
This expression is a generalization of Eq.~(\ref{powNN}) to the case $C > 1$. 
Its derivation can be found in Ref.~\cite{Weinberg:1992yk}. 
There is one subtlety here that needs to be addressed: 
according to Eq.~(\ref{pow_orig}), the chiral dimension $\nu$ for a given
process depends on the total number of nucleons in the system. For example,
one-pion exchange in the two-nucleon system corresponds to $N=2$, $L=0$,
$C=1$ and $\sum_i V_i \Delta_i =0$ and, therefore, contributes at order $\nu
=0$. On the other hand, the same process in the presence of a third
(spectator) nucleon leads, according to Eq.~(\ref{pow_orig}), to $\nu = -3$
since  $N=3$ and $C=2$.  The origin of this seeming discrepancy is due to the 
different normalization of the 2N and 3N states:
\beqa 
&2N:& \quad \langle \vec p_1 \, \vec p_2 | \vec p_1 {}' \,  \vec p_2 {}' 
\rangle = \delta^3 (\vec p_1 {} ' - \vec p_1 \, ) \, 
\delta^3 (\vec p_2 {}' - \vec p_2 \, ) \,,\nn
&3N:& \quad \langle \vec p_1 \, \vec p_2  \, \vec p_3 | \vec p_1 {}' \,  
\vec p_2 {}'  \,  \vec p_3 {}' \rangle = 
\delta^3 (\vec p_1 {} ' - \vec p_1 \, ) \, \delta^3 (\vec p_2 {}' - \vec p_2
\, ) \,\delta^3 (\vec p_3 {}' - \vec p_3 \, ) \,.
\eeqa
It can be circumvented by assigning a chiral dimension to the transition
operator rather than to its matrix elements in the $N$-nucleon space. 
Adding the factor $3N-6$  to the right-hand side of Eq.~(\ref{pow_orig}) in
order  to account for the normalization 
of the $N$-nucleon states and to ensure that the LO contribution to the nuclear
force appears at order $\nu = 0$ we obtain
\beq
\label{pow_mod}
\nu = -2 + 2 N + 2 (L - C) + \sum_i V_i \Delta_i \,.
\eeq
This expression provides a natural qualitative explanation of
the observed hierarchy of nuclear forces $V_{\rm 2N} \gg
V_{\rm 3N} \gg V_{\rm 4N} \ldots $ with  
\beqa
V_{\rm 2N} &=&  V_{\rm 2N}^{(0)} + V_{\rm 2N}^{(2)} + V_{\rm 2N}^{(3)} + V_{\rm 2N}^{(4)} + \ldots \,, \nn
V_{\rm 3N} &=&  V_{\rm 3N}^{(3)} + V_{\rm 3N}^{(4)} + \ldots \,, \nn 
V_{\rm 4N} &=&  V_{\rm 4N}^{(4)} + \ldots \,,
\eeqa
as shown in Fig.~\ref{hierarchy}. 

The form of power counting in Eq.~(\ref{pow_mod}) is still of less use for our purpose since the resulting 
chiral dimension is given it terms of the topological quantities such as $N$, $C$ and $L$ which is not 
appropriate for algebraic approaches such as the method of unitary transformation.    
Using certain topological identities, see \cite{Epelbaum:2007us}, Eq.~(\ref{pow_mod}) can be 
rewritten in a more suitable form:
\beq
\label{pow_fin}
\nu = -2 + \sum V_i \kappa_i \,, \quad \quad \kappa_i = d_i + \frac{3}{2} n_i + p_i - 4\,.
\eeq
The quantity $\kappa_i$ which enters this expression is nothing but the 
canonical field dimension of a vertex of type $i$ (up to the additional
constant $-4$) and gives the inverse mass dimension of the corresponding
coupling constant. In fact, this result can be obtained immediately by counting 
inverse powers of the hard scale $\Lambda_\chi$ rather than powers of the soft scale $Q$
(which is, of course, completely  equivalent). Indeed, since the only way 
for the hard scale to be generated is through the physics behind the LECs, 
the power $\nu$ is just the negative of the overall mass dimension of all LECs. The
additional factor $-2$ in  Eq.~(\ref{pow_fin}) is a convention 
to ensure that the contributions to the nuclear force start at $\nu = 0$. 
I encourage the reader to 
verify the equivalence of  Eqs.~(\ref{pow_fin}) and (\ref{pow_mod}) 
for specific diagrams. One immediately reads off from Eq.~(\ref{pow_fin})  that 
in order for perturbation theory to work, the effective Lagrangian must contain no renormalizable 
and super-renormalizable interactions with $\kappa_i =0$ and $\kappa_i < 0$, respectively, 
since otherwise adding new vertices would not increase or even lower the chiral dimension 
$\nu$.  This feature is guaranteed by the spontaneously broken chiral symmetry of QCD which 
ensures that only non-renormalizable interactions enter the effective Lagrangian. 

While Eq.~(\ref{pow_fin}) does not say much about the topology and is,
therefore, not particularly useful to
deal with diagrams, it is very convenient for algebraical calculations. In fact, 
it formally reduces the chiral expansion to the expansion in powers of the coupling constant, 
whose role is now played by the ratio $Q/\Lambda$. Applying the canonical transformation to the chiral 
Lagrangian and writing the resulting Hamiltonian in the form  
\begin{equation}
\label{n11}
H_I = \sum_{\kappa = 1}^{\infty} H^\kappa\,,
\end{equation}
the operator $A$ can be calculated by solving Eq.~(\ref{decoupling}) recursively,
\beqa
\label{n13}
A &=& \sum_{\alpha = 1}^\infty A^{(\alpha )}\,, \\ 
A^{( \alpha )} &=& \frac{1}{E_\eta - E_\lambda} \lambda \bigg[ H^{(\alpha )} + \sum_{i =
    1}^{\alpha -1} H^{(i)} A^{(\alpha -i)} - \sum_{i=1}^{\alpha -1} A^{(\alpha -i)} H^{(i)}
- \sum_{i = 1}^{\alpha -2} \; \sum_{j =1}^{\alpha - j - 1} A^{(i)} H^{(j)} A^{(\alpha -i-j)} 
\bigg] \eta\,. \nonumber
\eeqa
The expressions for the unitary operator and the effective potential then  
follow immediately by substituting Eqs.~(\ref{n11}) and (\ref{n13}) into
Eq.~(\ref{effpot}).

\subsection{The $1\pi$- and the leading $2\pi$-exchange potentials}

\begin{figure*}
\vspace{0.3cm}
\centerline{
\includegraphics[width=0.5\textwidth]{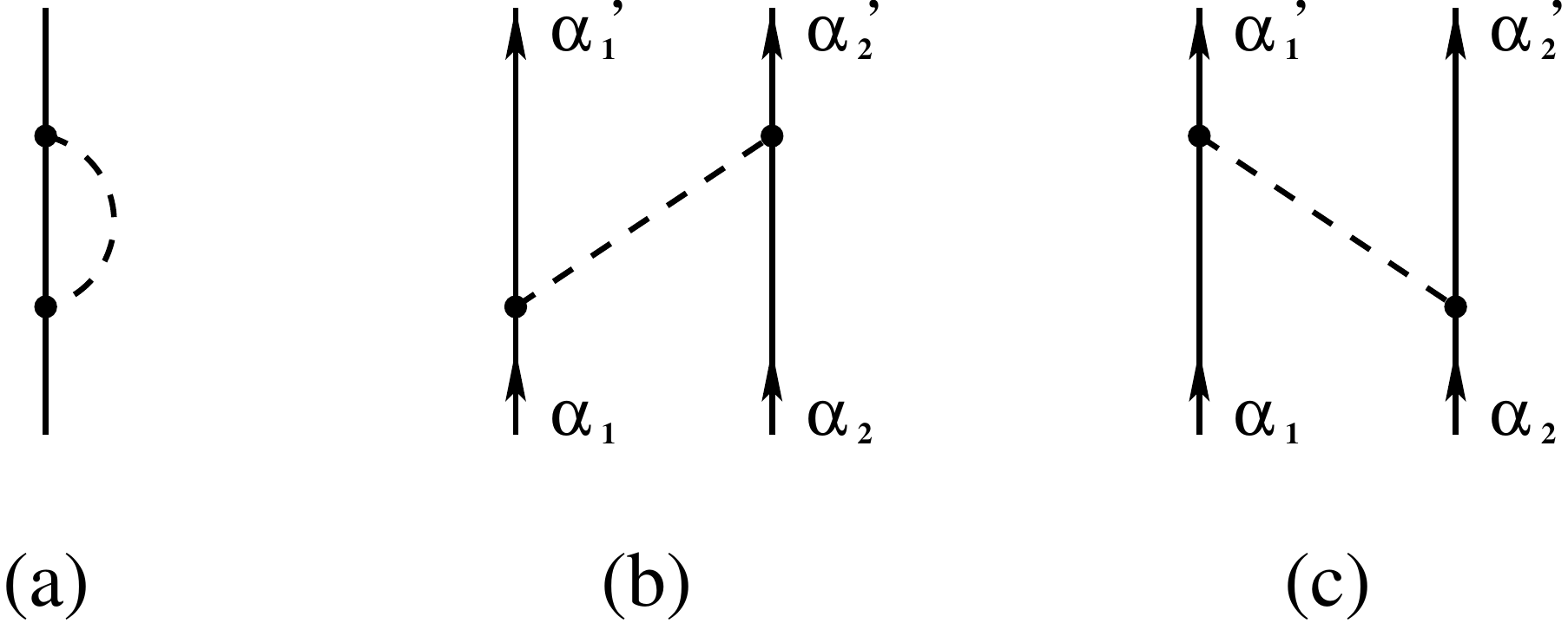}
}
\vspace{0.2cm}
\caption{\label{1pi} Diagrams that correspond to the operator in Eq.~(\ref{tempX1}).
Graph (a) yields a single-body contribution to the nuclear Hamilton operator while diagrams (b)
and (c) give rise to the $1\pi$-exchange NN potential. 
}
\vspace{0.2cm}
\end{figure*}

I now illustrate how the above ideas can be applied in practice. I begin with the simple 
case of the $1\pi$-exchange potential at leading order, i.e.~$\nu =0$. 
The only relevant contribution to the interaction Hamilton density is given by 
\beq
\label{vertex_ga}
\mathcal{H}^{(1)} = \frac{g_A}{2 F_\pi} N^\dagger \vec \sigma  \cdot ( \vec \nabla \fet \pi \cdot  \fet \tau ) N\,,
\eeq
where the superscript of $\mathcal{H}$ gives the canonical dimension $\kappa_i$ defined in Eq.~(\ref{pow_fin}). 
The relevant operator that contributes to the effective Hamiltonian after performing the unitary transformation 
is given by the first two terms in Eq.~(\ref{UTg4}):
\beq
\label{tempX1}
V_{\rm eff}^{\rm UT} = - \eta H^{(1)} \frac{\lambda^1}{\omega} H^{(1)} \eta\,,
\eeq
where $\omega$ denotes the pion free energy and 
I made use of the static approximation as appropriate at LO.\footnote{Corrections to the static $1\pi$-exchange
potential are suppressed by $Q^2/m^2$.}  Notice that $V_{\rm eff}^{\rm UT}$ in the above equation 
also gives rise to a one-body operator that contributes to the nucleon mass shift, see graph (a) in 
Fig.~\ref{1pi}. 
To compute the expression for the $1\pi$-exchange potential I first express   
the pion and nucleon fields in the interaction picture in terms of the creation and destruction operators:
\beqa
\label{2quant}
\pi_i ( x ) &=& \int \frac{d^3 k}{( 2 \pi )^{3 / 2}} \frac{1}{\sqrt{2 \omega_k}}
\left[ e^{- i k \cdot x} a_i ( \vec k \, ) + e^{i k \cdot x} a_i^\dagger ( \vec k \, ) \right] \,, \nn
N ( x) &=&  \sum_{t s} \int  \frac{d^3 p}{( 2 \pi )^{3 / 2}} \sum_{t s} e^{- i p \cdot x } \upsilon ( s )
\epsilon ( t )  b_t ( \vec p, \,\, s )  \;,  
\eeqa 
where $\omega_k =\sqrt {{\vec k}^2 + m_\pi^2}$ and  $\upsilon$ ($\epsilon$) denotes a  Pauli spinor (isospinor).  
$a_i ( \vec k \, )$ and $a_i^\dagger ( \vec k \, )$ denote a destruction and creation operator of 
a pion with isospin $i$. 
Further, $b_t ( \vec p, \,\, s )$ ($b_t^\dagger ( \vec p, \,\, s )$) is the 
destruction (creation) operator of a non-relativistic nucleon (i.e.~$p_0 = \vec p\, ^2/(2m)$) 
with the spin and isospin quantum numbers $s$
and $t$ and momentum $\vec{p}$. The creation and destruction operators of the pion (nucleon)
field satisfy the usual commutation (anti-commutation) relations. The $1\pi$-exchange potential can be 
calculated by  substituting the expressions in Eq.~(\ref{2quant}) for pion and nucleon 
fields into Eq.~(\ref{tempX1}) and evaluating the matrix element
\beq
\label{matr_el}
\langle \alpha_1 ' \alpha_2 ' |  - \eta H^{(1)} \frac{\lambda^1}{\omega} H^{(1)} \eta \, |  \alpha_1 \alpha_2 \rangle
\equiv \frac{1}{(2 \pi)^3} \delta^3 (\vec P \, ' - \vec P \, ) V_{2N}^{1\pi} \,. 
\eeq 
Here, $\vec P$ ($\vec P \,'$) denotes the total momentum of the nucleons before (after) 
the interaction takes place. Further, $\alpha_i$ and $\alpha_i '$ denote collectively 
the initial and final quantum numbers of the nucleon $i$ (momentum, spin and isospin). 
To keep the expressions for the potential compact, they are commonly given in the operator
form with respect to the spin and isospin quantum numbers using the corresponding Pauli matrices 
$\vec \sigma_i$ and $\fet \tau_i$ of a nucleon $i$. A straightforward 
calculation yields the final form of the $1\pi$-exchange potential: 
\beq
\label{1pi_res}
V_{2N}^{1\pi}  = -\frac{g_A^2}{4 F_\pi^2} \frac{\vec
  \sigma_1 \cdot \vec q \, \vec \sigma_2 \cdot \vec q}{\vec q \, ^2 +M_\pi^2} \fet \tau_1
\cdot \fet \tau_2\,.
\eeq
Clearly, this familiar result for the static $1\pi$-exchange potential can be obtained in a much 
simpler way by evaluating the corresponding Feynman diagram since it does not generate reducible topologies.    
One-loop corrections to the static $1\pi$-exchange potential and renormalization within the method of unitary 
transformation are discussed in detail in Ref.~\cite{Epelbaum:2002gb}. 
Notice further that when calculating the matrix element in Eq.~(\ref{matr_el}), I discarded 
the contributions corresponding to graph (a) in Fig.~\ref{tempX1} with one of the nucleons being a spectator 
and the contributions from diagrams (b) and (c) with the nucleon labels $\alpha_1 '$ and $\alpha_2 '$ being 
interchanged. The latter emerge automatically when the potential is inserted into the corresponding 
dynamical equation due to the antisymmetric nature of the two-nucleon wave function.

As a final example, I discuss  the leading $2\pi$-exchange potential arising 
at order $\nu = 2$ from the box and crossed box diagrams 
(the last two diagrams in the second raw in Fig.~\ref{hierarchy}. Again, the only vertex we need  
is given in Eq.~(\ref{vertex_ga}). The relevant operators that contribute to the effective nuclear Hamiltonian 
after performing the unitary transformation are listed in Eq.~(\ref{UTg4})
\beqa
\label{tempX2}
V_{\rm eff}^{\rm UT} &=& {} - \eta H^{(1)} \frac{\lambda^1}{\omega} H^{(1)}  \frac{\lambda^2}{\omega_1 + \omega_2} 
H^{(1)} \frac{\lambda^1}{\omega} H^{(1)} \eta + \frac{1}{2} \eta H^{(1)} \frac{\lambda^1}{\omega^2} H^{(1)}  \eta 
H^{(1)} \frac{\lambda^1}{\omega} H^{(1)} \eta \nn
&+& \frac{1}{2} \eta H^{(1)} \frac{\lambda^1}{\omega} H^{(1)}  \eta 
H^{(1)} \frac{\lambda^1}{\omega^2} H^{(1)}  \eta\,.
\eeqa
The contribution to the $2\pi$-exchange potential results from evaluating the  
matrix element $\langle \alpha_1 ' \alpha_2 ' | V_{\rm eff}^{\rm UT} | \alpha_1 \alpha_2 \rangle$
which can be computed along the same lines as above. Calculations of that kind can be optimized by 
using a diagrammatic approach and formulating a sort of ``Feynman'' rules. The building blocks 
are given by vertices and energy denominators that play the role of propagators in Feynman diagrams. 
Consider, for example, time-ordered box diagrams (b)-(g) in Fig.~\ref{fig4aa}. All these graphs  
have an identical sequence of non-commuting vertices generating exactly the same isospin-spin-momentum 
structure in the resulting potential. Thus, the energy denominators for different diagrams arising from 
the operators in Eq.~(\ref{tempX2}) can be added together yielding the result
\beq
2 \frac{\omega_1^2 + \omega_1 \omega_2 + \omega_2^2}{\omega_1^2 \omega_2^2 (\omega_1 + \omega_2)}\,.
\eeq
The same result but with an opposite sign is obtained for the sum of the energy denominators 
for the crossed-box diagrams.

\begin{minipage}{\textwidth}
\vskip 0 true cm
\rule{\textwidth}{.2pt}
{\it
Exercise: show that  the operators in Eq.~(\ref{tempX2}) do not give rise to the 
$2\pi$-exchange three-nucleon force. What would be the result for the  
three-nucleon force if one would employ time-ordered perturbation theory 
(in the static approximation) instead of the method of unitary transformation?
} \\
\vskip -0.8 true cm
\rule{\textwidth}{.2pt}
\end{minipage}

\medskip
The vertex in Eq.~(\ref{vertex_ga}) gives rise to the ``Feynman'' rule 
\beq
i \frac{g_A}{2 F_\pi} \tau_i^a \vec \sigma_i \cdot \vec q \frac{1}{\sqrt{2 \omega_q}}\,,
\eeq
for an incoming (outgoing) pion with momentum $\vec q$ ($- \vec q\, $) and the isospin quantum number $a$. 
Here, $i$ is the nucleon label. Putting everything together, we end up with the contribution 
from the box diagram of the form
\beqa
V_{2N}^{2\pi, \, \rm box} (\vec q \, ) &=& (2 \pi)^3 \left( \frac{g_A}{2 F_\pi } \right)^4 \fet \tau_1 \cdot \fet \tau_2
\, \fet \tau_1 \cdot \fet \tau_2 \int \frac{d^3 l_1}{(2 \pi)^3} \frac{d^3 l_2}{(2 \pi)^3} \, 
\vec \sigma_1 \cdot \vec l_1 \,  \vec \sigma_1 \cdot \vec l_2 \, 
 \vec \sigma_2 \cdot \vec l_1 \, \vec  \sigma_2 \cdot \vec l_2 \nn
&\times&  \frac{1}{2 \omega_{l_1}} \frac{1}{2 \omega_{l_2}} \, 
2 \frac{\omega_{l_1}^2 + \omega_{l_1} \omega_{l_2} + \omega_{l_2}^2}{\omega_{l_1}^2 \omega_{l_2}^2 
(\omega_{l_1} + \omega_{l_2})}\, 
\delta (\vec l_1 + \vec l_2 - \vec q \, ) \,, 
\eeqa
where the factor $ (2 \pi)^3$ in front of the integral 
is due to the normalization of the potential, see Eq.~(\ref{matr_el}). 
The contribution of the crossed-box diagrams can be written as
\beqa
V_{2N}^{2\pi, \, \rm cr.-box} (\vec q \, ) &=& - (2 \pi)^3 \left( \frac{g_A}{2 F_\pi } \right)^4 \sum_a \tau_1^a 
\, \fet \tau_1 \cdot \fet \tau_2 \, \tau_2^a \int \frac{d^3 l_1}{(2 \pi)^3} \frac{d^3 l_2}{(2 \pi)^3} \, 
\vec \sigma_1 \cdot \vec l_1 \,  \vec \sigma_1 \cdot \vec l_2 \, 
 \vec  \sigma_2 \cdot \vec l_2 \, \vec \sigma_2 \cdot \vec l_1  \nn
&\times&  \frac{1}{2 \omega_{l_1}} \frac{1}{2 \omega_{l_2}} \, 
2 \frac{\omega_{l_1}^2 + \omega_{l_1} \omega_{l_2} + \omega_{l_2}^2}{\omega_{l_1}^2 \omega_{l_2}^2 
(\omega_{l_1} + \omega_{l_2})}\, 
\delta (\vec l_1 + \vec l_2 - \vec q \, ) \,, 
\eeqa
Adding the two expressions together and performing straightforward simplifications one 
obtains the total contribution to the leading $2\pi$-exchange proportional to $g_A^4$:
\beq
\label{2pi_prom}
V_{2N}^{2\pi, \, \rm total} (\vec q \, ) = - \frac{g_A^4}{32 F_\pi^4 } 
\int \frac{d^3 l}{(2 \pi)^3} \left[ \fet \tau_1 \cdot \fet \tau_2 \left( \vec l\, ^2 - \vec q\, ^2 \right)^2 
+ 6 \, \vec \sigma_1 \cdot \vec q \times \vec l \,  \vec \sigma_2 \cdot \vec q \times \vec l \, \right]
 \frac{\omega_{+}^2 + \omega_{+} \omega_{-} + \omega_{-}^2}{\omega_{+}^3 \omega_{-}^3 
(\omega_{+} + \omega_{-})}\,,
\eeq
with $\omega_\pm \equiv \sqrt{(\vec q \pm \vec l)^2 + 4 M_\pi^2}$. The integrals appearing in the 
above expressions are ultraviolet divergent and need to be regularized. This can be achieved using standard 
methods such as e.g.~dimensional regularization. Cutoff regularization can be applied equally well. 
In the infinite-cutoff limit, $\Lambda \to \infty$, the regularized integrals can be decomposed into 
a \emph{finite} non-polynomial part (with respect to the momentum transfer $\vec q\,$) and polynomial 
in momenta terms that may diverge as $\Lambda$ goes to infinity. Such a decomposition follows from 
the local nature of the ultraviolet divergences and implies the uniqueness of the non-polynomial part  
(in the limit $\Lambda \to \infty$). This makes perfect sense from the physics point of view since 
the nonpolynomial part of the potential controls its long-range behavior which 
should not depend on the details of regularization at short distances. For the non-polynomial parts 
of the relevant integrals one obtains:
\beqa
I_1 \equiv \int \frac{d^3 l}{(2 \pi)^3} \frac{\vec l\, ^2}{\omega_+ \omega_- (\omega_+ + \omega_-)}
&=& \frac{1}{6 \pi^2} \left( 4 M_\pi^2 + q^2 \right) L (q) + \ldots \,, \nn
I_2 \equiv \int \frac{d^3 l}{(2 \pi)^3} \frac{\vec l \, ^4 + \vec q \, ^4}{\omega_+ \omega_- (\omega_+ + \omega_-)}
&=& -\frac{1}{60 \pi^2} \frac{512 M_\pi^6 + 384 M_\pi^4 q^2 + 156 M_\pi^2 q^4 +23 q^6}{4 M_\pi^2 + q^2} \, 
L (q) + \ldots \,,\nn
I_3 \equiv \int \frac{d^3 l}{(2 \pi)^3} \frac{(\vec q \cdot \vec l\, )^2}{\omega_+ \omega_- (\omega_+ + \omega_-)}
&=& + \ldots \,,
\eeqa
where $q \equiv | \vec q \, |$ and the ellipses refer to terms polynomial in $q$. 
Note that $I_3$ does not give rise to any non-polynomial terms. Further, 
I have introduced the loop function $L (q)$ defined as:
\beq
\label{Lq}
L(q) = \frac{1}{q}\sqrt{4 M_\pi^2 + q^2}\, 
\ln\frac{\sqrt{4 M_\pi^2 + q^2}+q}{2M_\pi}~.
\eeq
Using the identity
\beq
\label{nnn3}
\frac{\omega_+^2 + \omega_+ \omega_- + \omega_-^2}{\omega_+^3 \omega_-^3 (\omega_+ + \omega_-)}
= - \frac{1}{2}\, \frac{\partial}{\partial (M_\pi^2) } \,\frac{1}{\omega_+ \omega_- (\omega_+ + \omega_-)} \,,
\eeq
one can express all integrals entering Eq.~(\ref{2pi_prom}) in terms of $I_{1,2,3}$ as follows
\beqa 
\label{nnn4}
\int \, \frac{d^3 l}{(2 \pi)^3} \, \frac{\omega_+^2 + \omega_+ \omega_- + \omega_-^2}
{\omega_+^3 \omega_-^3 (\omega_+ + \omega_-)} \left( l^2 - q^2 \right)^2 &=&
- \frac{1}{2} \,\frac{\partial}{\partial (M_\pi^2) } \,\left( I_2 - 2 q^2 I_1 \right) \,, \\
\int \, \frac{d^3 l}{(2 \pi)^3} \, \frac{\omega_+^2 + \omega_+ \omega_- + \omega_-^2}
{\omega_+^3 \omega_-^3 (\omega_+ + \omega_-)} l_i l_j &=& \frac{1}{4} \frac{\partial}{\partial (M_\pi^2) }
\left\{ \left( - I_1 + \frac{1}{q^2} I_3 \right) \delta_{ij} + \left( \frac{1}{q^2} I_1 - 
\frac{3}{q^4} I_3 \right) q_i q_j \right\} \,. \nonumber
\eeqa
The final result for the $2\pi$-exchange potential $\propto g_A^4$ then takes the form:
\beqa
\label{pot_res}
V_{2N}^{2\pi, \, \rm total} (\vec q \, ) &=&{}
 - \frac{g_A^4}{384 \pi^2 F_\pi^4}\, \fet{\tau}_1 \cdot \fet{\tau}_2\, 
\left( 20 M_\pi^2 + 23 q^2 +  \frac{48 M_\pi^4}{4 M_\pi^2 + q^2} \right) 
L(q) \nn
&&{} - \frac{3 g_A^4}{64 \pi^2 F_\pi^4} \left(
\vec{\sigma}_1 \cdot\vec{q}\,\vec{\sigma}_2\cdot\vec{q} - q^2 \, 
\vec{\sigma}_1 \cdot\vec{\sigma}_2 \right) L(q)
+ \ldots \,.
\eeqa
The polynomial in momenta, divergent (in the limit $\Lambda \to \infty$) terms  
have the form of contact interactions that are anyway present in the potential at a given order 
and can be simply absorbed into an appropriate redefinition of the LECs $C_i$. 

\begin{minipage}{\textwidth}
\vskip 0 true cm
\rule{\textwidth}{.2pt}
{\it
Exercise:  verify the result for the non-polynomial part of $V_{2N}^{2\pi, \, \rm total}$ 
using dimensional regularization. Use  the equality 
\beq
\frac{1}{\omega_+ \omega_- (\omega_+ + \omega_-)} = \frac{2}{\pi} \int_0^\infty d\beta
\frac{1}{\omega_-^2 + \beta^2}\frac{1}{\omega_+^2 + \beta^2}\,,
\eeq
to get rid of the square roots in the integrand. 
The resulting integrals can be dealt with in the usual way by introducing the corresponding Feynman parameters.  
} \\
\vskip -0.8 true cm
\rule{\textwidth}{.2pt}
\end{minipage}

\begin{figure*}
\vspace{0.3cm}
\centerline{
\includegraphics[width=0.7\textwidth]{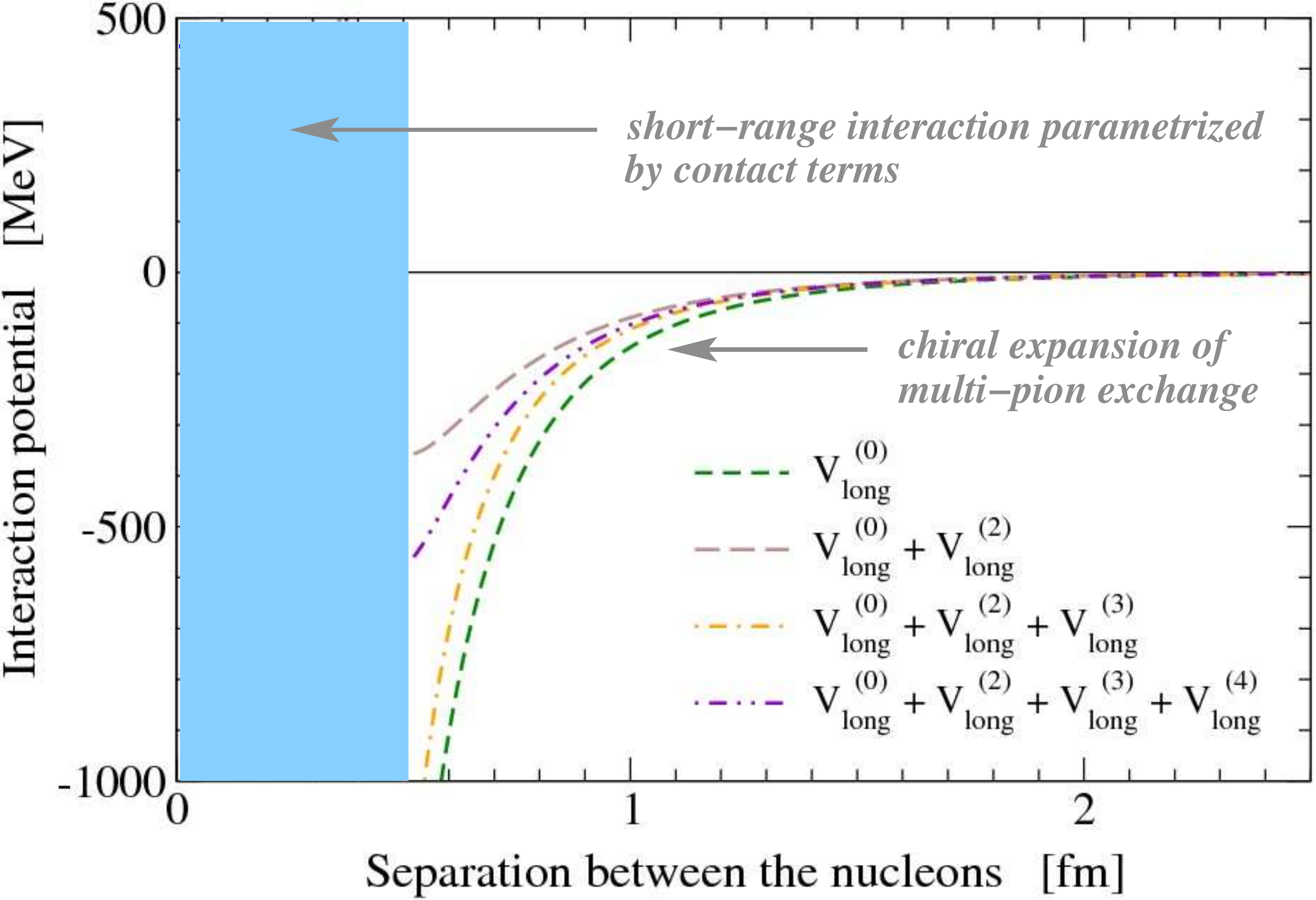}
}
\vspace{-0.2cm}
\caption[fig4aa]{\label{fig:poten} A schematic picture of the two-nucleon potential derived in chiral EFT
in a given partial wave.   
}
\vspace{0.2cm}
\end{figure*}

\medskip
In coordinate space, contact interactions have the form of the delta function 
at the origin, $\delta ( \vec r \, )$, and derivatives thereof.  In contrast,  
the nonpolynomial pieces give rise to the potential at finite distances. To see this 
let us take a closer look at the obtained expression for the $2\pi$-exchange potential. 
First, it should be emphasized that the Fourier transformation of the nonpolynomial 
terms alone is ill defined since they grow as $q$ goes to infinity. The potential 
$V_{2N}^{2\pi, \, \rm total} (\vec r \, )$ at a finite distance, $r \neq 0$, 
can be obtained from $V_{2N}^{2\pi, \, \rm total} (\vec q \, )$ via
\beq
V_{2N}^{2\pi, \, \rm total} (\vec r \, ) = \lim_{\Lambda \to \infty} \int \frac{d^3 q}{(2 \pi )^3 }\, 
e^{-i \vec q \cdot \vec r} \,
V_{2N}^{2\pi, \, \rm total} (\vec q \, ) \, F \left( \frac{q}{\Lambda} \right) \,,
\eeq
where $F \left( q/\Lambda \right)$ is an appropriately chosen regulator function 
such as e.g.~$F = \exp ( -q^2/\Lambda^2 )$. Alternatively, one can use a (twice-subtracted) dispersive representation by
expressing the potential $V_{2N}^{2\pi, \, \rm total} (\vec q \, )$ in terms of a continuous superposition 
of Yukawa functions. For example, the central part of the potential in Eq.~(\ref{pot_res}) can be written as 
\cite{Kaiser:1997mw,Epelbaum:2003gr}
\beq
V_{2N}^{2\pi, \, \rm central} (q ) =  \frac{2 q^4}{\pi} \int_{2 M_\pi}^\infty
d\mu \frac{1}{\mu^3} \frac{\rho (\mu )}{\mu^2 + q^2}\,,
\eeq
where the spectral function $\rho (q)$ is given by 
\beqa
\rho ( \mu ) &=& {\rm Im } \left[ V_{2N}^{2\pi, \, \rm central} ( 0^+ - i \mu ) \right]\nn
 &=& {} 
 - \frac{g_A^4}{768 \pi F_\pi^4}\, 
\left( 20 M_\pi^2 - 23 \mu^2 +  \frac{48 M_\pi^4}{4 M_\pi^2 - \mu^2} \right) 
\frac{\sqrt{\mu^2- 4 M_\pi^2}}{\mu}\,\fet \tau_1 \cdot \fet \tau_2\,.
\eeqa
The Fourier transformation can be easily carried out in this spectral representation by first 
integrating over $\vec q$ and then over the spectrum $\mu$. This leads to the central potential 
of the form
\beq
 V_{2N}^{2\pi, \, \rm central} ( r ) = -\frac{g_A^4 M_\pi}{128 \pi^3 F_\pi^4 r^4} \, \fet \tau_1 \cdot \fet \tau_2 \,
\left[
(23 + 12 x^2 ) K_1 ( 2 x) + x (23 + 4 x^2 ) K_0 ( 2 x)  \right]\,,
\eeq
where $K_i$ denote the modified Bessel functions and $x\equiv M_\pi r$.  
At large distances,
the potential behaves as $\exp (-2 M_\pi r )/r^{3/2}$. The expressions for the remaining components 
of the $2\pi$-exchange potential up to the chiral order $Q^3$, both in momentum and coordinate space, 
can be found in Ref.~\cite{Kaiser:1997mw}. The order-$Q^4$ contributions are given in 
Ref.~\cite{Kaiser:2001pc}. 
The expressions for  
pion exchange potentials derived in chiral EFT at large distances are controlled by low values 
of $\mu$ for which the chiral expansion is expected to converge. At shorter distances, the 
large-$\mu$ components in the spectrum start to contribute which cannot be computed reliably in 
chiral EFT. This is visualized in Fig.~\ref{fig:poten}. 
An extended discussion on the resulting theoretical uncertainty can be found 
in Ref.~\cite{Epelbaum:2003gr}.  

It is instructive to compare the toy models considered in section \ref{toy}
with the nucleon-nucleon potential derived in chiral EFT whose structure is symbolically 
illustrated in Fig.~\ref{fig:poten}.
The main conceptual difference is due to the lack of an exact (regular) expression for the long-range 
force in the realistic case of nucleon-nucleon interaction. Rather, it is represented in terms of the 
chiral expansion of the pion-exchange potential which is valid at large distances and behaves singular 
as $r \to 0$. This raised debates on the relative importance of the long- and short-range 
components in the potential and the most efficient way to organize the expansion for low-energy 
observables. There is little consensus on this issue in the literature (yet).

Chiral $2\pi$-exchange potential is, clearly, the most interesting new ingredient of the two-nucleon force from the 
chiral EFT point  of view: it is the next-longest-range contribution after the well established 
$1\pi$-exchange potential whose form is strongly constrained due to the chiral symmetry of QCD. 
Notice that three-pion exchange is already considerably less important for low-energy nuclear dynamics.  
The evidence of the chiral $2\pi$-exchange potential up to N$^2$LO has been verified in the 
Nijmegen PWA \cite{Rentmeester:1999vw}, see Ref.~\cite{Birse:2003nz} for a similar investigation. 
In their analysis, the Nijmegen group utilized the long-range interaction 
above some distance $b$ as input in order 
to constrain the behavior of high partial waves. The missing intermediate and short-range components 
are simulated by suitably chosen energy-dependent boundary conditions. The number of 
parameters entering the boundary conditions needed to achieve a perfect description 
of the data thus may be viewed as a measure of physics that is missing in the assumed 
long-range force. As demonstrated in Ref.~\cite{Rentmeester:1999vw}, adding the two-pion exchange potential 
derived at N$^2$LO in chiral EFT to the $1\pi$-exchange potential and the appropriate 
electromagnetic interactions allowed for a considerable reduction of parameters (from 31 to 23
for $b = 1.4$ fm in the case of proton-proton scattering) with even a slightly better 
resulting $\chi^2$. 
This is a big success of chiral EFT in the two-nucleon sector.

\section{Summary}
\def\theequation{\arabic{section}.\arabic{equation}}
\label{sec6}

In these lectures, I have outlined the foundations of chiral effective field theory  
and the application of this theoretical framework to the nuclear force problem. 
The method allows for a systematic derivation of nuclear forces with a direct connection 
to QCD via its symmetries. These lecture notes are mainly focused on the conceptual aspects and
do not cover applications of the novel chiral potentials to the few-nucleon problem 
and various related topics such as e.g.~isospin breaking effects, few-baryon systems with 
strangeness, electroweak and pionic probes in the nuclear environment, nuclear parity violation
and chiral extrapolations of few-baryon observables. For a discussion on these and other  
topics as well as for a detailed description of the structure of the two-, three- and four-nucleon 
forces in chiral EFT the reader is referred to recent review articles \cite{Epelbaum:2005pn,Epelbaum:2008ga},
see also \cite{Bedaque:2002mn}. There are many frontiers where future 
work is required. These include a better understanding of 
the power counting in the few-nucleon sector,
the consistent inclusion of electroweak currents, and the development
of chiral EFT with explicit $\Delta$(1232) degrees of freedom.   

\section*{Acknowledgments}
It is a great pleasure to thank Fa{\"i}{\c c}al AZAIEZ and other organizers of the 
International Joliot-Curie School 2009 for the superb organization and the pleasant 
atmosphere at the meeting. I also thank all my collaborators for sharing their insights into
the topics discussed here. Special thanks are due to Hermann Krebs and Ulf-G.~Mei{\ss}ner 
for a careful reading of the manuscript and their helpful comments. 
This work  was supported by funds provided from the Helmholtz Association 
to the young investigator group ``Few-Nucleon Systems in 
Chiral Effective Field Theory'' (grant  VH-NG-222), by the DFG (SFB/TR 16 ``Subnuclear Structure
of Matter''), 
and by the EU Integrated Infrastructure Initiative Hadron
Physics Project under contract number RII3-CT-2004-506078.

\setlength{\bibsep}{0.2em}
\bibliographystyle{h-physrev3}
\bibliography{/home/epelbaum/refs_h-elsevier3}

\end{document}